\documentclass[10pt]{iopart}
\usepackage{graphicx}
\usepackage{iopams,color,hyperref}
\usepackage[parfill]{parskip}
\usepackage{titlesec}

\titleformat{\subsubsection}
  {\normalfont\normalsize\itshape}{\thesubsubsection}{1em}{}
\titlespacing*{\subsubsection}{0pt}{3.25ex plus 1ex minus .2ex}{0ex plus .2ex}

\newcommand{\edd}{\varepsilon_{\rm dd}}
\newcommand{\bfrho}{\boldsymbol{\rho}}
\def\gappeq{\mathrel{ \rlap{\raise.5ex\hbox{$>$}}
                      {\lower.5ex\hbox{$\sim$}}  } }
\def\lappeq{\mathrel{ \rlap{\raise.5ex\hbox{$<$}}
                      {\lower.5ex\hbox{$\sim$}}  } }

\begin{document}

\topical{Vortices and vortex lattices in quantum ferrofluids}

\author{A. M. Martin$^1$, N. G. Marchant$^1$, D. H. J. O'Dell$^2$ and N. G. Parker$^{3,1}$} 
\address{$^1$ School of Physics, University of Melbourne, Victoria 3010, Australia}
\address{$^2$ Department of Physics and Astronomy, McMaster University, Hamilton, Ontario, L8S 4M1, Canada}
\address{$^3$ Joint Quantum Centre Durham--Newcastle, School of Mathematics and Statistics, Newcastle University, Newcastle upon Tyne, NE1 7RU, United Kingdom.}

\ead{martinam@unimelb.edu.au}

\begin{abstract}
The experimental realization of quantum-degenerate Bose gases made of atoms with sizeable magnetic dipole moments has created a new type of fluid, known as a quantum ferrofluid, which combines the extraordinary properties of superfluidity and ferrofluidity.  A hallmark of superfluids is that they are constrained to rotate through vortices with quantized circulation.  In quantum ferrofluids the long-range dipolar interactions add new ingredients by inducing magnetostriction and instabilities, and also affect the structural properties of vortices and vortex lattices. Here we give a review of the theory of vortices in dipolar Bose-Einstein condensates, exploring the interplay of magnetism with vorticity and contrasting this with the established behaviour in non-dipolar condensates.  We cover single vortex solutions, including structure, energy and stability, vortex pairs, including interactions and dynamics, and also vortex lattices. Our discussion is founded on the mean-field theory provided by the dipolar Gross-Pitaevskii equation, ranging from analytic treatments based on the Thomas-Fermi (hydrodynamic) and variational approaches to full numerical simulations.    Routes for generating vortices in dipolar condensates are discussed, with particular attention paid to rotating condensates, where surface instabilities drive the nucleation of vortices, and lead to the emergence of rich and varied vortex lattice structures.  We also present an outlook, including potential extensions to degenerate Fermi gases, quantum Hall physics, toroidal systems and the Berezinskii-Kosterlitz-Thouless transition.
\end{abstract}

\maketitle
\ioptwocol
\tableofcontents

\section{Introduction}
Ferrohydrodynamics  describes the motion of fluids comprised of particles with significant magnetic (or electric) dipole moments \cite{Rosensweig97,Hakim62}. The dipole-dipole (from henceforth: dipolar) inter-particle interaction causes magnetostriction (or electrostriction) and gives rise to spectacular instabilities such as the normal field instability  \cite{Cowley67} that can lead to complex pattern formation. Classical ferrofluids have been investigated since the 1960s,  the first ferrofluid having been invented at NASA with the intention of making a jet fuel whose flow could be directed in a zero-gravity environment using a magnetic field \cite{Papell66}.  They have subsequently  found a broad range of applications reaching from liquid seals around rotating shafts (such as in hard disks), where the fluid is held in place using magnets, to magnetically targeted drugs in medicine \cite{Alexiou02}.  However, the focus of this review is upon quantum ferrofluids as first realized in 2005 with the creation of a Bose-Einstein condensate (BEC) in a vapour of $^{52}$Cr atoms by the Stuttgart group \cite{Griesmaier05}. These atoms have a magnetic dipole moment of $6 \mu_{\mathrm{B}}$, six times larger than that found in the alkalis which are used in the majority of BEC experiments.  $^{52}$Cr atoms therefore have dipolar interactions which are 36 times larger than in standard BECs.  Other groups have also studied $^{52}$Cr BECs \cite{Beaufils08,Pasquiou11}, as well as BECs made of atoms with even larger magnetic dipoles such as $^{164}$Dy \cite{Lu11,Kadau16}, and $^{168}$Er \cite{Aikawa12}. Many of the signatures of ferrohydrodynamic behaviour have now been observed in these gases, including magnetostriction \cite{Lahaye07}\footnote{This paper coined the term ``quantum ferrofluid".}, collapse due to dipolar interactions \cite{Lahaye08,Koch08}, and the quantum analog of the Rosensweig instability \cite{Kadau16,Ferrier16}.  Recently, fully self-bound dipolar droplets have been reported \cite{chomaz_2016,schmitt_2016}.  Additionally, the production of ultracold fermionic $^{40}$K--$^{87}$Rb \cite{Ni08} polar molecules and the cooling of fermonic $^{161}$Dy \cite{Lu12} and  $^{167}$Er \cite{Aikawa14}, all with significant dipole moments, pave the way for a new generation of quantum degenerate Fermi gas experiments, where dipolar interactions dominate. In Fermi gas systems the partially attractive nature of the dipolar interaction opens up the possibility of Bardeen-Cooper-Schrieffer (BCS) pairing at sufficiently low temperatures \cite{You00,Baranov02,Baranov04,Baranov04a,Brunn08,Cooper09,Fregoso09,Zhao10,Levinson11}.  Excellent reviews of the field of ultracold dipolar gases can be found in Refs. \cite{Baranov02a,Baranov08,Lahaye09,Carr09,Baranov12}.   

Vortical structures have been generated experimentally in non-dipolar condensates in the form of single vortices \cite{Madison00,Freilich10}, vortex-antivortex pairs \cite{Neely10,Kwon15}, vortex rings \cite{Anderson01} and vortex lattices \cite{Hodby01,Aboshaeer01}, as well as disordered vortex distributions characteristic of quantum turbulence \cite{Henn09,Neely13,Kwon14}.  These excitations underpin a variety of rich phenomena, including vortex lattices, quantum turbulence, the Berezinskii-Kosterlitz-Thouless transition and Kibble-Zurek defect formation. In geometries approaching the one-dimensional limit, so-called solitonic vortices have been formed \cite{Becker13,Donadello14} which share properties between vortices and their one-dimensional analogs: dark solitons.  Several reviews exist which summarise the significant experimental and theoretical aspects of vortices and vortex lattices in non-dipolar BECs \cite{Fetter01,Kevrekidis08,Cooper08,Kasamatsu09,Fetter09,Fetter10}.  Vortices have yet to be observed in quantum ferrofluids, although numerical simulations suggest the formation of vortex rings in the dipolar collapse experiment of Ref. \cite{Lahaye08}, and the formation of vortex-antivortex pairs \cite{Bisset15} in the droplet experiment of Ref. \cite{Kadau16}.   

Here we establish the properties of vortices and vortex lattices in quantum ferrofluids, reviewing the theoretical progress that has been made over the last decade.  Whilst it is possible to also consider the properties of vortices and vortex lattices in dipolar Fermi gases, this review is confined to the bosonic case and only a brief discussion of fermionic systems will be given in the Summary and Outlook (Section \ref{sec:summary}).  The structure of this review is built upon the philosophy of taking the reader on a journey. This journey starts in Section \ref{sec:Clas_Ferro} where the properties of classical and strongly correlated quantum ferrofluids are briefly discussed. Sections \ref{sec:Quant_Ferro} and \ref{sec:vortex_free}  provide a brief introduction to the properties of dipolar BECs in the absence of vortices.  In Section \ref{sec:Quant_Ferro} we examine the mathematical form of the dipolar interaction in quantum ferrofluids, and present the most widely used model for quantum ferrofluids - the  dipolar Gross-Pitaevskii equation (GPE) - along with its hydrodynamical interpretation.  Section \ref{sec:vortex_free} builds on this theory to consider the stability of dipolar BECs. Specifically, we look at  stability in the Thomas-Fermi (hydrodynamic) regime, where interactions dominate, and more general dipolar GPE solutions, in three-dimensional and quasi-two-dimensional systems. Section  \ref{sec:single_vortices} focuses on the properties of single vortex lines in three-dimensional condensates and single vortices in quasi-two-dimensional systems. In Section   \ref{sec:dynamics} we consider vortex-vortex and vortex-antivortex dynamics in quasi-two-dimensional dipolar BECs, primarily focusing on solutions of the dipolar GPE. Section \ref{sec:gen_vortices} addresses the routes to vortex and vortex lattice formation. This focuses primarily on stationary solutions (in the rotating frame) and their dynamical stability, enabling us to ascertain under what conditions it might be expected that vortices will nucleate into the dipolar BEC. Section \ref{sec:lattices} analyses how dipolar interactions can induce changes to vortex lattice structures. This revisits previous work and presents some new variational calculations that elucidate the properties of vortex lattice structures in dipolar condensates. In Section \ref{sec:summary} we give a brief summary and provide an outlook to several topical aspects for future development which are not covered in the main body of the review. Prospects for quantum turbulence with quantum ferrofluids, which are not covered here, are discussed elsewhere \cite{Storm}.

\section{Classical Ferrofluids and Strongly Correlated Quantum Ferrofluids}
\label{sec:Clas_Ferro}
A classical ferrofluid can be formed from a suspension of small permanently magnetized particles, with a typical size of $0.01$ $\mu$m, in a non-magnetic solvent \cite{Rosensweig97}. Their most closely related electrical counterparts are known as electrorheological fluids which are suspensions of electrically polarizable particles, typically  $1-100$  $\mu$m in size, in an insulating solvent \cite{Winslow49,Gast89,Halsey92} (there are also magnetorheological fluids where small micelles of magnetizable fluid are suspended in a nonmagnetizable fluid \cite{Shulman82,Lemaire92}.)  

In the presence of an external field, a classical ferrofluid can form a zoo of different patterns including hexagonal cells \cite{Cowley67}, columns \cite{Halsey90},  stripe and bubble phases \cite{Halsey93}, and disordered stripe phases  producing labyrinthine structures \cite{Rosensweig97}. Some of these patterns also occur in quantum ferrofluids: stripe phases (density wave modulations) will be discussed in the quantum case in the absence of vortices in Section \ref{subsubsec:outsideTFregime} and in the  presence of vortices beginning in Section \ref{subsec:structureofvortex}. Vortex lattices in quantum fluids tend to form hexagonal patterns (Abrikosov lattices) even in the absence of dipolar interactions where the long-range logarithmic hydrodynamic interaction between vortices plays an important role. However, as we shall see in Sections \ref{sec:lattice_dipoles_aligned} and \ref{sec:lattice_dipoles_nonaligned}, the presence of dipolar interactions can change the lattice configuration to square and bubble geometries. 

The above patterns can all be captured to some degree by mean-field theory where no attempt is made to describe the fluid at the molecular level. However, it has long been predicted that with strong dipolar interactions ferrofluid molecules can form chains and rings \cite{Jacobs55,deGennes70,Clarke94}. In order to describe such strong correlation effects, which lie beyond the mean-field description, it is necessary to use computationally intensive Monte Carlo and Molecular Dynamics techniques \cite{Camp00,Teixeira00,Klapp05,Holm05}. These methods suggest a complex and rich phase diagram. In zero applied field at relatively low temperatures,  and as the density is increased from low to high, the molecules initially form rings which unbind into chains which then cluster into networks, which break down into a normal liquid, then form a ferroelectrically ordered liquid, followed by a possible ferroelectric columnar ordering, and finally form a ferroelectric solid. In the presence of an external field chains, columns, sheets, bent walls, lamellar, labyrinthine or worm-like structures, and hexagonal structures all appear \cite{Holm05}. Transitions from single chains to double chains (zig-zag) can also occur if initially strong transverse confinement is reduced \cite{Cartarius14}.

In the rest of this review we restrict ourselves to mean-field phenomena. It is important to mention, however, that there is a sizeable body of theoretical work in strongly correlated quantum dipolar systems in two dimensions. Strongly correlated dipolar gases do not yet exist in the laboratory but ideas to realize them include using ultracold molecules with very large dipole moments dressed by microwaves \cite{Buchler07} and Rydberg dressed ultracold gases \cite{Henkel10}. One of the main interests in these systems is the formation of so-called supersolids which are both crystalline and yet also have superfluid properties
\cite{Buchler07,Astrakharchik07,Cinti2010,Macia12,Macia14,Lu15}. Unlike the density wave structures we shall study later in this review, which have many atoms per wavelength, in the strongly correlated case the periodicity can be at the single atom or few atom length scale. Liquid crystal phases have also been identified \cite{Wu16}.

\section{Quantum Ferrofluids: Theory and Basic Properties \label{sec:Quant_Ferro}}
\begin{figure}[t]
\centering
	\includegraphics[width=0.4\textwidth,angle=0]{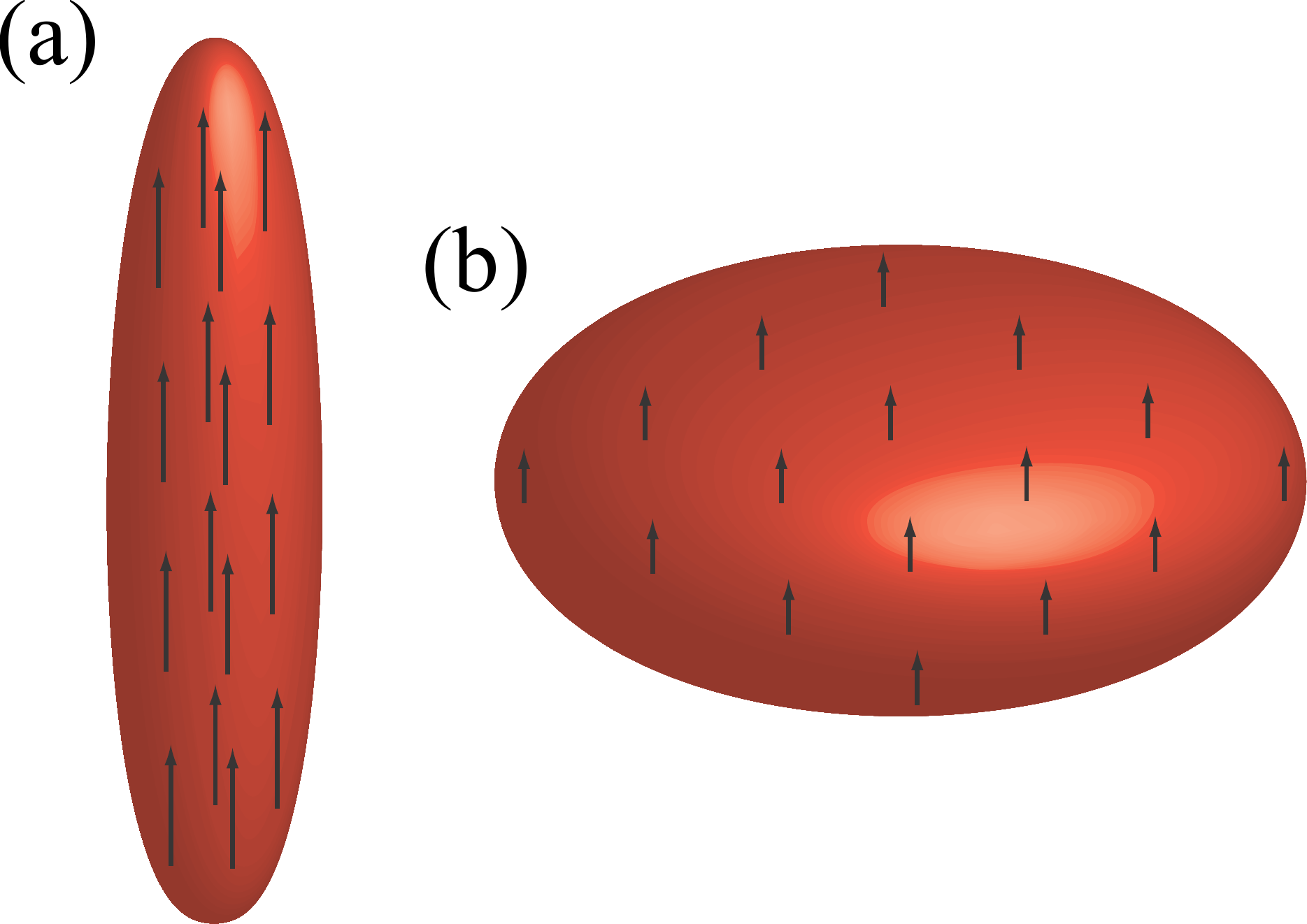}
		\caption{(a) For a prolate trapped condensate the net dipolar interaction is attractive. (b) For an oblate system the net dipolar interaction is repulsive.}
		\label{fig:shape}
\end{figure}

The successful Bose-Einstein condensation of gases of $^{52}$Cr atoms \cite{Griesmaier05,Beaufils08},  $^{164}$Dy \cite{Lu11,Kadau16} and $^{168}$Er \cite{Aikawa12} have realized BECs with significant dipolar interactions.  A basic property of these interactions is that their net effect depends on the \textit{shape} of the BEC, as illustrated in Figure~\ref{fig:shape}. For dipoles polarised along the long axis of a prolate (elongated) dipolar gas [Figure~\ref{fig:shape}(a)] the net contribution to the dipolar interaction is attractive. By contrast, for dipoles polarised along the the short axis of an oblate (flattened) dipolar gas  [Figure~\ref{fig:shape}(b)] the net contribution to the dipolar interaction is repulsive. Compared to BECs with interactions which are dominated by isotropic  {\it s}-wave scattering, a dipolar BEC will be elongated along the direction of the polarising field (magnetostriction) \cite{Santos00,Yi01,Yi02}. 

\subsection{The dipolar interaction}
\begin{figure}[t]
\centering
	\includegraphics[width=0.4\textwidth,angle=0]{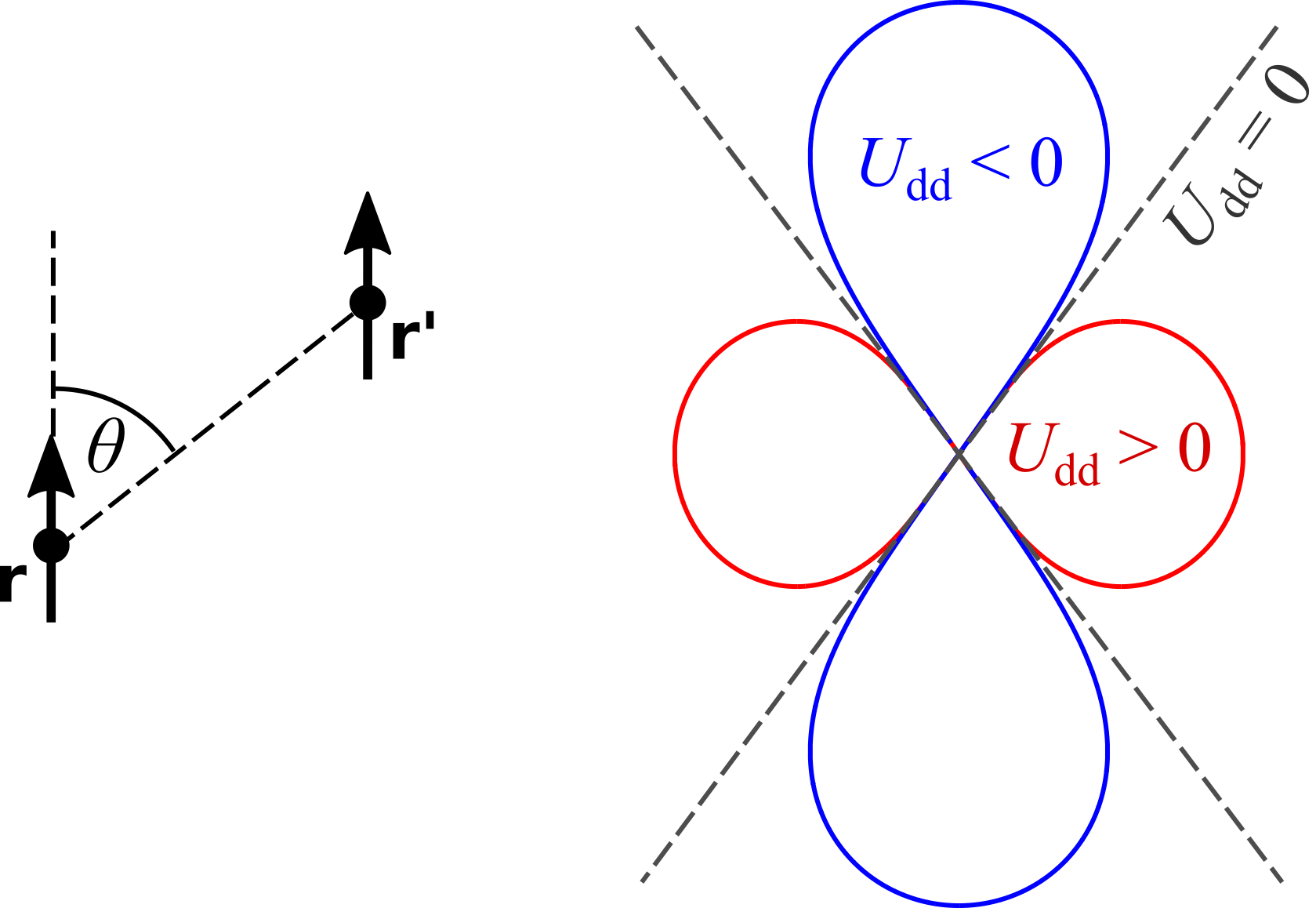}
		\caption{Illustration of the dipole-dipole interaction.  The magic angle, at which the dipole-dipole interaction reduces to zero, is indicated by the black dashed line.}
		\label{fig:dipolar_pot}
\end{figure}

At low energies, and far from any two-body bound states, the interatomic interactions can be described by an effective pseudo-potential which is the sum of a contact term originating from the van der Waals interactions and a bare dipolar term
\cite{Lahaye09,Yi01},
\begin{eqnarray}
U(\mathbf{r-r'})&=&U_{\rm vdW}(\mathbf{r-r'}) + U_{\rm dd}(\mathbf{r-r'}) \nonumber
\\
&=& g \delta(\mathbf{r}-\mathbf{r'}) + \frac{C_{\rm dd}}{4\pi}\frac{1-3\cos^{2}\theta}{|\mathbf{r-r'}|^3}, 
\label{eqn:atomic_interaction}
\end{eqnarray}
where we have assumed the dipoles are polarised by an external field with $\theta$ being the angle between the polarisation direction and the inter-atom vector  ${\bf r}-{\bf r'}$. The short-range interactions are characterized by the coupling constant $g=4 \pi \hbar^2 a_s/m$, where $a_s$ is the {\it s}-wave scattering length and $m$ is the mass of the atoms. The strength of the dipolar interactions is set by $C_{\rm dd}$. For magnetic dipoles $C_{\rm dd}=\mu_0 d^2$ \cite{Goral00}, where $\mu_0$ is the permeability of free space and $d$ is the magnetic dipole moment of the atoms. Equation~(\ref{eqn:atomic_interaction}) also holds for electric dipoles induced by a static electric field ${\bf E}=\hat{\mathbf{k}}E$, for which the coupling constant is $C_{\rm dd}=E^2 \alpha_{\rm p}^2/\epsilon_0$ \cite{Yi00,You98}, where $\alpha_{\rm p}$ is the static polarizability and $\epsilon_0$ is the permittivity of free space. The formation and cooling to degeneracy of polar molecules, with significant electric dipole moments, is proving to be challenging. However, progress is ongoing with $^{40}$K$^{87}$Rb \cite{Ospelkaus09} and $^{133}$Cs$^{87}$Rb \cite{Molony14}, which are expected to have a significant coupling constant for electric fields on the order of several hundred V/cm.  The dipolar interaction $U_{\rm dd}$, illustrated in Figure~\ref{fig:dipolar_pot}, is negative for $\theta=0$, representing the attraction of head-to-tail dipoles, and positive for $\theta=\pi/2$, representing the repulsion of side-by-side dipoles.  At the ``magic angle'', $\theta_{\rm m}=\arccos(1/\sqrt{3})\approx 54.7^\circ$, the dipolar interaction is  zero.

It is often convenient to work in momentum space.  The Fourier transform $\tilde{U}_{\rm dd}({\bf k})=\mathcal{F}[U_{\rm dd}]=\int U_{\rm dd}({\bf r}) e^{-i {\bf k}\cdot {\bf r}}{\rm d}{\bf r}$ of the dipolar interaction is \cite{Goral00},
\begin{equation}
\tilde{U}_{\rm dd}({\bf k})=C_{\rm dd} \left(\cos^2 \alpha - \frac{1}{3} \right),
\end{equation}
where $\alpha$ is the angle between ${\bf k}$ and the polarization direction.

The strength of the dipolar interactions is conveniently parameterized by the ratio \cite{Lahaye09},
 \begin{equation}
\edd=\frac{C_{\mathrm{dd}}}{3g},  
 \end{equation}
where $g$ can be tuned between $-\infty$ and $+\infty$ via a Feshbach resonance \cite{Feshbach58,Werner05}.  In effect $\edd$ gives the relative importance of the anisotropic, long-range dipole-dipole interactions to the isotropic, short-range van der Waals interactions.  It is defined with a factor of $3$ in the denominator so that the homogeneous dipolar condensate is unstable when $\edd>1$ - see Section \ref{sec:3d_stability}.  For $^{52}$Cr, $^{168}$Er and $^{164}$Dy, the natural value of $\edd$ is $0.16$ \cite{Griesmaier06}, $0.4$ \cite{Aikawa12} and $1.45$ \cite{Lu11,Tang15}, respectively.

While $C_{\rm dd}$ is conventionally positive and set to the natural value of the given atom, it is predicted to be possible to reduce $C_{\rm dd}$ below its natural value, including to negative values, by tilting the polarization direction off-axis and rotating it rapidly \cite{Giovanazzi02}.  Hence it is feasible to consider $-\infty < \varepsilon_{\rm dd} < \infty$, with both negative and positive $C_{\rm dd}$.  Note that for $C_{\rm dd}<0$ the dipole-dipole interaction becomes repulsive for head-to-tail dipoles and attractive for side-by-side dipoles. 

\subsection{The dipolar Gross-Pitaevskii equation}
In the mean-field limit, at zero temperature, a single wavefunction, $\Psi({\bf r},t)$, can be used to describe the condensate. The condensate density $n({\bf r},t)=|\Psi({\bf r},t)|^2$ is normalized such that
\begin{equation}
N=\int |\Psi|^2\,{\rm d} {\bf r},
\end{equation}
where $N$ is the number of atoms in the condensate. The wavefunction obeys the dipolar GPE \cite{Santos00,Goral00,Yi00},
\begin{equation}
i \hbar \frac{\partial \Psi}{\partial t}=\left [-\frac{\hbar^2}{2m}\nabla^2+V+g |\Psi|^2+\Phi \right ]\Psi,
\label{eqn:dgpe1}
\end{equation}
where $V = V({\bf r})$ is the external potential acting on the condensate (which in principle may also be time-dependent, but here we consider it to be static). The local term, $g|\Psi|^2$, arises from the van der Waals interactions and the non-local  term, $\Phi$, arises from the dipolar interactions \cite{Yi01}:
\begin{equation}
\Phi({\bf r},t)=\int U_{\rm dd}({\bf r}-{\bf r}')n({\bf r'},t)\,{\rm d}{\bf r}'.\label{dipole_a}
\end{equation}
If we take the dipoles to be polarized along the $z$-direction, then using identities from potential theory the dipolar potential can be expressed as \cite{ODell04,Eberlein05},
\begin{equation}
\Phi ({\bf r},t) = -3g \edd
\left(\frac{\partial^2}{\partial z^2} \phi({\bf r},t)
+\frac{1}{3} n({\bf r},t) \right), \label{eq:Phidd}
\label{dipole}
\end{equation}
where $\phi$ is a fictitious `electrostatic' potential defined as 
\begin{equation}
\phi({\bf r},t) = \frac{1}{4 \pi} \int \frac{
n({\bf r,t}^{\prime})}{\left|
{\bf r}-{\bf r}^{\prime} \right|}\,{\rm d} {\bf r^{\prime}}. \label{potential}
\end{equation}
This effectively reduces the problem of calculating the dipolar potential  $\Phi$ to one of calculating an electrostatic potential of the form (\ref{potential}) which is easier to compute because the Green's function $1/\left|{\bf r}-{\bf r}^{\prime} \right|$ has no angular dependence. Furthermore, hundreds of years of literature exists providing analytic methods for solving electrostatic and gravitational problems with this form of interaction \cite{Ferrers,Dyson,Routh,LevinMuratov}. Alternatively, $\Phi$ can  be evaluated in momentum space by exploiting the convolution theorem,
\begin{equation}
\Phi({\bf r},t)=\mathcal{F}^{-1} \left[\tilde{U}_{\rm dd}({\bf k}) \tilde{n}({\bf k},t)\right],   
\label{eqn:conv}
\end{equation}
where $\tilde{n}({\bf k},t)=\mathcal{F}[n({\bf r},t)]$.

In condensate experiments the external potential $V$ is typically harmonic with the general form,
\begin{equation}
V({\bf r})=\frac{1}{2}m \left(\omega_x^2 x^2 + \omega_y^2 y^2 + \omega_z^2 z^2 \right),
\end{equation}
where $\omega_j$ ($j=x, y, z$) are the trap's angular frequencies. In general, this gives rise to three harmonic oscillator lengths, $\ell_j=\sqrt{\hbar/m \omega_j}$, which are the characteristic length scales imposed by the trap on the wavefunction in the three different directions.   However, cylindrically-symmetric traps are common, defined as,
\begin{equation}
V({\bf r})=\frac{1}{2}m \left(\omega_\perp^2 \rho^2 + \omega_z^2 z^2 \right) = \frac{1}{2}m\omega_\perp^2\left(\rho^2+\gamma^2 z^2 \right),
\label{eqn:axisym_trap}
\end{equation}
where $\gamma=\omega_z/\omega_\perp$ is the so-called {\it trap ratio}.  When $\gamma \gg
1$ the BEC shape will typically be oblate (flattened) while for
$\gamma \ll 1$ it will typically be prolate (elongated). 

Time-independent solutions of the GPE satisfy,
\begin{equation}
\Psi({\bf r},t)=\psi({\bf r})e^{-i\mu t/\hbar},
\label{eqn:time-indep}
\end{equation}
where $\mu$ is the chemical potential\footnote{Throughout this review $\Psi$ ($\psi$) denotes the time-(in)dependent condensate wavefunction.}.  Inserting this into Eq.~(\ref{eqn:dgpe1}), the time-independent dipolar GPE for the time-independent wavefunction $\psi({\bf r})$ is
\begin{equation}
\mu \psi=-\frac{\hbar^2}{2m}\nabla^2 \psi + V\psi + g |\psi|^2 \psi + \Phi\psi.
\label{eqn:tidgp1}
\end{equation}
Solutions of the time-independent GPE are stationary solutions of the system, and the lowest energy of these is the ground state.  

The energy of the condensate is given by,
\begin{eqnarray}
E&=&\int \left[ \frac{\hbar^2}{2m}| \nabla \Psi |^2+V|\Psi|^2
+\frac{g}{2}| \Psi |^4 +\frac{\Phi}{2}|\Psi|^2 \right]\, {\rm d} {\bf r}  \nonumber
\\
&=& E_{\rm kin}+E_{\rm pot}+E_{\rm vdW}+E_{\rm dd}.
\label{eqn:energy}
\end{eqnarray}
The terms represent (from left to right)  kinetic energy $E_{\rm kin}$, potential energy $E_{\rm pot}$, the van der Waals interaction energy $E_{\rm vdW}$ and the dipolar interaction energy  $E_{\rm dd}$.  Provided that the potential $V$ is independent of time, then the total energy $E$ is conserved during the time evolution of the GPE.  

Comparing the relative size of the kinetic term and the net interaction term in the dipolar GPE in an untrapped ($V=0$) system defines a length scale termed the {\it healing length},
\begin{equation}
\xi= \frac{\hbar}{\sqrt{m \mu}},
\end{equation}
which may be interpreted as the minimum length-scale over which the wavefunction changes appreciably.

Efficient numerical methods for solving the dipolar GPE are available \cite{Bao10,Jiang14,Kumar15,Bao16,Loncar16} and progress has been made on extending this treatment to include finite temperature effects and quantum fluctuations  \cite{Ronen07a,Blakie09,Bisset12,Lima12}.

\subsection{Dipolar hydrodynamic equations \label{sec:hydro_eqs}}
There is a deep link between the GPE and fluid dynamics. 
Indeed, the condensate can be thought of as a fluid, characterised by its density and velocity distributions.
This is revealed by writing the condensate wavefunction in the Madelung form $\Psi({\bf r},t)=\sqrt{n({\bf r},t)} e^{i S({\bf r},t)}$, where  the local phase, $S({\bf r},t)$, defines the fluid velocity field ${\bf v}({\bf r},t)$,
\begin{equation}
{\bf v}({\bf r},t)=\frac{\hbar}{m} \boldsymbol{\nabla} S({\bf r},t).
\label{eqn:vel}
\end{equation}
Inserting the Madelung form into the GPE, and separating real and imaginary terms, yields two equations which together are exactly equivalent to the GPE. The first is the {\em continuity equation},
\begin{equation}
\frac{\partial n}{\partial t} + \boldsymbol{\nabla} \cdot(n {\bf v})=0.
\label{eq:continuity1}
\end{equation}
This embodies the conservation of the number of atoms.  The second equation is
\begin{eqnarray}
\displaystyle m \frac{\partial {\bf v}}{\partial t} =- \boldsymbol{\nabla} \left(\frac{1}{2}mv^2+ V+ gn +\Phi - \frac{\hbar^2}{2m} \frac{\nabla^2 \sqrt{n}}{\sqrt{n}}\right). \nonumber \\
 \label{eqn:quasiEuler2}
\end{eqnarray}
The $\nabla^2 \sqrt{n} / \sqrt{n}$-term is the {\it quantum pressure}, arising from the zero-point kinetic energy of the atoms. It can be dropped when the interactions and external potential dominate the zero-point motion, leading to the \textit{Thomas-Fermi approximation}.  In this regime, and in the absence of dipolar interactions,  Eqs.~(\ref{eq:continuity1}) and (\ref{eqn:quasiEuler2}) resemble the continuity and Euler hydrodynamical equations for inviscid fluids. As such  they are often referred to as the superfluid hydrodynamic equations 
\cite{Stringari96,Pines&NozieresBook,Pethick&Smithbook,Pitaevskii&StringariBook,Barenghi&ParkerBook}.  Equations (\ref{eq:continuity1}) and (\ref{eqn:quasiEuler2}) have been extended to include dipolar interactions and are referred to as the dipolar superfluid hydrodynamic equations.

\section{Vortex-Free Solutions and Stability \label{sec:vortex_free}}
Before discussing vortices, we next describe the solutions and stability of the dipolar condensates themselves, in homogeneous and trapped systems, and introduce some key analytical tools and physical concepts.  

\subsection{Homogeneous condensate}
\subsubsection{Three-dimensional case \label{sec:3d_stability}}
For $V({\bf r})=0$ (uniform condensate of infinite extent), the stationary solution is,
\begin{equation}
\psi=\sqrt{n_0}, \qquad \mu=n_0 g  \left(1- \varepsilon_{\mathrm{dd}} \right),
\label{eqn:homog_solution}
\end{equation}
i.e. a state of uniform density $n_0$. The two contributions to the chemical potential $\mu$ are the uniform mean-field potentials generated by the van der Waals and the dipolar interactions, respectively.  In the absence of dipolar interactions ($\edd=0$), the corresponding solution has chemical potential $\mu=n_0 g$. By comparison, the homogeneous dipolar system is akin to a non-dipolar system but with an effective coupling
\begin{equation}
g_{\rm eff} = g \left( 1 - \varepsilon_{\mathrm{dd}} \right).
\label{eq:g_eff}
\end{equation}

For a three-dimensional homogeneous dipolar condensate, the Bogoliubov dispersion relation between the energy $E_{\rm B}$ and momentum $\mathbf{p}$ of a perturbation is given by,
\begin{eqnarray}
E_{\rm B}(\mathbf{p})&=&\sqrt{c^2(\theta) \, p^2 + \left(\frac{p^2}{2m} \right)^2},
\label{eqn:bog}
\end{eqnarray}
where $c(\theta)$ is the speed of sound, 
\begin{equation}
c(\theta)=\sqrt{\frac{gn_0}{m} \left[ 1+ \varepsilon_{\mathrm{dd}}\left(3 \cos^{2} \theta-1 \right) \right]}.
\end{equation}
The angle $\theta$ is that between the excitation momentum $\mathbf{p}$ and the polarization direction.  For low momenta the spectrum is linear $E_{\rm B}\approx c(\theta) \, p$ and characteristic of phonons with a phase velocity  $c(\theta)$ that depends on direction.  For higher momenta the relation becomes quadratic in $p$ which is characteristic of free-particle excitations.  

The amplitude of a mode specified by momentum $\mathbf{p}$ evolves in time as $\exp(-i E_{\rm B}(\mathbf{p}) t/\hbar)$. If $E_{\rm B}(\mathbf{p})$ should become imaginary the relevant amplitude grows exponentially, signifying a dynamical instability. In the case of the three-dimensional homogeneous dipolar BEC considered in this Section, with $E_{\rm B}(\mathbf{p})$ provided by Eq.~(\ref{eqn:bog}), such an instability arises for small $p$, i.e.\ long wavelengths, and this is known as the {\em phonon instability}, familiar from non-dipolar attractive ($g<0$) condensates \cite{Pethick&Smithbook}.  Examining the parameter space over which Eq.\ (\ref{eqn:bog}) is real-valued indicates that the three-dimensional homogeneous system is stable against the phonon instability in the range $-0.5 \leq \edd \leq 1$ for $g>0$, and $\edd \leq -0.5, \edd > 1$ for $g<0$.

\subsection{Trapped dipolar condensates}
A full theoretical treatment of a trapped BEC involves solving the dipolar GPE, given in Eq.\ (\ref{eqn:dgpe1}) \cite{Pethick&Smithbook,Pitaevskii&StringariBook}. The non-local nature of the mean-field potential describing dipolar interactions means that this task is more challenging than for purely {\it s}-wave BECs.  Moreover, the stability of the condensate becomes non-trivial, becoming dependent on the geometry of the trap and the number of atoms (in addition to the dipole strength). Additionally, a dipolar condensate can suffer from a density-wave instability associated with a novel type of excitation called a roton in analogy with a similar type of excitation in superfluid helium \cite{ODell03,Santos03,Giovanazzi04}.  To characterise the stability of a dipolar condensate we first derive and examine  the Thomas-Fermi ground state solutions for a dipolar BEC.

\subsubsection{Thomas-Fermi solutions}
The problem of finding the ground state solution (as well as low-energy dynamics) is greatly simplified by making use of the Thomas-Fermi  approximation, whereby density gradients in the GPE (or, equivalently, the hydrodynamic equations) are ignored, allowing  analytic solutions \cite{Stringari96}.  For a non-dipolar condensate, with repulsive van der Waals interactions, this is valid for $N a_s/ \bar{\ell}\gg 1$, where $\bar{\ell}=(\ell_x \ell_y \ell_z)^{1/3}$ is the geometric mean of the harmonic oscillator lengths \cite{Pethick&Smithbook,Pitaevskii&StringariBook}. This regime is relevant to many experiments.  

In the dipolar case, the Thomas-Fermi approximation is valid when the \emph{net} interactions are repulsive and the number of atoms is large; rigorous criteria have been established for certain geometries in Ref. \cite{Parker08}. Although the governing equations for a dipolar BEC contain the non-local potential $\Phi(\mathbf{r})$, exact solutions known from the pure $s$-wave case hold, in modified form, in the dipolar case too \cite{ODell04, Eberlein05}, and we make extensive use of them throughout this review.   

Consider a dipolar condensate polarized in the $z$-direction, with repulsive van der Waals interactions ($a_s>0$), and confined by a cylindrically-symmetric trap of the form of Eq.~(\ref{eqn:axisym_trap}).  We limit the analysis to the regime of $-0.5 \leq \edd \leq 1$, where the Thomas-Fermi approach predicts that stationary solutions are stable \cite{ODell04}. Outside of this regime the condensate becomes prone to collapse \cite{Parker09,Ticknor08}. Under the Thomas-Fermi approximation the time-independent GPE (\ref{eqn:tidgp1}) reduces to,
\begin{equation}\label{time_indep_GPE_harmonicV}
\frac{1}{2}m\omega_\perp^2\left(\rho^2+\gamma^2 z^2 \right)+ \Phi({\bf r})+ g n({\bf r}) =
\mu.
\label{eqn:tf}
\end{equation}
Making use of the electrostatic formulation given in Eqs.\ (\ref{eq:Phidd}) and (\ref{potential}), exact solutions of Eq.\ (\ref{eqn:tf}) can be obtained for any general parabolic trap, as proven in Appendix A of
Ref. \cite{Eberlein05}. In particular, the solutions for the density profile
 take the form,
\begin{eqnarray}
n({\bf r})=n_{\rm cd}\left(1-\frac{\rho^2}{R_\perp^2}-\frac{z^2}{R_z^2}\right)
\,\,\,\, {\rm for} \,\,\, n({\bf r}) \ge 0, \label{eqn:tf2}
\end{eqnarray}
where $n_{\rm cd}=15N/(8 \pi R_\perp^2 R_z)$\footnote{Usually the central density of the Thomas-Fermi profile is denoted $n_0$. However to avoid confusion, later in the review, we have used $n_{\rm cd}$ to define the central density of the Thomas-Fermi profile.} is the central density, and $R_z$ and $R_\perp$ are the Thomas-Fermi radii of the condensate in the axial and transverse directions.
\begin{figure}[b]
\centering
\includegraphics[width=0.35\textwidth]{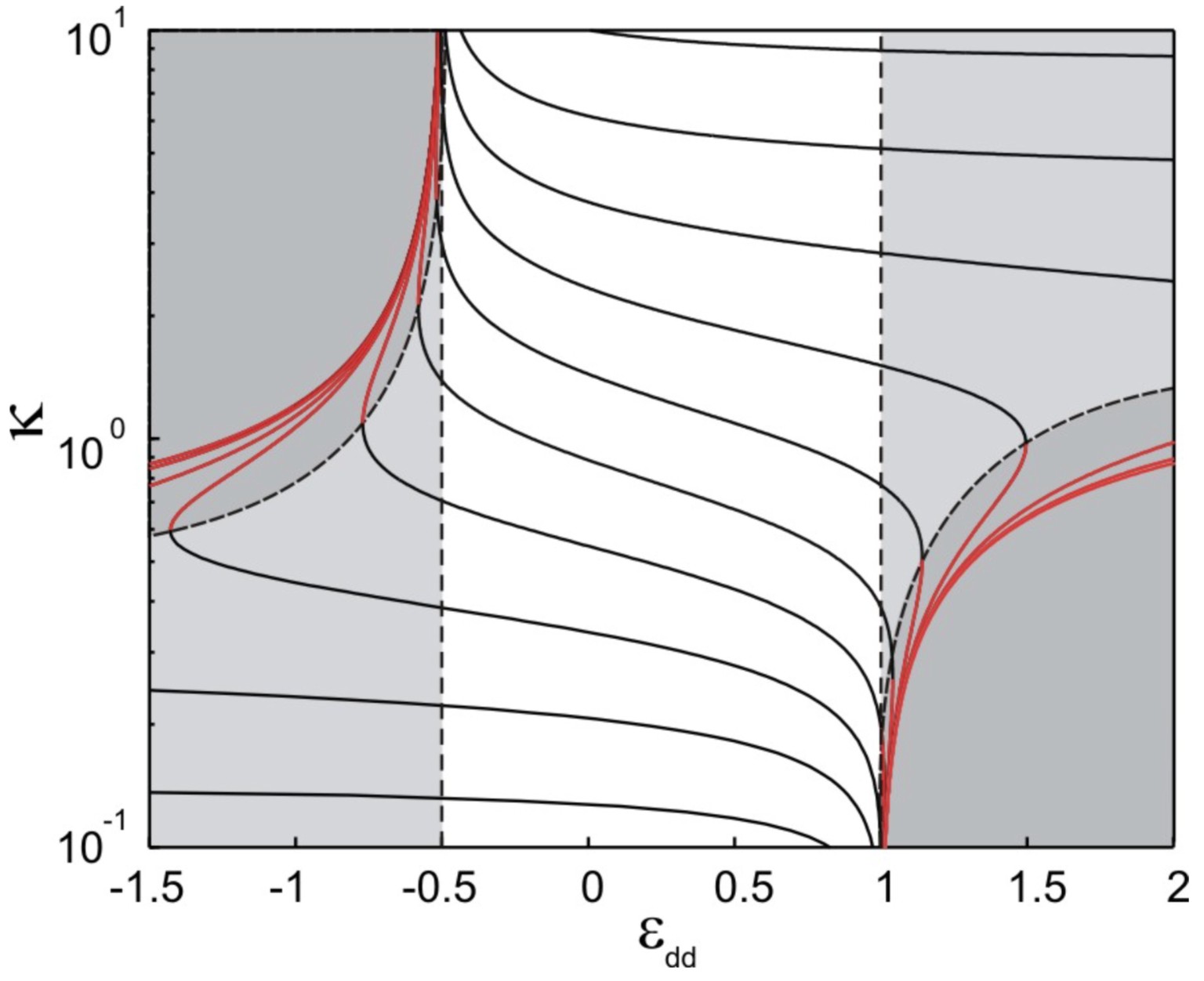}\\
	\includegraphics[width=0.35\textwidth,angle=0]{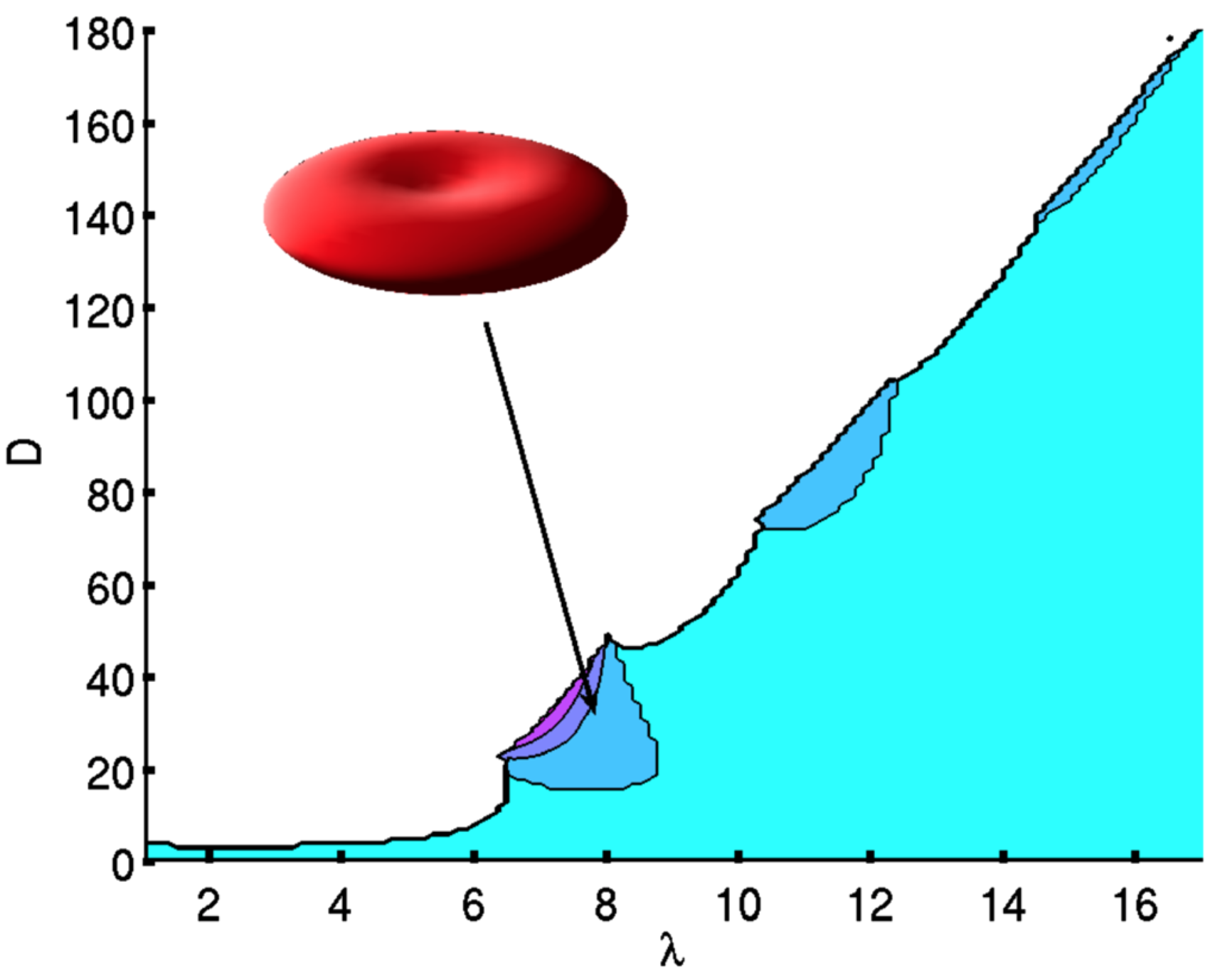}
\caption{Top: The aspect ratio, $\kappa$, (solid curves) of harmonically trapped, cylindrically-symmetric, dipolar condensates in the Thomas-Fermi regime. Each line corresponds to an equally spaced (on a logarithmic scale) trap aspect ratio, $\gamma$ ($\gamma = [0.1, 10]$). White, light grey and dark grey shading correspond to regimes of  global, metastable and unstable solutions respectively. Reprinted figure with permission from \cite{Bijnen10}. Copyright 2010 by the American Physical Society. Bottom: Stability diagram of the purely dipolar harmonically-trapped condensate (ground state), as a function of the trap aspect ratio $\lambda \equiv \gamma=\omega_z/\omega_\perp$  and the dipolar interaction parameter is $D= NmC_{\rm dd}/(4\pi \hbar^2 l_\perp)$. The shaded region denotes stability against collapse.  The dark shaded regions indicate biconcave condensates.  Reprinted figure with permission from \cite{Ronen07}. Copyright 2007 by the American Physical Society.} 
\label{fig:magnetostriction}
\end{figure}

Remarkably, the inverted parabolic density profile, Eq. (\ref{eqn:tf2}), is of the same form as that found in non-dipolar BECs \cite{Pethick&Smithbook,Pitaevskii&StringariBook}.  However, whereas non-dipolar BECs have the same aspect ratio, $\kappa=R_\perp/R_z$, as the trap, for dipolar BECs $\kappa \ne \gamma$ (in general) and must be evaluated using the following transcendental equation \cite{ODell04,Eberlein05},
\begin{eqnarray}
0&=&3 \kappa^2 \edd \left[\left(\frac{\gamma^2}{2}+1 \right)\frac{f(\kappa)}{1-\kappa^2}-1 \right] \nonumber \\
&+& (\edd-1)(\kappa^2-\gamma^2),
\label{eq:transendental}
\end{eqnarray}
where,
\begin{equation}
f(\kappa)=\frac{1+2\kappa^2}{1-\kappa^2}-\frac{3\kappa^2 {\rm arctanh} \sqrt{1-\kappa^2}}{(1-\kappa^2)^{3/2}},
\end{equation}
which takes the value $f=1$ at $\kappa=0$, and monotonically
decreases  towards $f= -2$ as $\kappa \rightarrow \infty$, passing through zero at
$\kappa=1$.
This is a robust feature: the same transcendental equation is recovered using a variational approach based on a gaussian ansatz for the condensate wave function \cite{Yi01,Yi02}.
For a non-dipolar ($\edd=0$) condensate one finds the expected result that $\kappa=\gamma$.  However, the presence of dipolar interactions leads to magnetostriction of the condensate, such that $\kappa < \gamma$ for $\edd>0$ and $\kappa > \gamma$ for $\edd < 0$.   This behaviour is shown in Figure~\ref{fig:magnetostriction} (top) \cite{Bijnen10}.  Note that, within the range $-0.5\leq \edd\leq1$ these are global solutions; elsewhere the solutions are either metastable (light grey shading) or unstable (dark grey shading).  For conventional dipoles ($C_{\rm dd}>0$, $\edd>0$), the condensate is least stable in prolate ($\gamma<1$) traps; here the dipoles lie predominantly in the attractive head-to-tail configuration and undergo collapse when $\edd$ becomes too large.  By contrast, in oblate ($\gamma>1$) traps stability is enhanced since the dipoles lie predominantly in the repulsive side-by-side configuration.  Meanwhile the opposite is true for anti-dipoles ($C_{\rm dd}<0$, $\edd<0$).  Away from the instabilities, these solutions agree well with numerical solutions of the full dipolar GPE in the Thomas-Fermi regime \cite{Parker09}.  Close to the instabilities, zero-point kinetic energy (neglected within the Thomas-Fermi approach) can enhance the stability of the solutions.  

Once the BEC aspect ratio $\kappa$ is found from the transcendental equation, the Thomas-Fermi radii are determined by the expressions,
\begin{eqnarray}
R_\perp&=&\left[ \frac{15 N g \kappa}{4 \pi m \omega_\perp^2} \left\{1+\edd\left(\frac{3}{2}\frac{\kappa^2 f(\kappa)}{1-\kappa^2}-1 \right) \right\}\right]^{1/5}, \label{eq:RperpTF} \\
R_z&=&\frac{R_\perp}{\kappa},
\end{eqnarray}
and the total energy is given by,
\begin{eqnarray}
E_{\mathrm{TF}}  & = &  \frac{N}{14} m \omega_{x}^{2} R_{x}^{2}
\bigg(2 + \frac{\gamma^{2}}{\kappa^{2}} \bigg)  \nonumber \\
  &+& \frac{15}{28 \pi}
\frac{N^{2}g}{R_{x}^{2}R_{z}}\left[1-
\varepsilon_{\mathrm{dd}}f(\kappa) \right].
\label{eq:energyfunctionalTF}
\end{eqnarray}
The first term corresponds to the trapping energy and the second to the $s$-wave and dipolar interaction energies. Finally, the dipolar potential inside the condensate can be explicitly obtained as \cite{Eberlein05},
\begin{eqnarray}
\Phi_{\mathrm{TF}} (\rho,z)  &=&  \frac{ n_{\rm cd}
C_{\mathrm{dd}}}{3}\bigg[\frac{\rho^2}{ R_{x}^2}  -\frac{2 z^2}{
R_{z}^2} \nonumber \\ &-&  f(\kappa)\left(1-\frac{3}{2}\frac{\rho^2 -
2 z^2}{ R_{x}^2- R_{z}^2}\right) \bigg].
\label{eq:phiddinside}
\end{eqnarray}
This is generally a saddle-shaped function that reflects the anisotropic nature of the dipolar interactions and drives the elongation of the BEC along the polarization direction. A more general version of $\Phi_{\mathrm{TF}} (\mathbf{r})$ for the case of a dipolar BEC without cylindrical symmetry is given later in Eq.\ (\ref{eq:Phiellipsoid}). 

\subsubsection{Outside the Thomas-Fermi regime: rotons and density oscillations \label{subsubsec:outsideTFregime}}
According to the Thomas-Fermi approach, a trap which is sufficiently oblate ($\gamma\gappeq 5.2$) is stable to collapse even in the limit $\edd \rightarrow \infty$.  However, numerical solutions reveal a different fate, whereby the condensate undergoes instability even for $\gamma \rightarrow \infty$ \cite{Santos03}.  This is associated with the development of a roton minimum in the dispersion relation \cite{ODell03,Santos03,Fischer06},  reminiscent of rotons in superfluid helium \cite{Donnelly91}. For certain parameters, this minimum can approach zero energy, triggering an instability at finite $k$ known as the roton instability.  The Thomas-Fermi approach, which is limited to the class of inverted parabolic solutions, is unable to account for this phenomenon.  The roton is a strict consequence of the non-local interactions, and does not arise for conventional condensates.  

The effect of this in trapped, purely dipolar condensates was revealed by Ronen {\it et al.}, with the stability diagram shown in Figure~\ref{fig:magnetostriction} (bottom) \cite{Ronen07}.  When the dipolar interaction parameter $D=NmC_{\rm dd}/(4\pi \hbar^2 l_\perp)$\footnote{In the original work by Ronen {\it et al.} \cite{Ronen07} $D$ was defined as $D=(N-1)mC_{\rm dd}/(4\pi \hbar^2 l_\perp)$. However, the derivation of the dipolar GPE requires $N\gg1$. Hence, in this review, we have defined $D=NmC_{\rm dd}/(4\pi \hbar^2 l_\perp)$.} exceeds a critical value, for any trap ratio, the system is unstable to collapse.  The condensate becomes unstable to modes with increasingly large number of radial and angular modes as the trap aspect ratio increases, signifying that collapse proceeds on a local, rather than  global, scale \cite{Wilson09}. Of particular interest is the appearance, close to the instability boundary and under oblate traps, of ground state solutions with a biconcave, red blood cell-like, shape [see Figure~\ref{fig:magnetostriction} (bottom)] \cite{Ronen07,Dutta07b}.  Subsequent works confirmed these density oscillations as being due to the roton, which, for certain parameters, mixes with the ground state of the system \cite{Wilson08}.  More generally, when van der Waals interactions are included \cite{Abad09,Lasinio13}, both biconcave and dumbbell shapes can arise \cite{Yi10}.  Under box-like potentials, which have been realized in recent years \cite{Gaunt13,Chomaz15}, density oscillations associated with the roton can arise at the condensate edge \cite{Yi01}.  

An intuitive interpretation of the roton in an oblate trap was put forward by Bohn {\it et al.} \cite{Bohn09}.  As the dipole strength is increased, it is energetically favourable for the dipoles to locally move out of the plane and align head-to-tail perpendicular to the plane, thereby taking advantage of this attractive configuration. This leads to a periodic density in the plane, with a wavenumber corresponding to that of the roton minimum.  In this geometry it is also interesting to note that quantum depletion of the condensate is predicted to diverge at the roton instability \cite{Fischer06}.  

\subsection{Quasi-two-dimensional dipolar Bose-Einstein condensate}
For a condensate strongly confined in one dimension it is possible to reduce the effective dimensionality of the system to form a quasi-two-dimensional condensate.  This offers a simplified platform to study vortices and vortex lattices in dipolar condensates, while still retaining the key physics.
\begin{figure}[b]
\centering
	\includegraphics[width=0.4\textwidth,angle=0]{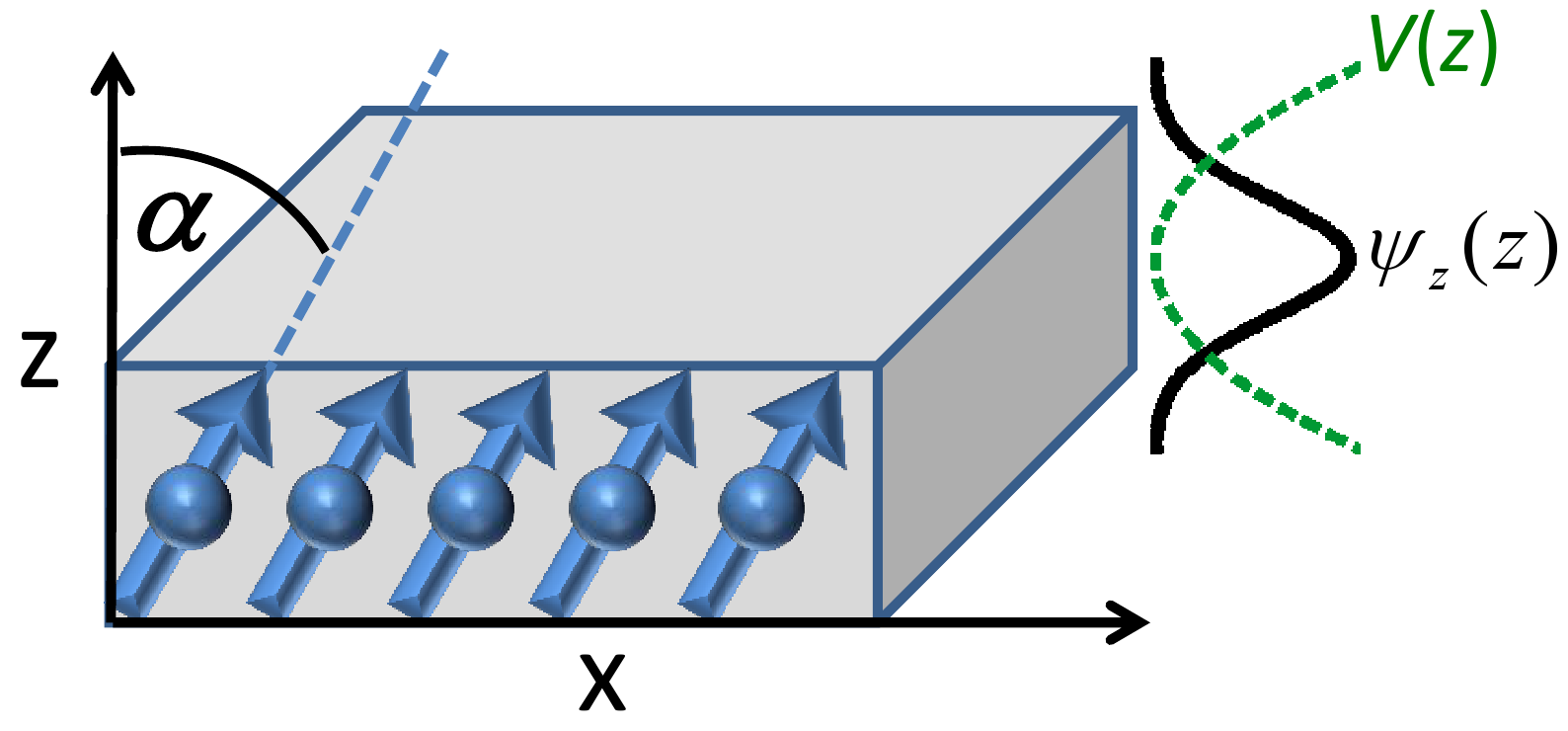}
		\caption{Schematic of the quasi-two-dimensional dipolar condensate, with strong harmonic trapping along $z$. The dipoles are taken to be polarized at angle $\alpha$ to the $z$-axis in the $x-z$ plane.  The condensate is assumed to follow the static ground harmonic oscillator state along $z$.  Figure reproduced from Ref. \cite{Mulkerin14} under a \href{https://creativecommons.org/licenses/by/3.0/}{CC BY licence}.}
		\label{fig:schematic}
\end{figure}

Consider the dipoles to be polarized at an angle $\alpha$ to the $z$-axis, lying in the $x-z$ plane, and strong harmonic confinement $V(z)=\frac{1}{2}m\omega_z^2 z^2$ in the $z$-direction which satisfies $\hbar \omega_z \gg \mu$, i.e. the trapping energy dominates over the condensate energy scale.  This set-up is illustrated in Figure~\ref{fig:schematic} \cite{Mulkerin14}.  In this regime, one can approximate the wavefunction by the ansatz,
\begin{equation}
\Psi(\bfrho,z,t)=\Psi_\perp(\brho,t) \psi_z(z).
\label{eq:quasi_2d_psi}
\end{equation}
Axially, the condensate is taken to be {\it frozen} into the axial ground harmonic oscillator state $\psi_z(z)=(\pi \ell_z^2)^{-1/4} e^{-z^2/2 \ell_z^2}$.  The dynamics then become planar, parametrised by the two-dimensional time-dependent wavefunction, $\Psi_\perp(\bfrho,t)$.  Note that $\Psi_\perp$ is normalized to the number of atoms, i.e. $N=\int |\Psi_\perp|^2{\rm d}\bfrho$.  Inserting this ansatz into the dipolar GPE, Eq. (\ref{eqn:dgpe1}), and integrating out the axial direction then leads to the effective two-dimensional dipolar GPE \cite{Ticknor11}, 
\begin{eqnarray}
i \hbar \frac{\partial \Psi_\perp}{\partial t}=\left [-\frac{\hbar^2}{2m}\nabla_{\perp}^2+V(\bfrho)+\frac{g}{\sqrt{2\pi}l_z} |\Psi_\perp |^2+\Phi_\perp\right ]\Psi_\perp. \nonumber \\
\label{eqn:dgpe2}
\end{eqnarray}
The $g/\sqrt{2 \pi}l_z$ coefficient characterises the effective van der Waals interactions in the plane, and $\Phi_\perp$ is the effective planar dipolar potential, 
\begin{equation}
\Phi_\perp(\bfrho,t)=\int U^\perp_{\rm dd}(\bfrho-\bfrho')~n_\perp(\bfrho',t)\,{\rm d}\bfrho',
\label{eqn:2d_phi}
\end{equation} 
where $n_\perp=|\Psi_\perp|^2$ is the two-dimensional density.
The real-space form of the effective two-dimensional dipolar interaction potential $U^\perp_{\rm dd}$ is given elsewhere \cite{Cai10}, while in this review  its Fourier transform is used \cite{Ticknor11,Fischer06},
\begin{equation}
\tilde{U}^\perp_{\rm dd}(\mathbf{{\tilde q}})=\frac{4\pi C_{\rm dd}}{9 \sqrt{2 \pi}l_z} \left[F_{\parallel}\left (\mathbf{{\tilde q}} \right ) \sin^2\alpha +F_{\perp}\left (\mathbf{{\tilde q}} \right )\cos^2\alpha\right],
\label{eq:dipolar_interaction_2D}
\end{equation}
where  $F_{\parallel}({\bf {\tilde q}})=-1+3\sqrt{\pi}\frac{{\tilde q}_{x}^{2}}{{\tilde q}}e^{{\tilde q}^2} \mathrm{erfc}({\tilde q})$, $F_{\perp}({\bf {\tilde q}})=2-3\sqrt{\pi}{\tilde q} e^{{\tilde q}^2} \mathrm{erfc}({\tilde q})$ and $\mathbf{{\tilde q}}=\mathbf{q}l_z/\sqrt{2}$ with ${\bf q}$ being the projection of ${\bf k}$ onto the $x-y$ plane, i.e. the reciprocal space analogue of $\bfrho$\footnote{Throughout this review ${\bf k}$ is the reciprocal lattice vector in three spatial dimensions ${\bf r}$ and  ${\bf q}$ is the reciprocal lattice vector in two spatial dimensions ${\bfrho}$.}.  From this momentum space representation $\Phi_\perp$ can then be evaluated using the convolution theorem as in Eq. (\ref{eqn:conv}).  An important parameter is the ratio $\sigma=l_z/\xi$, where $\xi$ is the healing length.  The two-dimensional approximation requires $\sigma < 1$.  

Under a cylindrically-symmetric harmonic trap and for dipoles polarized along $z$, the above ``two-dimensional mean-field regime'' is formally entered when $N a_s(1+2 \edd)l_z^3/l_\perp^4 \ll 1$.  In the opposing regime, when $N a_s(1+2 \edd)l_z^3/l_\perp^4 \gg 1$, the system enters the three-dimensional Thomas-Fermi regime \cite{Parker08}.  A more general analysis of flattened condensates in Ref. \cite{Baillie15}, has established the validity of the two-dimensional mean-field regime for arbitrary polarization direction.

\begin{figure}[b]
\centering
	\includegraphics[width=0.5\textwidth,angle=0]{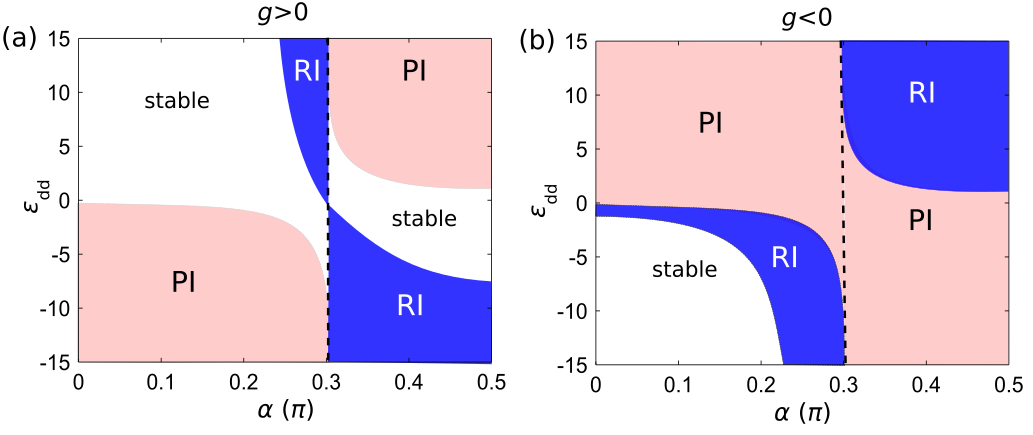}
		\caption{The stability diagram  in $\varepsilon_{\rm dd}-\alpha$ space, for (a) $g>0$ and (b) $g<0$, of a homogeneous dipolar BEC in the quasi-two-dimensional regime ($\sigma=0.5$).  Shown are the regions of stability (white), phonon instability (pink) and roton instability (blue).  The vertical dashed line indicates the magic angle $\alpha_{\rm m}$.  For $\alpha>\pi/2$ the results are the mirror image of the presented region.  Figure adapted from Ref. \cite{Mulkerin14}. }
		\label{fig:phase_diagram}
\end{figure}

In the absence of any planar trapping potential [$V(\bfrho)=0$], the stationary solution of the quasi-two-dimensional dipolar condensate is the homogeneous state \cite{Ticknor11},
\begin{equation}
\psi_{\perp}=\sqrt{n_0}, \quad \mu=\frac{n_0}{\sqrt{2\pi}l_z} \left[g+\frac{C_{\rm dd}}{3} (3\cos^2\alpha-1)\right],
\end{equation}
where $n_0$ is the uniform two-dimensional density. This system undergoes the phonon instability when the net local interactions become attractive in the plane, i.e. when $g+C_{\rm dd}[3\cos^2\alpha-1]/3<0$.  The phonon unstable regions in the $\edd-\alpha$ plane are shown in Figure~\ref{fig:phase_diagram}. These can be understood by considering the trade-off between the van der Waals and the dipolar interactions \cite{Mulkerin14,Mulkerin13}.  Note the divergent behaviour at the magic angle $\alpha_{\rm m}$, across which the planar dipolar interactions switch between repulsive and attractive.
 
The roton instability also arises in this quasi-two-dimensional BEC \cite{Santos03}, as indicated in Figure~\ref{fig:phase_diagram}(a) and (b) (blue shaded regions).
For $g>0$ the roton instability is induced by the attractive part of the dipolar interaction and is only possible for $\alpha \neq 0$; for $\alpha=0$ the condensate cannot probe the attractive part of $U_{\rm dd}^{\perp}$  \cite{Fischer06} (this is true only in the strict quasi-two-dimensional limit).  For $g<0$ the roton exists for all $\alpha$ \cite{Klawunn09a} (excluding the magic angle).  For small $\alpha$ it is induced by the attractive van der Waals interactions, while for larger $\alpha$ it is driven by the attractive axial component of the dipolar interactions.    

The extent of the dipolar BEC in the $z$-direction ($\sigma$) effects the stability of the roton. Specifically, as the condensate becomes narrower ($\sigma$ decreases) the out-of-plane component of $U_{\rm dd}^{\perp}$ decreases \cite{Mulkerin14} and the regimes of roton instability (blue regions in Figure~\ref{fig:phase_diagram}) shrink. 

\section{Single Vortices \label{sec:single_vortices}}
Quantized vortices are a consequence of the condensate's phase coherence. To preserve the single-valuedness of the condensate wavefunction, the total change in phase around some closed path $C$ must be $2\pi q_{\rm v}$, with $q_{\rm v}=0, \pm1, \pm 2,...$.  If $q_{\rm v} \neq 0$ then there exists one or more phase singularities within $C$.  These singularities are the quantized vortices, and $q_{\rm v}$ is the {\it vortex charge}.

Consider an isolated vortex at the origin in a uniform system, which is straight along $z$.  The condensate phase about the vortex follows the azimuthal angle, $S(\rho,\theta,z)=\theta = \arctan(y/x)$.  From Eq. (\ref{eqn:vel}) this gives rise to a circulating azimuthal flow with speed,
\begin{equation}
v=\frac{q_{\rm v}\hbar}{m \rho}.
\label{eq:vortexvelocity}
\end{equation}
{\it Within the fluid} the flow is irrotational with zero vorticity, i.e. $\nabla \times {\bf v}=0$.  This can also be seen directly from the definition of the fluid velocity, which is curl-free $\boldsymbol{\nabla} \times {\bf v}=0$ and thus very different to the classical solid-body rotation, for which $v \propto \rho$.  At the point of the singularity, however, the vorticity takes the finite value $q_{\rm v} h/m$.  Also, at this point the density is zero, preventing the unphysical scenario of infinite mass current.  The cylindrically-symmetric flow associated with the velocity field given in Eq.\ (\ref{eq:vortexvelocity}) carries a total angular momentum $L_{z}=N \hbar q_{\rm v}$.

\subsection{Energetics of vortex formation \label{subsec:energetics}}
Even single vortices are giant excitations involving a considerable fraction of the entire BEC.  The energy associated with the formation of a vortex $E_{\rm v} \equiv E-E_{0}$, where $E_{0}$ is the energy of the non-rotating (vortex-free) state,  is generically much larger than the energy of elementary excitations described by the Bogoliubov spectrum given in Eq.\ (\ref{eqn:bog}).  In a frame rotating at angular frequency $\Omega$, the total energy of the system is shifted to $E-\mathbf{\Omega} \cdot \mathbf{L}$, where $\mathbf{L}$ is the angular momentum in the laboratory frame, and hence  it only becomes energetically favourable to form a vortex if the angular momentum is such that $\vert \mathbf{\Omega} \cdot \mathbf{L} \vert > E_{\rm v}$ leading to a critical rotation frequency,
\begin{equation}
\Omega_{\rm v}=\frac{E_{\rm v}}{N \hbar q_{\rm v}} .
\label{eqn:criticalfrequency}
\end{equation}
$E_{\rm v}$ can be computed analytically in certain situations as we discuss below. Before we do so, it is important to point out that Eq.\ (\ref{eqn:criticalfrequency}) considers the energetics but not the kinetics of vortex formation. Both theory and experiment reveal that the true value of $\Omega_{\rm v}$ is often considerably higher than predicted by Eq.\ (\ref{eqn:criticalfrequency}). This is because the vortex-free state can remain a local energy minimum separated by an energy barrier from the global minimum corresponding to the vortex state. The kinetics of vortex formation are examined in Section \ref{sec:gen_vortices}.

In the simplest case of an infinite system with a vortex the  condensate wavefunction can be written,
\begin{equation}
\psi(\rho,\theta,z)=f(\rho)e^{i q_{\rm v} \theta}.
\label{eq:vortexwf}
\end{equation}
Substituting into the dipolar GPE (\ref{eqn:dgpe1})  an equation for the amplitude $f$ about the vortex is obtained,
\begin{eqnarray}
\mu f=-\frac{\hbar^2}{2m} \frac{1}{\rho} \frac{\partial}{\partial\rho} \left(\rho \frac{\partial f}{\partial\rho}\right)
+\frac{\hbar^2 q_{\rm v}^2}{2 m \rho^2} f 
+g f^3+\Phi f, 
\label{eqn:vortex_profile}
\end{eqnarray}
where the Laplacian has been expressed in its cylindrically symmetric form,
\begin{equation}
\nabla^2=\frac{1}{\rho}\frac{\partial }{\partial \rho}\left(\rho\frac{\partial}{\partial \rho} \right)+\frac{1}{\rho^2}\frac{\partial^2}{\partial \theta^2}+\frac{\partial^2}{\partial z^2}.
\label{eqn:vortex_eqn}
\end{equation}
The second term on the right-hand side of Eq. (\ref{eqn:vortex_profile}) is the only difference from the non-vortex case, and is associated with the kinetic energy of the circulating flow, giving rise to a centrifugal barrier. Note that $|q_{\rm v}|>1$ vortices are energetically unstable compared to multiple singly-charged vortices, and rarely arise unless engineered; for this reason we will  consider only $|q_{\rm v}|=1$. 

In any system uniform along the polarization direction ($z$) the dipolar potential reduces to a local potential $\Phi(\mathbf{r}) \rightarrow -g \edd n({\bf r}) $. This can be seen from Eq.\ (\ref{eq:Phidd}), which gives the relationship between the fictitious electrostatic potential $\phi(\mathbf{r})$ and the dipolar potential $\Phi(\mathbf{r})$, and noting that $\partial^2 \phi/ \partial z^2$ must equal zero\footnote{If the density profile $n(\mathbf{r})$ is uniform along $z$ the electrostatic potential $\phi(\mathbf{r})$ it generates, as given by Eq.~(\ref{potential}), must also be uniform along $z$ and hence $\partial^2 \phi/ \partial z^2$ vanishes.}. Thus, the last two terms in Eq.\ (\ref{eqn:vortex_profile}) can be combined into a single contact term $g (1- \edd) f^3$.  Results that hold for the usual $s$-wave case in this context therefore also hold for the dipolar case provided one replaces $g$ by $g (1- \edd)$.  For example, analysis of Eq.\ (\ref{eqn:vortex_profile}) for the $s$-wave case reveals that the centrifugal barrier term dictates that the density relaxes as $1/\rho^2$ to the asymptotic background value $\sqrt{n_0}$ as $\rho \to \infty$,  and for $\rho \to 0$ the density tends to zero as $\rho^{2|q_{\rm v}|}$ \cite{Pethick&Smithbook}. Furthermore, although Eq.\ (\ref{eqn:vortex_profile}) can not be solved in terms of known functions, it can be solved numerically, and with appropriate scaling the result for $f(\rho)$ is universal. This solution for $f(\rho)$ can then be used in the energy functional given in Eq.\ (\ref{eqn:energy}) and the extra energy per unit length due to the introduction of a vortex evaluated. 
In the pure $s$-wave case one obtains \cite{Pethick&Smithbook},
\begin{equation}
\epsilon_{\rm v}= \pi n_0 \frac{\hbar^2}{m} \log \left( 1.464 \frac{b}{\xi} \right) .
\label{eq:GinzburgPitaevskii}
\end{equation}
This result was originally obtained for superfluid $^4$He by Ginzburg and Pitaevskii in 1958 \cite{GinzburgPitaevskii}. In this expression the healing length $\xi$, which gives the size of the vortex core, forms a lower cutoff and the length $b$, which could be the system radius, is the upper cutoff.  Although $b \gg \xi$, their finite values avoid a logarithmic singularity that originates from the centrifugal barrier term. The only place that interactions enter this expression for $\epsilon_{\rm v}$ is through the healing length $\xi=\hbar /\sqrt{m \mu}$ which can immediately be adapted to the dipolar case using $\mu= n_0 g (1 - \edd)$. The critical rotation frequency for this case can now be evaluated by replacing $E_{\rm v}$ by $\epsilon_{v}$ and $N$ by $N/L$ (number of atoms per unit length) in Eq.\ (\ref{eqn:criticalfrequency}).

\subsection{General features of a vortex  in a quantum ferrofluid \label{subsec:structureofvortex}}
In order to obtain analytic results for the untrapped system it can sometimes be useful to use the following approximate solution for $f(\rho)$ which incorporates the correct behaviour for $\rho \rightarrow 0$ and $\rho \rightarrow \infty$,
\begin{equation}
f^2(\rho)=n(\rho,z)=n_0 \frac{\rho'^2}{1+\rho'^2},
\label{eqn:vortex_profile1}
\end{equation}
where $\rho'=\rho/\xi$ and $n_0$ is the density at infinity.  It will be convenient later to write this in the form of a homogeneous background density $n_0$ and a \textit{negative} vortex density $n_{\rm v}$,
\begin{equation}
n(\rho,z)=n_0+ n_{\rm v},
\label{eqn:vortex_profile2}
\end{equation}
where $n_{\rm v}=-n_0/(1+\rho'^2)$.

\begin{figure}[t]
\centering \includegraphics[width=0.4\textwidth]{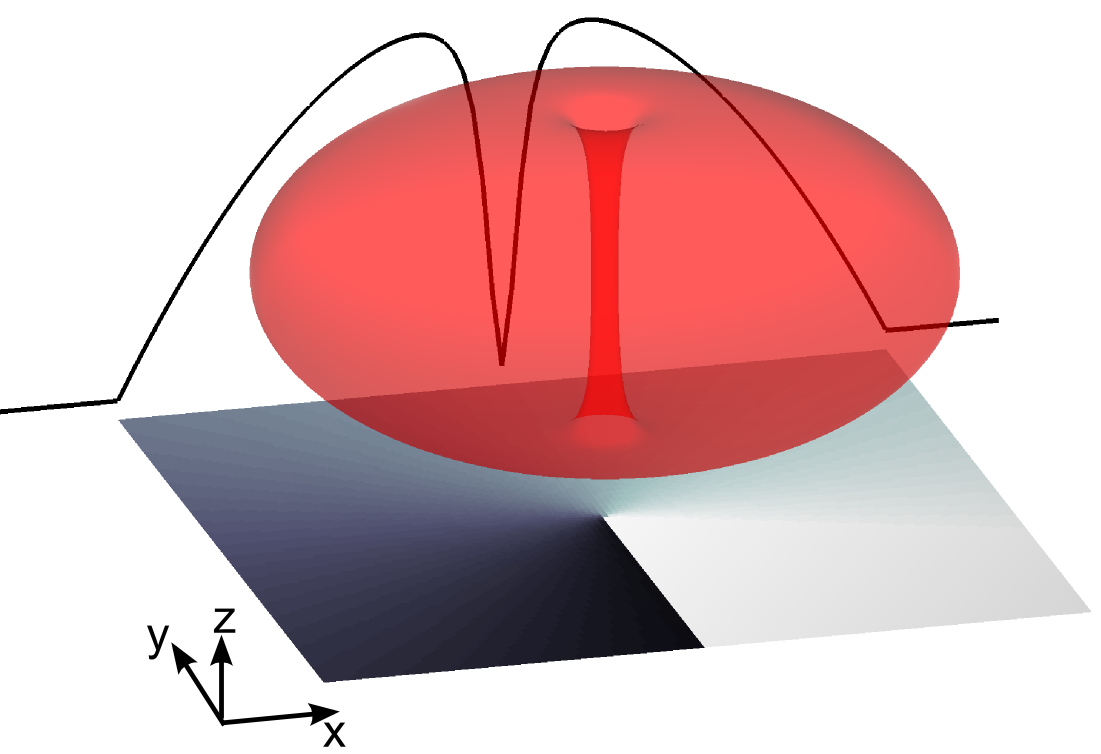}
\caption{Schematic of a three-dimensional density (red iso-surface plot) of a trapped non-dipolar condensate featuring a vortex line along the $z$-axis.  The corresponding two-dimensional phase profile (grey scale plot at the base of the figure, with white corresponding to a condensate phase of $0$ and black corresponding to a condensate phase of $2\pi$) and central one-dimensional density profile (solid black curve) are also depicted.}
\label{fig:parker_fig2}
\end{figure}

A schematic of a straight singly-charged vortex line through a non-dipolar harmonically-trapped condensate is shown in Figure~\ref{fig:parker_fig2}.  The vortex has a well defined core, with zero density and a phase singularity at its centre, relaxing to the background condensate density over a distance given by the conventional healing length $\xi=\hbar/\sqrt{m\mu}$. This is typically of the order of $0.1-1\mu$m but can be tuned by means of a Feshbach resonance.   

In three dimensions the vortices may bend, e.g. into tangles and rings, carry linear or helical Kelvin wave excitations and undergo reconnections.  However, under strong axial confinement of the condensate, the dynamics become effectively two-dimensional.  Being topological defects, vortices can only disappear via annihilation with an oppositely-charged vortex (the two-dimensional analog of a reconnection) or by exiting the condensate at a boundary.  In trapped condensates, an off-centre vortex precesses about the trap centre; this can be interpreted in terms of a Magnus force \cite{Freilich10}.  Thermal dissipation causes a precessing vortex to spiral out of a trapped condensate \cite{Jackson09,Rooney10,Allen13,Gautum14a}.  Acceleration of a vortex (or an element of a three-dimensional vortex line) leads to emission of phonons, analogous to the Larmor radiation from an accelerating charge, although under suitable confinement these phonons can be reabsorbed to prevent net decay of the vortex \cite{Parker04}.    

Optical absorption imaging of the vortices is typically preceded by expansion of the cloud to enlarge the cores  \cite{Madison00,Raman01}.  This method has been extended to provide real-time imaging of vortex dynamics \cite{Freilich10}.  While this imaging approach detects density only, the vortex circulation is detectable via gyroscopic techniques \cite{Powis14}. 

\begin{figure}[t]
\centering
	\includegraphics[width=0.4\textwidth,angle=0]{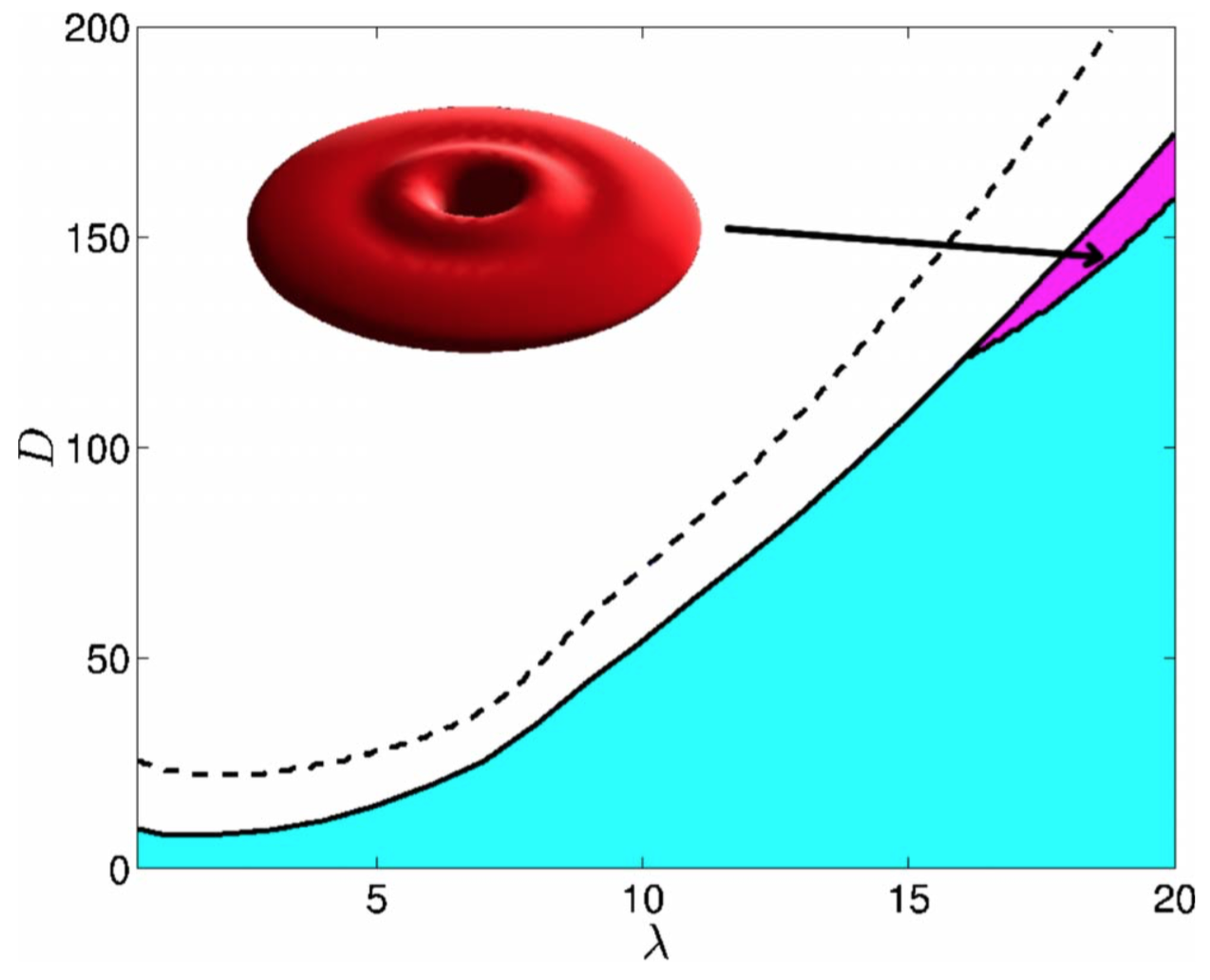}\\
		\includegraphics[width=0.4\textwidth,angle=0]{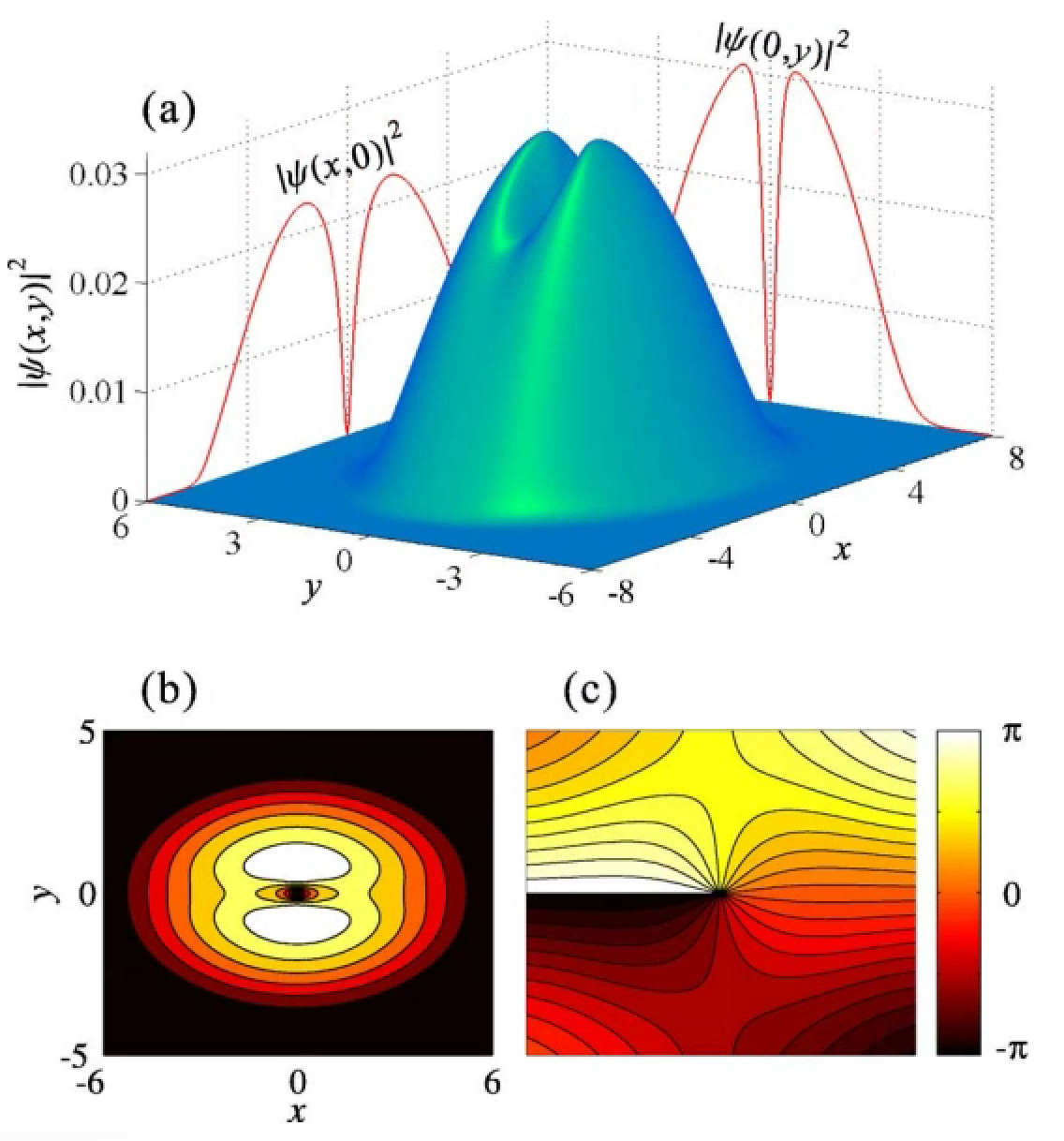}
		\caption{Top: Stability diagram of the trapped purely dipolar condensate, with dipoles polarized along the $z$-axis and featuring an axial vortex, as a function of the trap ratio ($\lambda\equiv \gamma=\omega_z/\omega_{\perp}$) and dipolar interaction strength [$D=NmC_{\rm dd}/(4\pi \hbar^2 l_{\perp})$].  Below the solid line the condensate is dynamically stable.  Ripples about the vortex arise in the pink region.  The inset shows an isosurface of the density for such a solution.  Reprinted figures with permission from \cite{Wilson09}. Copyright 2009 by the American Physical Society.  Bottom:  Density (a-b) and phase (c) profiles of a trapped quasi-two-dimensional condensate with dipoles polarized along $x$.  Reprinted figures with permission from \cite{Pu06}. Copyright 2006 by the American Physical Society. 
}
		\label{fig:wilson_yi}
\end{figure}

Yi and Pu \cite{Pu06} performed the first study of vortices in a dipolar BEC, obtaining numerical solutions for a quasi-two-dimensional trapped dipolar condensate featuring a vortex. For dipoles polarized perpendicular to the plane they found the striking result that density ripples form about the vortex core for trap ratios $\gamma \sim 100$ and \emph{attractive} van der Waals interactions. These ripples are not contained in the simple ansatz given above in Eq.\ (\ref{eqn:vortex_profile1}) and seem to be a rather special feature associated with non-local interactions. Indeed, they had previously been seen in numerical simulations of vortices in superfluid $^4$He where non-local potentials are employed \cite{Dalfovo92,Ortiz,Sadd,Berloff}. For purely dipolar ($g=0$), oblate  condensates, Wilson {\it et al.} \cite{Wilson09,Wilson08} numerically found vortex ripples for moderate trap ratios $\gamma \sim 17$, see Figure~\ref{fig:wilson_yi} (top) \cite{Wilson09}, and established the link to roton mixing into the vortex solution, similar to the biconcave structure that they found was induced in vortex-free dipolar condensates (energetic favourability of dipoles aligning head-to-tail). Vortex ripples have since been studied in other works \cite{Abad09,Mulkerin14,Mulkerin13}, and similar ripples arise in the presence of other localized density depletions, such as due to localized repulsive potentials \cite{Wilson08,Ticknor11} and dark solitons \cite{Bland15,Edmonds16}.  The presence of the vortex slightly reduces the stable parameter space for the condensate relative to the vortex-free condensate \cite{Wilson09,Pu06}.  For dipoles tilted perpendicular to the axis of the vortex, the vortex core becomes elliptical, see Figure~\ref{fig:wilson_yi} (bottom) \cite{Pu06} due to the anisotropic dipolar potential in the plane.  

The properties of an off-axis straight vortex line in a trapped dipolar condensate have been considered in \cite{Yuce10,Yuce11} in the Thomas-Fermi regime, showing that the dipolar interaction lowers (raises) the precession speed in an oblate (prolate) trap.   In the presence of thermal dissipation, making the dipolar interactions partially attractive by changing the polarization direction leads to a reduction in the condensate size and a faster decay rate of the precessing vortex \cite{Gautam14b}. 

For a general vortex line in three dimensions, the vortex elements interact with each other at long-range via the dipolar interactions \cite{Klawunn08}, as well as the usual hydrodynamic interaction \cite{Svidzinsky00,Koens12}.  This modifies the Kelvin-wave (transverse) modes of the vortex line, and can support a roton minimum in their dispersion relation.  For large dipolar interactions, the Kelvin waves can undergo a roton instability, leading to novel helical or snake-like configurations \cite{Klawunn09b}.

With $C_{\rm dd}<0$ and tight axial trapping, stable two-dimensional bright solitons have been predicted \cite{Pedri05,Nath07}, i.e. wavepackets which are self-trapped by interactions in two dimensions. This idea was extended by Tikhonenkov {\it et al.} \cite{Tikhonenkov08} to predict stable two-dimensional vortex solitons, which may be considered as a two-dimensional bright soliton carrying a central vortex.

\subsection{Vortex in a trapped dipolar Bose-Einstein condensate in the hydrodynamic regime  \label{subsec:vortexTF}}
In the hydrodynamic (Thomas-Fermi) regime appropriate for large condensates, the problem of a  dipolar BEC with single vortex in a three-dimensional trap  can be tackled semi-analytically \cite{ODell07}. In this case the energy associated with the curvature of the density due both to the trapping and the vortex core can be ignored in comparison to each of the rotational, interaction, and trapping energies.  These remaining energies can be evaluated analytically by assuming a density  profile much like the one given in Eq.\ (\ref{eqn:vortex_profile1}), i.e.\ an unperturbed background piece plus a negative vortex ``density''. The difference is that we now take the unperturbed background density to be the inverted parabola $n_{\mathrm{TF}}(\rho,z)$ given in Eq.\ (\ref{eqn:tf2}) which is an exact solution of the vortex-free Thomas-Fermi problem. Thus, the density profile reads,
\begin{equation}
n(\rho,z)=n_{\mathrm{TF}}(\rho,z)+n_{\rm v}(\rho,z)
\end{equation}
where,
\begin{equation}
n_{\rm v}(\rho,z)= -n_{\mathrm{TF}}(\rho,z) \frac{\beta^2}{\beta^2 +\rho^2}.
\end{equation}
The length scale $\beta$ parameterizes the size of the vortex core and is one of three variational parameters $\{\beta, R_{\perp}, \kappa \}$ with respect to which the total energy functional must be minimized in order to find their stationary values. Notice that this ansatz does not include ripples which are beyond the Thomas-Fermi approximation.
\begin{figure*}
\centering
	\includegraphics[width=0.8\textwidth,angle=0]{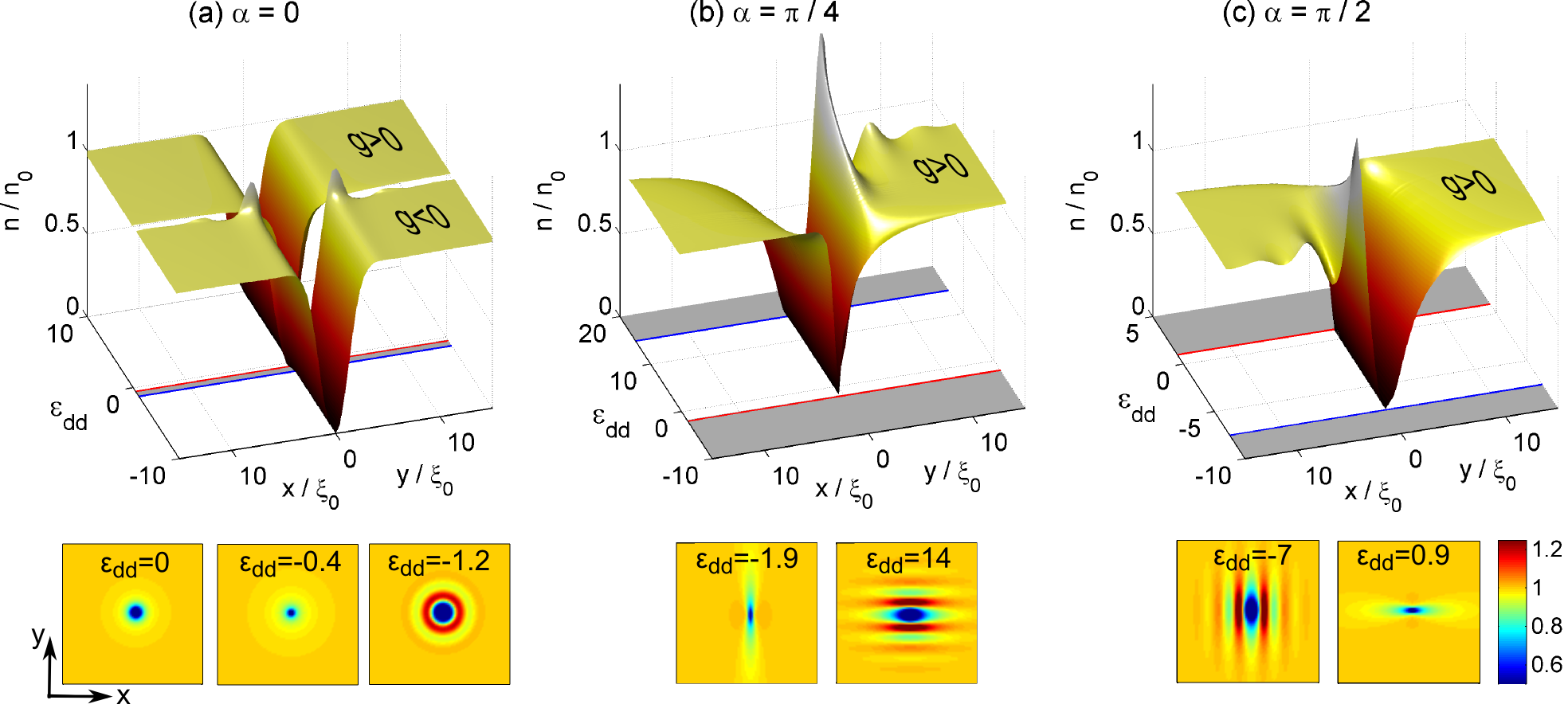}
		\caption{Vortex solutions in an infinite dipolar condensate, as a function of $\edd$, in the quasi-two-dimensional regime ($\sigma=0.5$).  Along $x$ (for $y=0$) the normalised density profile ($n/n_0$) is shown on the left-hand side of the main plots.  Along the right-hand side the normalised density profile is plotted  along $y$ (with $x=0$). (a) Dipoles polarized along $z$ ($\alpha=0$).   (b) Dipoles polarized off-axis at $\alpha=\pi/4$.  (c) Dipoles polarized off-axis at $\alpha=\pi/2$.  In each of the main plots grey bands indicate the unstable regimes of $\edd$.  Insets: normalised density profile  over an area $(40\xi)^2$ for indicated values of $\edd$.  Figure reproduced from Ref. \cite{Mulkerin14} under a \href{https://creativecommons.org/licenses/by/3.0/}{CC BY licence}.
}
		\label{fig:single_vortices}
\end{figure*}

The energy functionals for the rotational, $s$-wave interaction, and trapping energies can all be evaluated analytically, albeit laboriously,  using the above density profile. The results are given in Ref. \cite{ODell07} and will not be repeated here. Obtaining an analytic result for the dipolar interaction energy $E_{\mathrm{dd}}=(1/2)  \int \Phi(\mathbf{r}) n(\mathbf{r}) \mathrm{d} \mathbf{r}$ is more difficult. However, in the hydrodynamic regime we have $\beta \ll R_{\perp}$, implying that the contribution to $\Phi(\mathbf{r})$ from $n_{\rm v}$ is negligible in comparison to that from $n_{\mathrm{TF}}$. Thus, to a very good approximation we can write,
\begin{equation}
E_{\mathrm{dd}} \approx \frac{1}{2} \int  \Phi_{\mathrm{TF}}(\mathbf{r}) \ \left[ n_{\mathrm{TF}}(\mathbf{r}) 
+ n_{\mathrm{v}}(\mathbf{r}) \right]
 \mathrm{d} \mathbf{r}  
\end{equation}
i.e.\ replace the true $\Phi(\mathbf{r})$ by that purely due to the unperturbed background  $\Phi_{\mathrm{TF}}(\mathbf{r})$ which is known analytically and is given in Eq.\ (\ref{eq:phiddinside}). Since $\Phi_{\mathrm{TF}}(\mathbf{r})$ is a quadratic function of the coordinates this integral can be done exactly \cite{ODell07}. Finally, to find the energy $E_{\rm v}$ associated with exciting a vortex it is necessary to subtract from $E$ the energy $E_{0}$ of the vortex-free state, but this latter energy is also known analytically and is given as $E_{\mathrm{TF}}$ in Eq.\ (\ref{eq:energyfunctionalTF}). 

In this way $E_{\rm v}$, and hence $\Omega_{\rm v}$, can be computed for the trapped dipolar BEC. It is found that dipolar interactions increase $\Omega_{\rm v}$ in prolate traps and lower it in oblate traps when compared to the pure $s$-wave case  \cite{ODell07,Abad10b}. Intuitively, this makes sense because in the prolate case 
dipolar interactions tend to reduce $R_{\perp}$ but in the oblate case they increase it. The rotational energy density $(1/2) n(\rho,z) v^2(\rho)$ is \emph{lower} at larger radii because $v \propto 1/\rho$ and hence $E_{\rm v}$ is lowered if atoms are moved to larger radii, like in the oblate case, and vice-versa in the prolate case. This interpretation is backed up by the following expression for the critical rotation frequency derived in the pure $s$-wave case in the Thomas-Fermi limit \cite{Lundh97},
\begin{equation}
\Omega_{\rm v}=\frac{5}{2} \frac{\hbar}{mR_{\perp}^{2}} \ln \frac{0.67 R_{\perp}}{\xi} .
\label{crit_freq}
\end{equation}
The numerical factors 5/2 and 0.67 arise from the inverted parabola of the Thomas-Fermi density profile: since the parabolic profile is maintained in the dipolar case we expect a similar expression to hold there. If $R_{\perp}$ in Eq.\ (\ref{crit_freq}) is replaced by its dipolar version as given in Eq.\ (\ref{eq:RperpTF}), the resulting prediction for $\Omega_{\rm v}$ is in very close agreement with the variational calculation described above \cite{ODell07}. In principle, the healing length should also be changed to its dipolar version but this only leads to a logarithmic correction. 

\subsection{Vortex in a quasi-two-dimensional dipolar Bose-Einstein condensate}
We now review the vortex solutions in the simpler context of the homogeneous quasi-two-dimensional dipolar condensate, following the work of Refs. \cite{Mulkerin14,Mulkerin13}. Figure~\ref{fig:single_vortices} \cite{Mulkerin14} plots these solutions, found by numerical solution of the quasi-two-dimensional dipolar GPE, as a function of $\edd$.  

The vortex profile has a non-trivial dependence on the polarization angle and $\edd$. For $\alpha=0$, the vortex density is axisymmetric.  For $\edd=0$ the condensate is non-dipolar and the vortex takes the standard form [see left inset of Figure~\ref{fig:single_vortices}(a)]  of a circularly-symmetric core of vanishing density at the centre that monotonically returns to its background value over a healing length $\xi=\hbar/\sqrt{m\mu}$ \cite{Pethick&Smithbook}.  Since the length scaling applied in Figure~\ref{fig:single_vortices} is the dipolar healing length, the vortex core structure shown in Figure~\ref{fig:single_vortices}(a) for $\edd \neq 0$ remains, for the most part, similar to that for $\varepsilon_{\rm dd}=0$, i.e.\ the main effect of dipolar interactions is to rescale the size of the vortex core, but not change its structure. The exceptions to this are close to the phonon and roton instabilities.  As  the phonon instability boundary at $\varepsilon_{\rm dd}=-0.5$ is approached from the stable side, the vortex core takes on a funnel-like profile [middle inset of Figure~\ref{fig:single_vortices}(a)]. This is associated with the cancellation of explicit $s$-wave van der Waals interactions in the system, i.e.\ the van der Waals interactions cancel the contact contribution from the dipolar interactions.  The right inset of  Figure~\ref{fig:single_vortices}(a) also shows that as the roton instability is approached density ripples emerge around the vortex core. Moving away from the vortex core these ripples decay. The maximum amplitude of the density ripples is  $\sim 20\%$ of $n_0$, and their wavelength is of order the roton wavelength $\approx 4 \xi$.

For $\alpha \neq 0$, see Figures~\ref{fig:single_vortices}(b,c), the vortex profile becomes anisotropic.  In particular, as the roton instability is approached, density ripples again form, but now aligned in the direction of the attractive dipolar interactions (along the polarization direction for $C_{\rm dd}>0$ and perpendicular for $C_{\rm dd}<0$).  These anisotropic ripples are related to the anisotropic mixing of the roton into the ground state \cite{Ticknor11}.    

\subsubsection{Dipolar mean-field potential due to a vortex: giant anti-dipoles \label{subsubsection:vortexpotential}}
\begin{figure}[b]
\centering	\includegraphics[width=0.47\textwidth,angle=0]{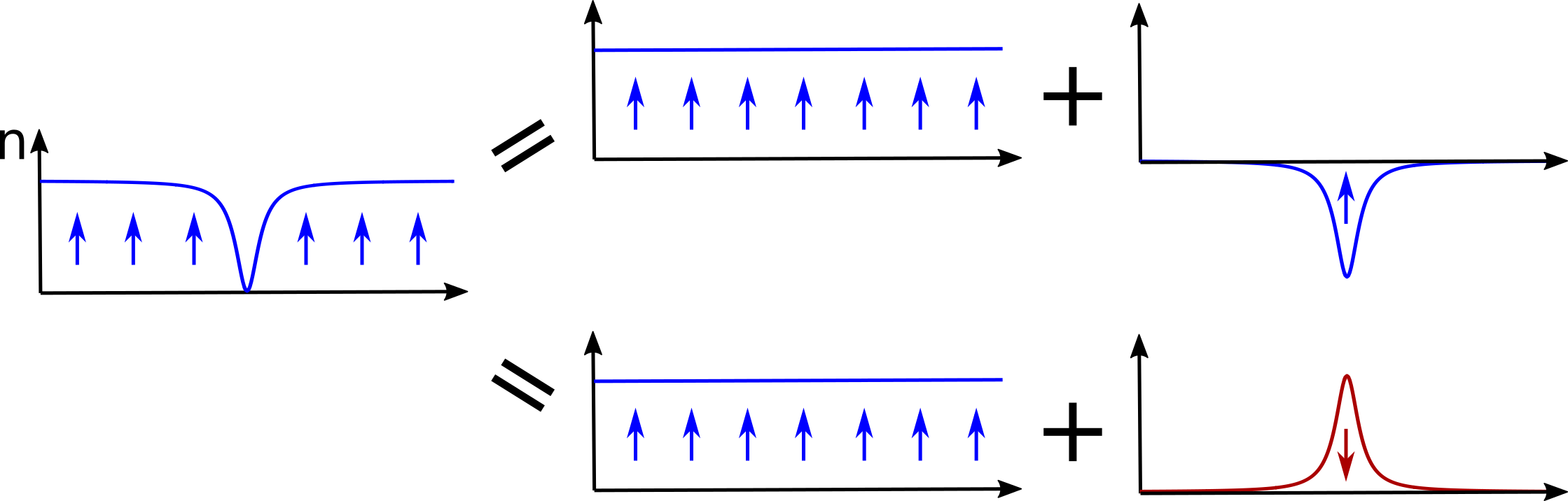}
		\caption{Schematic of the decomposition of a density featuring a vortex into a uniform density $n_0$ and a negative vortex density $n_{\rm v}$. }
		\label{fig:vortex_decomposition}
\end{figure}

Considered as a density defect in a homogenous background, the vortex gives rise to its own dipolar mean-field potential. Furthermore, because the creation of a vortex core displaces a large number of atomic dipoles, the vortex can be treated as a single giant anti-dipole \cite{Klawunn08,ODell07}.   Take the case of a vortex within an otherwise uniform background of density $n_0$, with the density expressed using the decomposition as in Eq. (\ref{eqn:vortex_profile2}).  Then, by noting that the dipolar potential $\Phi$ is a linear functional of density, the total dipolar potential can be written as,
\begin{eqnarray}
\Phi[n]({\bf r})&=&\Phi[n_0+n_{\rm v}]({\bf r}) \nonumber \\
&=&\Phi[n_0]({\bf r})+\Phi[n_{\rm v}]({\bf r})=\Phi_0+\Phi[n_{\rm v}]({\bf r}),
\end{eqnarray}
i.e.\ a contribution $\Phi_0$, which is just a constant, from the uniform background $n_0$ and a spatially dependent contribution $\Phi[n_{\rm v}]$ from the hole created by the vortex core, i.e.\ the giant anti-dipole.  This decomposition, illustrated in Figure~\ref{fig:vortex_decomposition}, assists in understanding the interaction between vortices in dipolar systems.

Consider a vortex in the quasi-two-dimensional dipolar condensate.  For $\alpha=0$ and away from the roton and phonon instabilities, the vortex profile is well-approximated by the non-dipolar ansatz given by Eqs.\ (\ref{eqn:vortex_profile1}) and (\ref{eqn:vortex_profile2}), where the healing length is taken to be the dipolar healing length $\xi=\hbar/\sqrt{m\mu}$.  The dipolar potential generated by the density defect can be determined via the convolution result $\Phi_{\perp}(\boldsymbol{\rho},t)=\mathcal{F}^{-1} \left[\tilde{U}^\perp_{\rm dd}({\bf q}) \tilde{n}_{\perp}({\bf q},t)\right]$.  Due to the cylindrical symmetry of the $\alpha=0$ case, one can perform the Fourier transforms through Hankel transforms.  The Hankel transform of the two-dimensional equivalent of Eq. (\ref{eqn:vortex_profile2}) is,
\begin{eqnarray}
\tilde{n}_{\perp}(\mathbf{q})=n_{0} \left(\frac{\delta (q)}{q}+\frac{K_0( q \xi / \sqrt{2})}{2}
\right),
\end{eqnarray}
where $K_0(\cdot)$ is a modified Bessel function of the second kind. Expanding the dipolar interaction potential  $\tilde{U}^\perp_{\rm dd}(\mathbf{q})$ given in Eq.~(\ref{eq:dipolar_interaction_2D}) with $\alpha=0$  as a series in the condensate width parameter, $\sigma$, gives
\begin{eqnarray}
\tilde{U}^\perp_{\rm dd}(\mathbf{q}) = \frac{4\pi C_{\rm dd} }{9 \sqrt{2 \pi} l_z}\left(2-\sqrt{\frac{9 \pi }{2}} q l_z \sigma \right) +\mathcal{O}\left(\sigma ^2\right).
\label{eq:U_2d_expansion}
\end{eqnarray}
Then, to first order in $\sigma=l_z/\xi$ and third order in $1/\rho'=\xi/\rho$ \footnote{The first (third) order expansion in $\sigma$ ($1/\rho'$) allows the leading long-range dipolar contribution to $\tilde{U}^\perp_{\rm dd}(\mathbf{q})$ to be evaluated ($\propto \sigma$).}, the dipolar potential due to a vortex is \cite{Mulkerin13,Mulkerin14},
\begin{equation}
\Phi_\perp(\mathbf{\rho})= \Phi_0 \left[1 -\frac{1}{\rho'^{2}}+\left(\frac{A \ln \rho' + B}{\rho'^{3}}\right) \sigma \right], \label{eq:analytic_phidd}
\end{equation}
with constants $A=-\sqrt{9\pi/8} \approx -1.88$, $B=(\ln 2-1)A\approx 0.577$, and $\Phi_0$ the dipolar potential at infinity.  
\begin{figure}[b]
\centering
	\includegraphics[width=0.48\textwidth]{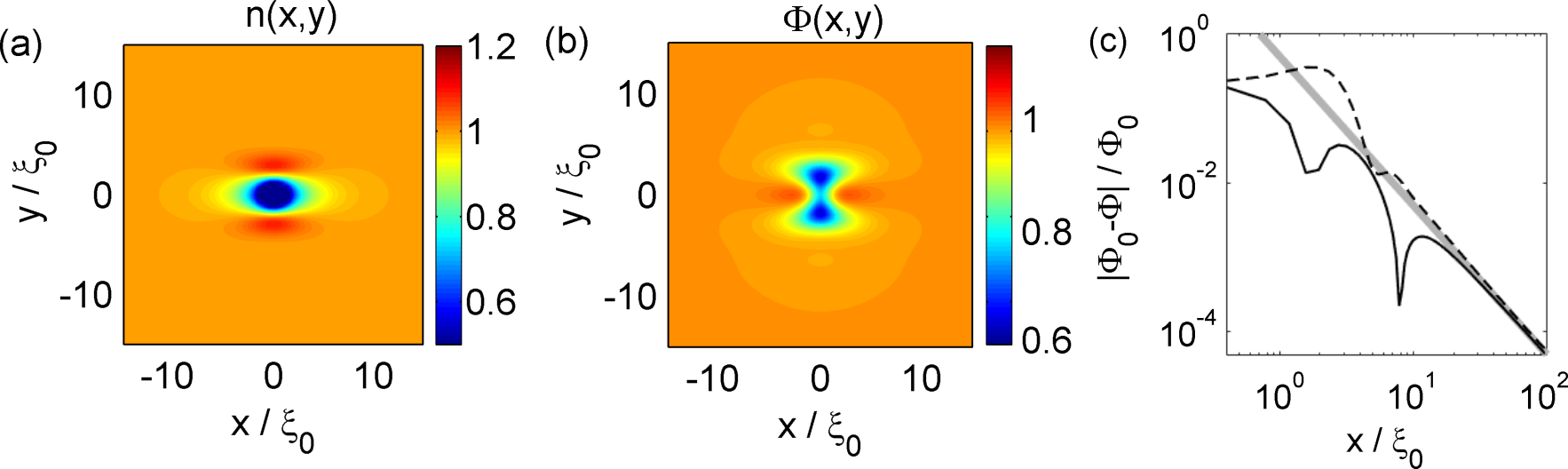}
\caption{Vortex in the quasi-two-dimensional dipolar condensate, with the dipoles polarized at $\alpha=\pi/4$ along $x$ ($\varepsilon_{\rm dd}=5$ and $\sigma=0.5$).  (a) Density profile $n_{\perp}(x,y)$, expressed in terms of the density at infinity $n_0$. (b) Dipolar potential $\Phi_{\perp}(x,y)$, rescaled by the homogeneous value $\Phi_0$. (c) The decay of $\Phi_{\perp}$ along $x=0$ (dashed black line) and $y=0$ (solid black line) recovers the $1/\rho'^2$ scaling at large distance (grey line).  Figure adapted from Ref. \cite{Mulkerin14}.}
	\label{fig:vortex_dipolar_potential2}
\end{figure}

This result  indicates that the vortex causes a reduction in the mean-field dipolar potential which is consistent with the reduced density of dipoles in the vicinity of the vortex.  One also sees from Eq.~(\ref{eq:analytic_phidd}) that the dipolar potential generated by the vortex relaxes predominantly as $1/\rho'^2$ to the background value $\Phi_0$, and this is confirmed by numerical solutions,  as shown in Figure~\ref{fig:vortex_dipolar_potential2}(c) \cite{Mulkerin14}.  This dependence arises because the vortex density itself relaxes as $1/\rho'^2$, and the leading contribution to $\Phi_{\rm v}$ is from the local density.  Indeed, the mean-field potential due to van der Waals interactions from the vortex also scales in proportion to the local density with a $1/\rho'^2$ dependence \cite{Wu94}.  The long-range contribution to  $\Phi_{\rm v}$ can be interpreted as arising from effective anti-dipoles in the vortex core, see Figure~\ref{fig:vortex_decomposition}. This non-local contribution is represented in Eq.~(\ref{eq:analytic_phidd}) by terms linear in $\sigma$. In the limit $\sigma \rightarrow 0$ the volume of the anti-dipoles in the vortex core vanishes and hence this long-range contribution also vanishes. Unlike the topological potential associated with quantised superfluid flow around a vortex core the dipolar potential due to the vortex core is not topological, i.e. it depends on the volume of the vortex core. The dominant contribution to the non-local vortex potential scales as  $\ln \rho'/\rho'^3$, i.e. the absence of dipoles in the vortex core can not be considered to be point-like. This is unsurprising since in the vicinity of the vortex core the density scales as a power law \cite{Pethick&Smithbook}. 

For dipoles tilted away from the vertical ($\alpha \neq 0$), the vortex core and its dipolar potential become anisotropic and an analytic treatment is challenging.  Figure~\ref{fig:vortex_dipolar_potential2}(a,b) shows an example numerical solution of the vortex density and dipolar potential for $\alpha \neq 0$.  The dipolar potential is indeed anisotropic about the vortex.  Remarkably, the modification to the dipolar potential induced by the vortex mimics the dipole-dipole interaction itself, with an angular dependence which resembles $1-3\cos^2 \theta$, being reduced (attractive) along $y$ and increased (repulsive) along $x$.  Thus, at least in its angular dependence, the vortex shares qualitatively the characteristics of a mesoscopic dipole.  Like the above $\alpha=0$ case, the dipolar potential is found to decay at long-range as $1/\rho'^2$ to the background value, as seen in Figure~\ref{fig:vortex_dipolar_potential2}(c).  At short range, $\rho \lappeq 10 \xi$, the dipolar potential is dominated by the core structure.  

\section{Vortex Pairs: Interactions and Dynamics \label{sec:dynamics}}
\subsection{Interaction between vortices}
Two vortices (or indeed two elements of the same three-dimensional vortex line \cite{Klawunn08,Svidzinsky00,Koens12}) have a well-known hydrodynamic interaction due to the kinetic energy associated with the mutual cancellation/reinforcement of their velocity fields.  Consider a cylindrical condensate of radius $R$ and height $L$, featuring two straight vortices at planar positions $\bfrho_1$ and $\bfrho_2$, a distance $d$ apart.  The vortices have charge $q_1$ and $q_2$, and individual velocity fields ${\bf v}_1$ and ${\bf v}_2$, respectively.  The net velocity field of the two vortices is ${\bf v}_1+{\bf v}_2$.  The energy of the vortices can be estimated by integrating the total kinetic energy across the system,
\begin{eqnarray}
E_{\rm kin}=L\int \frac{1}{2}  m n(\bfrho) |{\bf v_1}(\bfrho)+{\bf v_2}(\bfrho)|^2\,{\rm d}\bfrho.
\label{eqn:kin}
\end{eqnarray}
For simplicity, one can ignore the vortex core density and set $n(\bfrho)=n$.  Assuming $\xi \ll d \ll R$ then the (kinetic) energy of the pair is,
\begin{equation}
E_{\rm kin}=
  \frac{L \pi n  \hbar^2}{m} \left[q_1^2 \ln \frac{R}{\xi}+q_2^2 \ln \frac{R}{\xi}+ 2 q_1 q_2 \ln \frac{R}{d}  \right].
\end{equation}
Note that, to avoid singularities in the velocity field the integration region excludes a disc of radius of one healing length about each vortex centre.  The first two terms are the energies of the individual vortices if they were isolated, and the third term is the pair interaction energy.  For a vortex-antivortex pair ($q_1=-q_2$) the interaction energy is negative, whilst for a co-rotating pair ($q_1=q_2$) it  is positive. This is explained physically by the fact that for a vortex-antivortex pair the flow fields tend cancel out in the bulk, reducing the net kinetic energy in the bulk, whilst for a co-rotating pair the flow fields tend to reinforce, increasing the total kinetic energy.  

\begin{figure}[b]
	\includegraphics[width=1\columnwidth,angle=0]{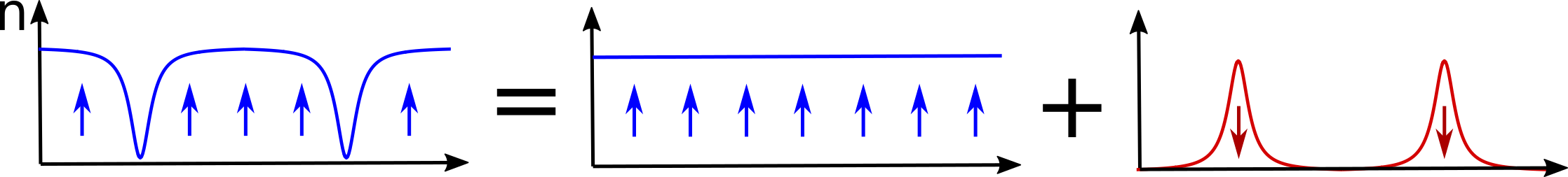}
		\caption{The dipolar interaction between two vortices may be interpreted as the interaction between two collections of anti-dipoles.  
}
		\label{fig:vortex_decomposition2}
\end{figure}

\begin{figure}[t]
\centering	\includegraphics[width=0.48\textwidth]{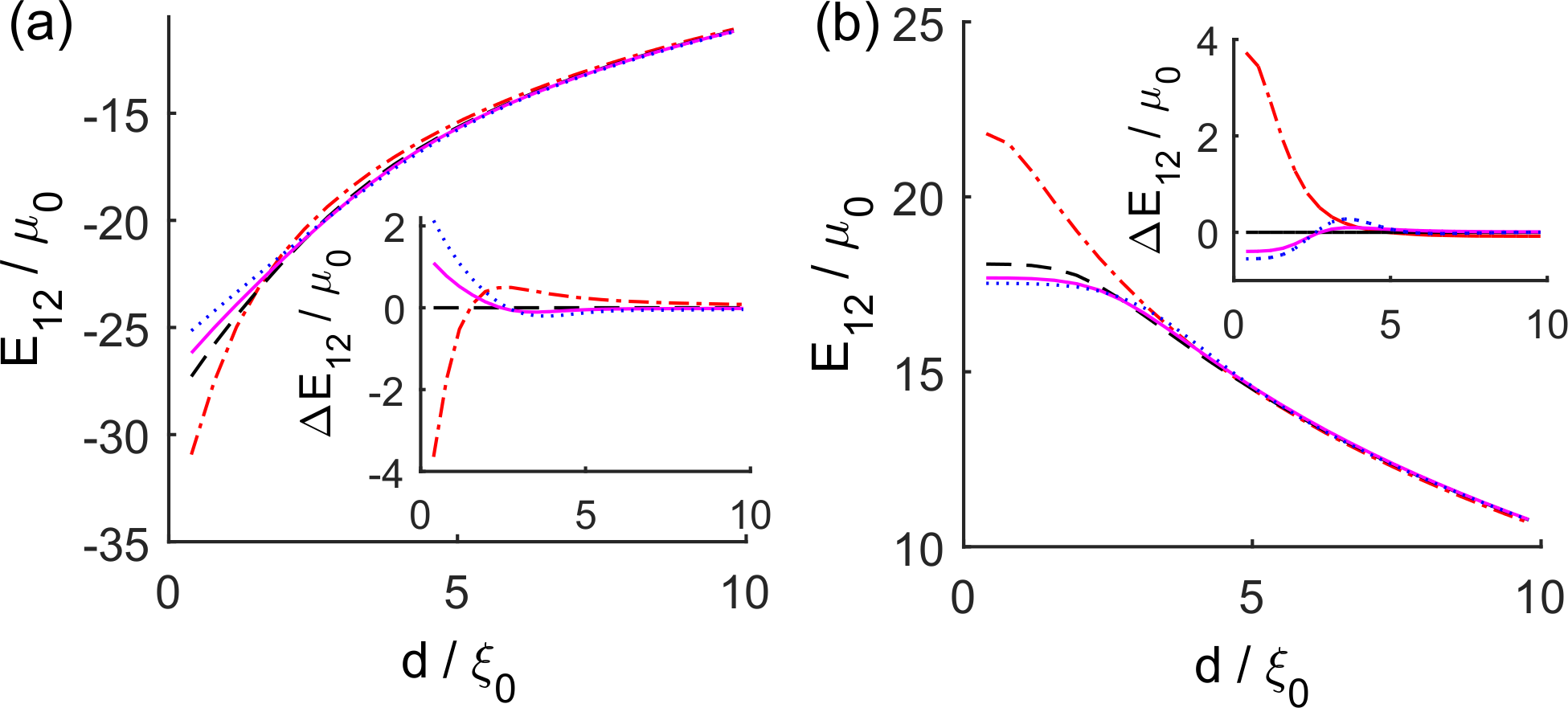}
	\caption{Vortex interaction energy $E_{12}$ versus separation $d$ for (a) vortex-antivortex and (b) vortex-vortex pairs in the quasi-two-dimensional dipolar condensate, with dipoles polarized along $z$ ($\alpha=0$), shown for various values of $\edd$: $\edd=0$ (dashed black line), $\edd=-1.4$ (dotted blue line), $\edd=-0.45$ (dot-dashed red line) and $\edd=5$ (solid magenta line).  Insets show deviation from the non-dipolar value.  Figure adapted with permission from \cite{Mulkerin13}. Copyright 2013 by the American Physical Society. }
\label{fig:ints}
\end{figure}

In the presence of dipolar interactions the vortices feature an additional long-range interaction.  This interaction can be pictured as the interaction between two lumps of anti-dipoles in empty space, as illustrated in Figure~\ref{fig:vortex_decomposition2}.  
Before we review how two vortices interact in the presence of dipolar interactions, we first make more precise the definition of the vortex energy introduced in Section \ref{sec:single_vortices}, and hence allow the identification of vortex-vortex interaction energy.  In non-dipolar condensates, the vortex energy is conventionally defined as the energy difference between a system with and without a vortex, where both systems have the same number of particles \cite{Pethick&Smithbook}.   Imagine first a system (quasi-two-dimensional) with a vortex and $N$ particles covering an area $A$.  If the asymptotic density is $n_0$, then the number of particles in this system can be expressed as,
\begin{equation}
N= A n_0 - \int_A \left(n_0 - |\psi_{\perp}|^2 \right)\,{\rm d} \boldsymbol{\rho}.
\label{eqn:normV}
\end{equation}
Now consider the system without a vortex, but with the same number of particles.  It has constant density $n_0=N/A$ and its energy is
\begin{equation}
E_0=\frac{N^2}{2\sqrt{2\pi} l_zA} g_{\rm eff},
\label{eqn:energy2}
\end{equation}
where $g_{\rm eff}=g+(C_{\rm dd}/3)[3 \cos^2 \alpha-1]$. Inserting Eq. (\ref{eqn:normV}) into Eq. (\ref{eqn:energy2}) gives,
\begin{equation}
E_0 \approx \frac{g_{\rm eff}}{2\sqrt{2\pi}l_z} \left ( A n_0^2 - 2n_0 \int \limits_A  \left (n_0 - |\psi_{\perp}|^2 \right)\,{\rm d}\boldsymbol{\rho}\right),
\end{equation}
where a term, negligible in the limit $A \gg \xi^2$, has been omitted. Defining the energy of a single vortex as $E_1=E-E_0$,  the interaction energy between two vortices must be \cite{Pethick&Smithbook},
\begin{equation}
E_{12}(\boldsymbol{\rho}_1-\boldsymbol{\rho}_2)=E_2(\boldsymbol{\rho}_1,\boldsymbol{\rho}_2)-E_1(\boldsymbol{\rho}_1)-E_1(\boldsymbol{\rho}_2),
\end{equation}
where $E_2(\boldsymbol{\rho}_1,\boldsymbol{\rho}_2)$ is the energy of the 2-vortex system.  This is plotted in Figure~\ref{fig:ints} for $\alpha=0$ and for (a) vortex-antivortex (VA) pairs and (b) vortex-vortex (VV) pairs, as a function of their separation $d$, based on numerical dipolar GPE two-vortex solutions \cite{Mulkerin13}.
\begin{figure}[b]
\centering	\includegraphics[width=0.48\textwidth]{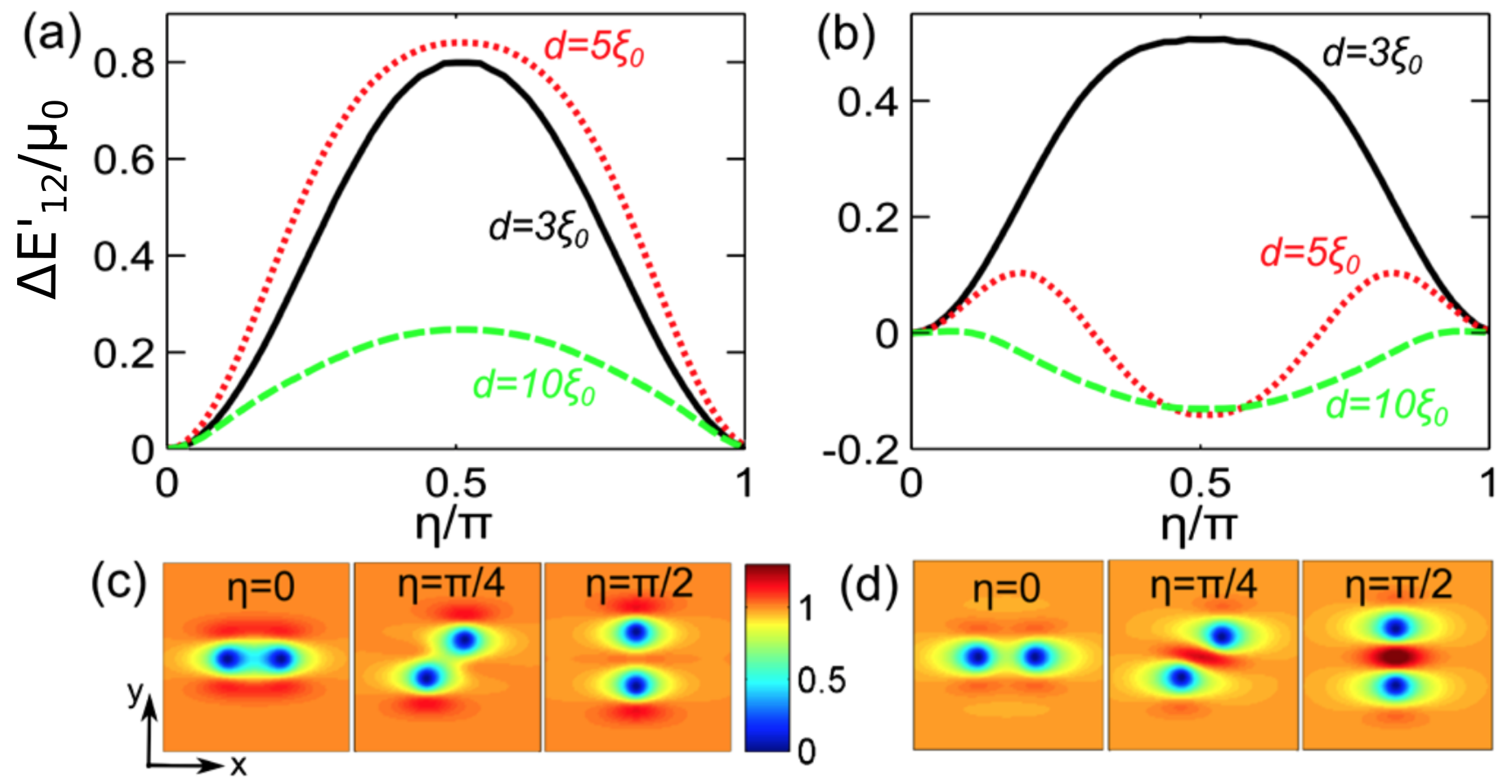}
	\caption{Angular dependence of the vortex interaction energy $\Delta E'_{12}(\eta)=E_{12}(\eta)-E_{12}(\eta=0)$, for (a) VA and (b) VV pairs, for various separations  $d$.  Parameters: $\alpha=\pi/4$, $\sigma=0.5$ and $\edd=5$.   Example (c) VA and (d) VV pair density profiles for $d=5\xi$ over an area $(20\xi)^2$.  Reprinted figure with permission from \cite{Mulkerin13}. Copyright 2013 by the American Physical Society.}
\label{fig:angle}
\end{figure}

In the absence of dipoles, $E_{12}$ increases with $d$ for the VA pair and decreases for the VV pair.  For $d \gg \xi$, $E_{12}$ closely follows the logarithmic scaling of the hydrodynamic prediction, while for $d \lappeq \xi$ the overlap of the cores causes a breakdown of the logarithmic behaviour.  With dipoles, $E_{12}$ is significantly modified at short and  intermediate length scales up to $d \approx 5 \xi$, but at larger scales the effects of the dipoles are small in comparison to the hydrodynamic effects.  The modification due to the dipoles arises from a non-trivial combination of the modified density profile and non-local interactions.

When the dipoles are tilted in the plane, $E_{12}$ becomes dependent on the in-plane angle of the pair relative to the polarization direction, $\eta$.  This is illustrated for VA and VV pairs by the examples in Figure~\ref{fig:angle} \cite{Mulkerin13}.   For small separations, the angular dependence is dominated by local effects arising from the density profile of the pairs, particularly by any density ripples.  However, for $d \gg \xi$, $\Delta E'_{12}=E_{12}(\eta) - E_{12}(\eta=0)$ approaches a sinusoidal dependence on $\eta$, analogous to the dipolar interaction itself.

\subsection{Dynamics of vortex pairs}
In a two-dimensional system, a vortex co-moves with the local fluid velocity.  Thus, for a vortex pair, each vortex is carried along by the flow field of the other vortex.  
  For a VA pair this means that the vortices move in the same direction, perpendicular to the inter-vortex axis.   This solitary wave has speed $v = \hbar/m d$ for well-separated vortices.   
\begin{figure}[b]
\centering
	\includegraphics[width=0.5\textwidth]{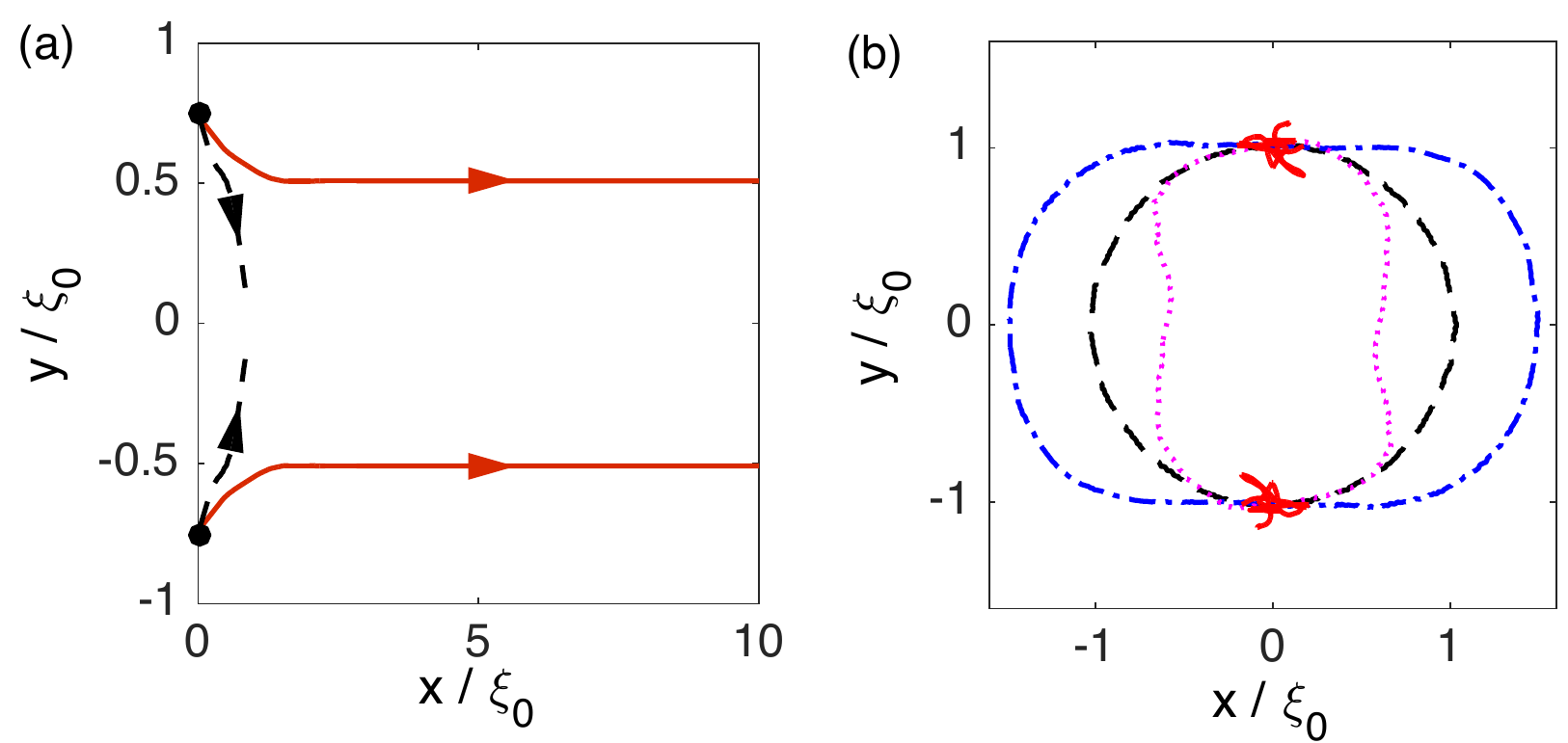}
	\caption{(a) For a given initial separation, a non-dipolar VA pair (dashed black lines) annihilates while dipolar interactions ($\varepsilon_{\rm dd}=-0.4$, $\alpha=0$, solid red lines) support stable pair propagation (red lines).  (b) A non-dipolar VV pair co-rotates in a circular path (dashed black line).  Off-axis ($\alpha=\pi/4$) dipolar interactions lead to anisotropic paths (dot-dashed blue line: $\varepsilon_{\rm dd}=-1.5$; dotted magenta line: $\varepsilon_{\rm dd}=5$) and  suppression of co-rotation altogether (solid red lines: $\varepsilon_{\rm dd}=10$).  Reprinted figure with permission from \cite{Mulkerin13}. Copyright 2013 by the American Physical Society. }
\label{fig:pairs}
\end{figure}

When the VA separation is small, $d \sim \xi$, the vortex and anti-vortex are susceptible to annihilation, an event which results in a burst of density waves.  Numerical simulations \cite{Mulkerin13} show that dipoles modify this separation threshold [Figure~\ref{fig:pairs}(a)].  Moreover, since for $\alpha \neq 0$ the speed of sound varies with angle \cite{Ticknor11}, one can expect the pair speed to be anisotropic in space.

For a VV pair, the flow which carries each vortex now acts in opposite directions (again, perpendicular to the line separating the vortices and with the above speed), resulting in the co-rotation of the vortices about their mid-point.  Viewed another way, the vortices follow a path of constant energy; since the interaction energy in the absence of dipoles depends only on $d$, this path is circular. The same is true for axisymmetric dipoles, $\alpha=0$. However, for $\alpha \neq 0$, the vortices co-rotate on an anisotropic path, as shown in Figure~\ref{fig:pairs}(b) \cite{Mulkerin13}, due to the anisotropic interaction energy of the pair [refer to Figure~\ref{fig:angle}]. Moreover, for extreme cases where the vortex is highly elliptical with significant ripples, co-rotation can be prevented altogether, with the vortices being localized [Figure~\ref{fig:pairs}(b), red solid lines].  In this limit, the vortices act as extended, highly-elongated objects, with effective geometrical restrictions on their motion past each other, reminiscent of the smectic phase of liquid crystals.

Gautam \cite{Gautam14b} numerically considered the  dynamics of a corotating VV pair in a dipolar BEC in the presence of dissipation. For symmetric configurations in the trap, the vortices decay with equal decay times, while for asymmetric initial configurations the decay is modified with one vortex decaying slower at the expense of the other.  

\section{Generation of Vortices \label{sec:gen_vortices}}
\subsection{Summary of vortex generation methods}
In conventional condensates, vortices have been generated through several mechanisms.  Below we list the main ones, as well as relevant considerations in the presence of dipoles.

\begin{itemize}
\item The most common and intuitive approach to generate vortices is via mechanical rotation of the system \cite{Madison00,Hodby01,Aboshaeer01}, analogous to the ``rotating bucket'' experiments  in Helium II \cite{Donnelly91}.  Both the thermodynamic threshold for vortices to be favoured, as well as the process by which vortices nucleate \cite{Bijnen07,Martin08,Bijnen09} into the condensate, are sensitive to dipolar interactions; this will be analysed in detail below.

\item Motion of a localized obstacle or potential (as generated by a tightly-focussed blue-detuned laser beam) through a condensate (or, equivalently, motion of the condensate relative to a static obstacle) leads to the nucleation of vortices above a critical relative speed \cite{Neely10,Kwon15,Raman01}, forming a quantum wake downstream of the obstacle.   The critical speed is related to the Landau criterion which predicts the formation of elementary excitations in the fluid for relative speeds exceeding $v_{\rm c}={\rm min}[\omega({\bf k})/k]$ \cite{Pethick&Smithbook,Pitaevskii&StringariBook}.  Ticknor {\it et al} \cite{Ticknor11} examined this process in a quasi-two-dimensional dipolar condensate.  For dipoles tilted into the plane ($\alpha >0$), the critical speed becomes anisotropic, a consequence of Landau's criterion and the anisotropic dispersion relation in the plane.  The critical velocity for vortex nucleation can also be derived by considering the energetics of a vortex-antivortex pair \cite{Pines&NozieresBook}, implying that the aniostropic critical velocity is directly related to the anisotropic vortex interaction energy.

\item The phase of the condensate can be directly engineered via optical imprinting to produce vortex phase singularities, as employed to generate both singly- and multiply-charged vortices \cite{Leanhardt02}. This mechanism is independent of the dipoles themselves.  

\item Following a rapid quench through the transition temperature for the onset of Bose-Einstein condensation, the growth of local phase-coherent domains leads to the entrapment of phase singularities and hence vortices (i.e. the Kibble-Zurek mechanism \cite{Kibble76,Zurek85}) \cite{Freilich10,Donadello14,Weiler08}.  A relevant consideration is the effect of the dipoles on the critical temperature.  This shift is sensitive to the shape of the trap, relative to the polarization direction, but is only up to a few percent for $^{52}$Cr \cite{Glaum07}.  However, this shift may be more significant in $^{168}$Er and $^{164}$Dy condensates.

\item Dark solitons are dimensionally unstable to decay into vortex pairs or vortex rings \cite{Anderson01,Scott03,Scott04} via the so-called snake instability, previously established in nonlinear optics \cite{Proukakis04}.  As shown by Nath {\it et al.} \cite{Nath08} the nonlocal character of dipolar interactions can stabilize the dark soliton against this instability.

\item Instead of rotating the system it is possible to introduce a time-reversal symmetry breaking synthetic magnetic field \cite{Lin09}. Since the atoms in an atomic BEC are charge-neutral it may at first seem counter-intuitive to consider the effects of  a synthetic magnetic field in these systems. However, time-reversal symmetry-breaking in charge-neutral systems can be overcome through mechanical rotation and exploiting the equivalence of the  Coriolis force and the Lorentz force to create a synthetic vector potential which gives rise to a synthetic magnetic field. This case of mechanical rotation is analysed in detail below. It is also possible to realise an optically synthesised vector potential field for BECs using  spatially dependent optical coupling between internal states of the atoms in the condensate. This spatially dependent coupling can yield Berry phases \cite{Berry84} sufficient to create large synthetic magnetic fields.  As in the case of mechanical rotation, vortex nucleation is dependent on the properties of the stationary solutions and can be analyzed in the Thomas-Fermi approximation \cite{Taylor11}.  Numerical investigations of the dipolar GPE, carried out by Zhao and Gu \cite{Zhao15}, find that the nucleation of vortices depends on the dipole strength, the strength of the synthetic magnetic field, the potential geometry, and the orientation of the dipoles, with anisotropic interactions significantly altering vortex nucleation.
\end{itemize}

\subsection{Stationary solutions of rotating dipolar condensates in elliptical traps \label{sec:dip_ellip}}
The most common approach for generating vortices and vortex lattices in trapped condensates is via rotation.  Since a cylindrically-symmetric trap set into rotation applies no torque to the condensate, the trap is made anisotropic in the plane of rotation.  In the simplest case, this leads to a trap which is weakly elliptical in the plane of rotation \cite{Madison00,Hodby02}, with a potential of the form,
\begin{equation}
V({\bf r})= \frac{1}{2}m\omega_\perp^2\left[(1-\epsilon)x^2 + (1+\epsilon)y^2+\gamma^2 z^2 \right],
\label{eqn:harmonicV}
\end{equation}
where rotation is performed about the $z$-axis.

For typical parameters in the absence of dipolar interactions, vortices become energetically favourable in harmonically-trapped condensates for rotation frequencies $\Omega \sim 0.3 \omega_{\perp}$.  Surprisingly, in non-dipolar BEC experiments the observed nucleation of vortices occurs at considerably larger rotation frequencies $\Omega \sim 0.7 \omega_\perp$.  Theoretical analysis based on the hydrodynamic equations reveals the important role of collective modes.  Specifically, for $\Omega \lesssim 0.7 \omega_\perp$ low-lying collective modes are excited via elliptical deformation. The seeding of vortices, at higher rotation frequencies, arises when one or more of these modes becomes unstable  \cite{Recati01,Sinha01}.  Evidence for this comes from comparison between experiments \cite{Hodby02,Madison01} and full numerical simulations of the GPE \cite{Lundh03,Lobo04,Parker06,Corro07}. 

The hydrodynamic description of condensates in rotating elliptical traps can be extended to include dipolar interactions.  For rotation about the $z$-axis, described by the rotation vector ${\bf \Omega}$ where $\Omega=|{\bf \Omega}|$ is the rotation frequency, the Hamiltonian  in the rotating frame is given by,
\begin{equation}\label{Heff}
H_{{\rm eff}} = H_0 - \boldsymbol{\Omega} \cdot \hat{L},
\end{equation}
where the Hamiltonian in the absence of rotation is $H_0$  and the quantum
mechanical angular momentum operator is $\hat{L} = -i\hbar ({\bf r} \times \boldsymbol{\nabla})$. Applying this result for $\boldsymbol{\Omega}=(0,0,\Omega)$, the dipolar GPE in the rotating frame is \cite{Leggett_Book,Leggett00},
\begin{eqnarray}
i \hbar \frac{\partial \Psi}{ \partial t}&=&
\left[-\frac{\hbar^2}{2m}\nabla^2 + V+
\Phi+g|\Psi|^2 \right. \nonumber \\
&~& \left. - \Omega\frac{\hbar}{i} \left(x \frac{\partial}{\partial y}
- y \frac{\partial}{\partial x} \right)\right]
\Psi,\label{GPE_rotating_d}
\end{eqnarray}
where in the rotating frame the trapping potential $V$, given by Eq.\ (\ref{eqn:harmonicV}), is stationary. The spatial coordinates ${\bf r}$ are those of the rotating frame, with the  momentum coordinates expressed in the laboratory frame \cite{Leggett_Book,Leggett00,Lifshitz}.

Using the Madelung transform, as per Section \ref{sec:hydro_eqs}, leads to the following dipolar hydrodynamical equations in the rotating-frame
\begin{eqnarray}\label{cont_eq_d}
\frac{\partial n}{\partial t} = -\boldsymbol{\nabla} \cdot \left[
n\left({\bf v} - \boldsymbol{\Omega} \times
{\bf r}\right) \right],\\
m \frac{\partial {\bf v}}{\partial t} = -\boldsymbol{\nabla} \left(
\frac{1}{2} m v^2 + V + g n+
\Phi - m {\bf v} \cdot \left[ {\bf \Omega}
\times {\bf r} \right] \right),\nonumber \\ \label{motion_eq_d}
\end{eqnarray}
where the quantum pressure is assumed to be small and is neglected, i.e.\  the Thomas-Fermi limit.

Stationary solutions of Eqs.\
(\ref{cont_eq_d}) and (\ref{motion_eq_d}) satisfy the
equilibrium conditions,
\begin{equation}\label{stationary_solutions}
\frac{\partial n}{\partial t} = 0 \hspace{0.5cm} {\rm and}  \hspace{0.5cm} \frac{\partial
{\bf v}}{\partial t} = 0.
\end{equation}
Assuming the irrotational ($\nabla \times {\bf v}=0$) velocity field ansatz \cite{Recati01}
\begin{equation}\label{velocity_ansatz}
{\bf v} = \alpha_{\rm v} \nabla(x y)
\end{equation}
permits us to examine the rotating solutions in terms of the velocity field amplitude $\alpha_{\rm v}$. A physical interpretation of velocity field amplitude, $\alpha_{\rm v}$, can be gained from  the continuity equation (\ref{cont_eq_d}). Specifically, it can be written as \cite{Pitaevskii&StringariBook}
\begin{equation}
\alpha_{\rm v} = - \mathcal{D} \Omega, \label{eq:alphadeformation}
\end{equation} 
where $\mathcal{D}$ is the deformation of the BEC in the $x-y$ plane,
\begin{equation}
\mathcal{D}=\frac{\langle y^2 - x^2 \rangle}{\langle y^2 + x^2 \rangle} = \frac{\kappa_{y}^{2}-\kappa_{x}^2}{\kappa_{y}^{2}+\kappa_{x}^{2}},
\end{equation}
where $\langle \ldots \rangle$ denotes the expectation value in the stationary state and  $\kappa_x=R_x/R_z$ and $\kappa_y=R_y/R_z$ represent the aspect ratios of the BEC along $x$ and $y$ with respect to $z$.

Combining Eqs.\ (\ref{motion_eq_d}) and (\ref{velocity_ansatz}) gives,
\begin{eqnarray}
\mu=\frac{m}{2} \left(\tilde{\omega}_x^2 x^2 +\tilde{\omega}_y^2 y^2
+\omega_z^2 z^2
\right)+gn({\bf r})+\Phi({\bf r}), \label{mu}
\end{eqnarray}
where the \textit{effective} trap frequencies $\tilde{\omega}_x$ and
$\tilde{\omega}_y$ are given by,
\begin{eqnarray}\label{omax}
\tilde{\omega}_x^2 = \omega_{\perp}^2(1 - \epsilon) + \alpha_{\rm v}^2 - 2
\alpha_{\rm v} \Omega, \\
\tilde{\omega}_y^2 = \omega_{\perp}^2(1 + \epsilon) + \alpha_{\rm v}^2 + 2
\alpha_{\rm v} \Omega.
\end{eqnarray}
The breaking of cylindrical symmetry means that the BEC has an ellipsoidal shape and in the Thomas-Fermi approximation it adopts an inverted parabolic density profile of the form,
\begin{eqnarray}
n({\bf r})=n_{\rm cd}\left(1-\frac{x^2}{R_x^2}-\frac{y^2}{R_y^2}-\frac{z^2}{R_z^2}\right)
\,\,\,\, {\rm for} \,\,\, n({\bf r}) \ge 0, \label{eqn:tf_general}
\end{eqnarray}
where $n_{\rm cd}=15N/(8\pi R_xR_yR_z)$. This is an exact solution of the stationary dipolar hydrodynamic equations given in Eqs.\ (\ref{cont_eq_d}) and (\ref{motion_eq_d}) with velocity field (\ref{velocity_ansatz}), and it only remains to find the radii $\{R_{x},R_{y},R_{z} \}$ and velocity amplitude $\alpha_{\rm v}$. The exact dipolar potential due to a parabolic density distribution of  general ellipsoidal symmetry, with dipoles aligned in the $z$-direction, i.e.\ the generalized version of that given in Eq.\ (\ref{eq:phiddinside}), is derived in the Appendices of Refs. \cite{Eberlein05} and \cite{Bijnen10}, and is given by,
\begin{eqnarray}
\Phi=3g\edd \times \nonumber
\\
 \left(\frac{n_{\rm cd} \kappa_x
\kappa_y}{2}\left[\beta_{001}-\frac{x^2\beta_{101}+y^2\beta_{011}+3z^2\beta_{002}}{R_z^2}\right]
- \frac{n}{3} \right), \nonumber \\
\label{eq:Phiellipsoid}
\end{eqnarray}
where  the coefficients $\beta_{ijk}$ are,
\begin{eqnarray}
\beta_{ijk}=\int_0^{\infty}\frac{1}{\left(\kappa_x^2+s\right)^{i+\frac{1}{2}}
\left(\kappa_y^2+s\right)^{j+\frac{1}{2}}
\left(1+s\right)^{k+\frac{1}{2}}}\,{\rm d}s, \nonumber \\
\end{eqnarray}
for integer-valued $i$, $j$ and $k$. Thus,  Eq.\ (\ref{mu}) can be rearranged to obtain an expression for the density profile \cite{Bijnen07,Martin08,Bijnen09},
\begin{eqnarray}
n&=&\frac{\mu -\frac{m}{2} \left(\tilde{\omega}_x^2 x^2 +\tilde{\omega}_y^2 y^2 +\omega_z^2 z^2 \right)}{g\left(1-\edd\right)} \nonumber \\
&+&\frac{3g\edd\frac{n_{\rm cd}\kappa_x \kappa_y}{2R_z^2}\left[x^2\beta_{101}+y^2\beta_{011}+3z^2\beta_{002}-R_z^2\beta_{001}\right] }{g\left(1-\edd\right)}. \nonumber \\
\label{rho} 
\end{eqnarray}
By equating the $x^2$, $y^2$ and $z^2$  terms in Eq.\ (\ref{eqn:tf_general}) and Eq.\ (\ref{rho}) three self-consistency conditions are found. These  conditions define the size and shape of the condensate,
\begin{eqnarray}
\kappa_{x}^2&=&\left(\frac{\omega_z}{\tilde{\omega}_{x}}\right)^2
\frac{1+\edd\left(\frac{3}{2}\kappa_x^3 \kappa_y
\beta_{101}-1\right)}{\zeta} \label{kx},
\\
\kappa_y^2&=&\left(\frac{\omega_z}{\tilde{\omega}_y}\right)^2 \frac{1+\edd\left(\frac{3}{2}\kappa_y^3\kappa_x \beta_{011}-1\right)}{\zeta} \label{ky}, \\
R_z^2&=&\frac{2gn_{\rm cd}}{m\omega_z^2}\zeta, \label{size}
\end{eqnarray}
where $\zeta=1-\edd\left[1-\frac{9 \kappa_x \kappa_y}{2}\beta_{002}\right]$. Furthermore, by inserting Eq.~(\ref{rho}) into Eq.~(\ref{cont_eq_d}), the stationary solutions are seen to satisfy the condition \cite{Bijnen07,Martin08,Bijnen09},
\begin{eqnarray}
0&=&\left(\alpha_{\rm v}+\Omega\right)\left(\tilde{\omega}_x^{2}-\frac{3}{2}\edd \frac{\omega_{\perp}^2\kappa_x\kappa_y \gamma^{2}}{\zeta} \beta_{101}\right) \nonumber \\
&+&\left(\alpha_{\rm v}-\Omega\right)\left(\tilde{\omega}_y^{
2}-\frac{3}{2}\edd \frac{\omega_{\perp}^2\kappa_x
\kappa_y \gamma^{2}}{\zeta} \beta_{011}\right). \label{alpha}
\end{eqnarray}
Equation~(\ref{alpha}) gives the velocity field amplitude $\alpha_{\rm v}$ for a given $\edd$, $\Omega$ and trap geometry. For $\edd=0$, $\alpha_{\rm v}$ is independent of the {\it s}-wave interaction strength $g$ and the trap ratio $\gamma$. Dipolar interactions qualitatively alter this scenario with $\alpha_{\rm v}$ becoming dependent on $\edd$ and $\gamma$. The  solutions to Eq.~(\ref{alpha}), as a function of $\Omega$, have significantly different properties depending on whether the traps  are circular ($\epsilon=0$) or elliptical ($\epsilon>0$) in the $x-y$ plane.  We restrict our analysis to  $\Omega<\omega_{\perp}$, since the static solutions are known to disappear for  $\Omega \sim \omega_{\perp}$ due to centre of mass instabilities \cite{Recati01}. Below, the circular ($\epsilon=0$) and elliptical ($\epsilon>0$) cases are considered.

\subsubsection{Circular trapping in the $x-y$ plane: $\epsilon=0$ \label{sec:circular_trap}}
\begin{figure}[b]
\centering
\includegraphics[width=6cm]{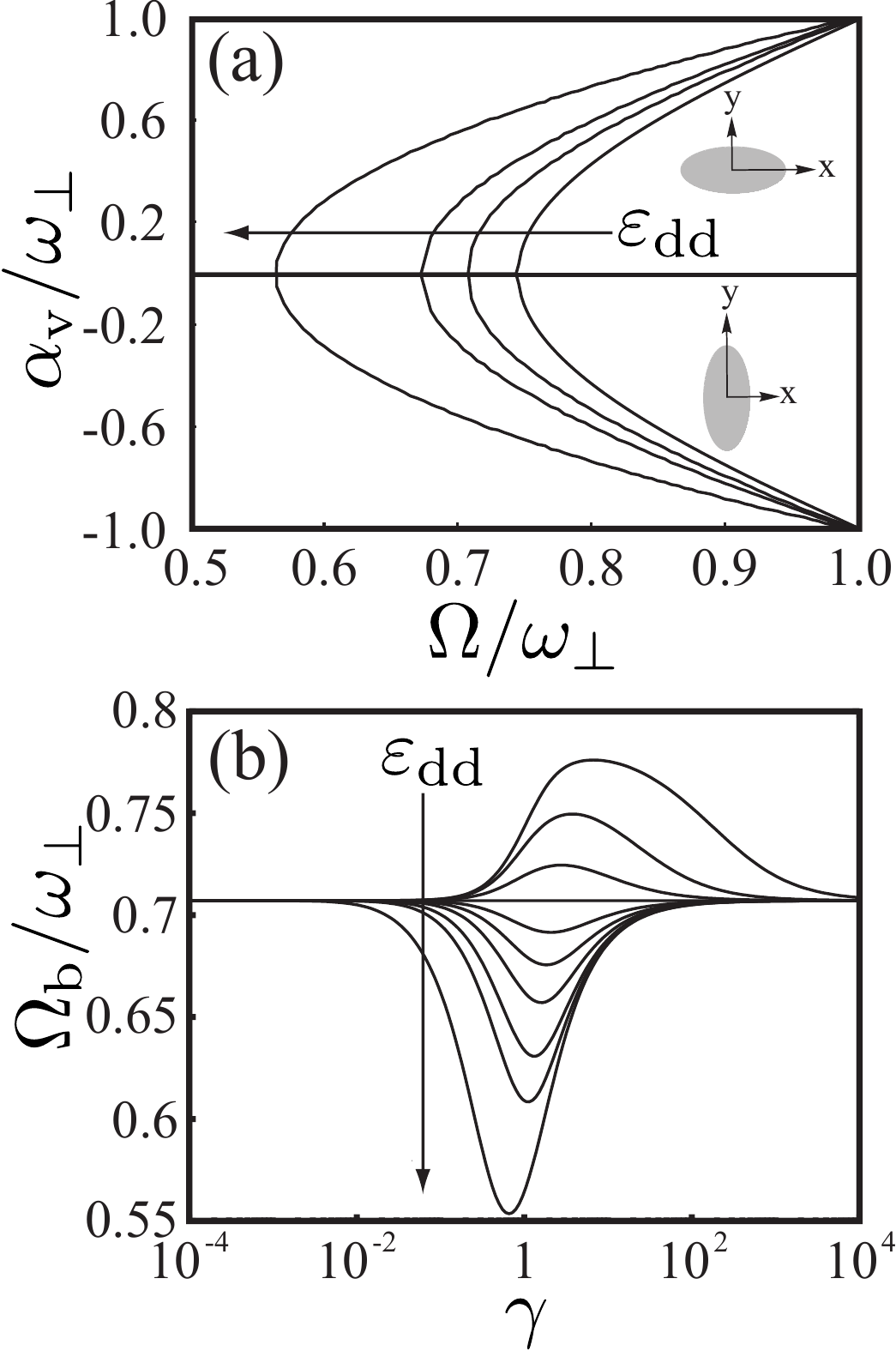}
\caption{(a) Irrotational velocity field amplitude $\alpha_{\rm v}$ of the rotating frame stationary solutions as a function of the rotation frequency of the trap, $\Omega$, with $\gamma=1$ and $\epsilon=0$ and $\edd=-0.49$, $0$, $0.5$ and $0.99$. Insets illustrate the deformation of the condensate in the $x-y$ plane, for $\alpha_{\rm v}>0$ and $\alpha_{\rm v}<0$. (b) The bifurcation frequency, $\Omega_{\rm b}$, calculated from Eq.\ (\ref{eq:Bif}) as a function of the trap aspect ratio, $\gamma$. Results are plotted for $\edd=-0.49$, $-0.4$, $-0.2$, $0$, $0.2$, $0.4$, $0.6$, $0.8$, $0.9$ and $0.99$. In both (a) and (b) $\edd$ increases as indicated by the arrow. Reprinted figure with permission from \cite{Bijnen09}. Copyright 2009 by the American Physical Society.} \label{Fig1}
\end{figure}
In Figure~\ref{Fig1}(a) \cite{Bijnen09} the solutions of Eq.~(\ref{alpha}) are plotted as a function of rotation frequency $\Omega$ for a spherically-symmetric trap, $\gamma=1$ and $\epsilon=0$, for various values of $\edd$. For a given value of $\edd$ the solutions have the same qualitative structure. Specifically, only one solution exists ($\alpha_{\rm v}=0$) up to some critical frequency. Two additional solutions ($\alpha_{\rm v}>0$ and $\alpha_{\rm v}<0$) bifurcate from this single solution at the critical rotation frequency, denoted as the bifurcation frequency $\Omega_{\rm b}$.

As expected, when  $\edd=0$  the results of Refs.~\cite{Recati01,Sinha01} are reproduced with $\Omega_{\rm b}=\omega_\perp/\sqrt{2}$ and $\alpha_{\rm v}=\pm \sqrt{2 \Omega^2-\omega_\perp^2}$ for the two additional solutions when $\Omega > \Omega_{\rm b}$. The critical frequency $\Omega_{\rm b}$ is associated with the spontaneous excitation of quadrupole modes. Specifically, in the Thomas-Fermi regime, the surface excitation dispersion is given by (see page 183 of \cite{Pitaevskii&StringariBook})
\begin{eqnarray}
\omega_{l}^2=(q_{l}/m) \nabla_{R} V =(q_{l} \omega_{\perp}^{2}/2) \nabla_{R} (R^{2}),
\label{eq:ang_dispers}
\end{eqnarray} 
 where $R$ is the radius of the BEC. For a surface excitation with angular momentum $\hbar l =\hbar q_l R$, in the absence of rotation, Eq.~(\ref{eq:ang_dispers}) reduces to $\omega_{l}=\sqrt{l}\omega_\perp$. Rotation shifts the mode frequency by $-l\Omega$, see Eq.~(\ref{Heff}). Hence for a rotating BEC the quadrupole surface excitation ($l=2$) frequency is $\omega_2(\Omega)=\sqrt{2}\omega_\perp-2\Omega$ \cite{Pitaevskii&StringariBook}. The bifurcation frequency, $\Omega_{\rm b}=\omega_{\perp}/\sqrt{2}$, occurs at the same rotation frequency at which the energy of the quadrupole mode is zero. Within the context of the Thomas-Fermi approximation, this critical frequency is independent of the strength of the contact interactions ($g$) and for  $\Omega \ge \omega_\perp/\sqrt{2}$ the two additional solutions arise from the excitation of the quadrupole mode.  

Referring back to Eq.~(\ref{eq:ang_dispers}), the inclusion of non-local dipolar interactions  implies that the force $-\nabla V$ no longer has a simple dependance on $R$ \cite{ODell04}. Hence, there is no reason to suspect that the condition to excite the quadrupole mode will be independent of $\edd$. Indeed, the dependence on $\edd$ can be seen in Figure~\ref{Fig1}(a) where the introduction of dipolar interactions leads to a shift in $\Omega_{\rm b}$. Specifically, for $\edd>0$ ($\edd<0$) the bifurcation frequency is increased (decreased). It is possible to evaluate $\Omega_{\rm b}$ analytically by realising that $\kappa_x = \kappa_y = \kappa$, for $\alpha_{\rm v} = 0$. In this limit the aspect ratio $\kappa$ is determined by the transcendental Eq.~(\ref{eq:transendental}) \cite{ODell04,Eberlein05}. As $\alpha_{\rm v} \rightarrow 0_+$,  the first order corrections to $\kappa_x$ and $\kappa_y$ with respect to $\kappa$ from Eqs.\ (\ref{kx}) and (\ref{ky}) can be calculated and inserted into Eq.\ (\ref{alpha}). Solving for $\Omega$ (noting that  $\Omega \rightarrow \Omega_{\rm b}$ as $\alpha_{\rm v}\rightarrow 0_{+}$) gives,
\begin{eqnarray}
\label{Omega_b}
\frac{\Omega_{\rm b}}{\omega_{\perp}}=\sqrt{\frac{1}{2}+\frac{3}{4}\kappa^2\edd\gamma^2\frac{\kappa^2\beta_{201}-\beta_{101}}{1-\edd\left(1-\frac{9}{2}\kappa^2\beta_{002}\right)}}
\label{eq:Bif},
\end{eqnarray}
which is plotted in Figure~\ref{Fig1}(b)  as a function of $\gamma$ for various values of $\edd$. When $\edd=0$ then  $\Omega_{\rm b}=\omega_x/\sqrt{2}$, which is independent of $\gamma=\omega_z/\omega_x$ \cite{Recati01,Sinha01}. However, for  $\edd \ne 0$ the value of $\gamma$ for which $\Omega_{\rm b}$ reaches a minimum changes. Specifically, as $\edd$ is increased from $-0.5$ the minimum value of $\Omega_{\rm b}$ changes from occurring at values of $\gamma$  where the trap shape is oblate ($\gamma > 1$) to ones where it is prolate ($\gamma < 1$). Fixing  $\gamma$ and increasing  $\edd$ leads to a monotonic decrease in $\Omega_{\rm b}$. Increasing $\edd$ can lead to a significant reduction in the bifurcation frequency, i.e. for  $\edd=0.99$, $\Omega_{\rm b}$ is reduced by $20\%$ compared to its non-dipolar value to $ \approx 0.55\omega_{\perp}$. It is tempting to consider the case where $\edd>1$ to induce shifts in $\Omega_{\rm b}$ to even lower values, however, in these regimes the premise of the calculation, i.e. the Thomas-Fermi approximation, may not be valid. 

\subsubsection{Elliptical trapping in the $x-y$ plane: $\epsilon > 0$\label{sec:ellipt}}
\begin{figure}[b]
\centering
\includegraphics[width=6cm]{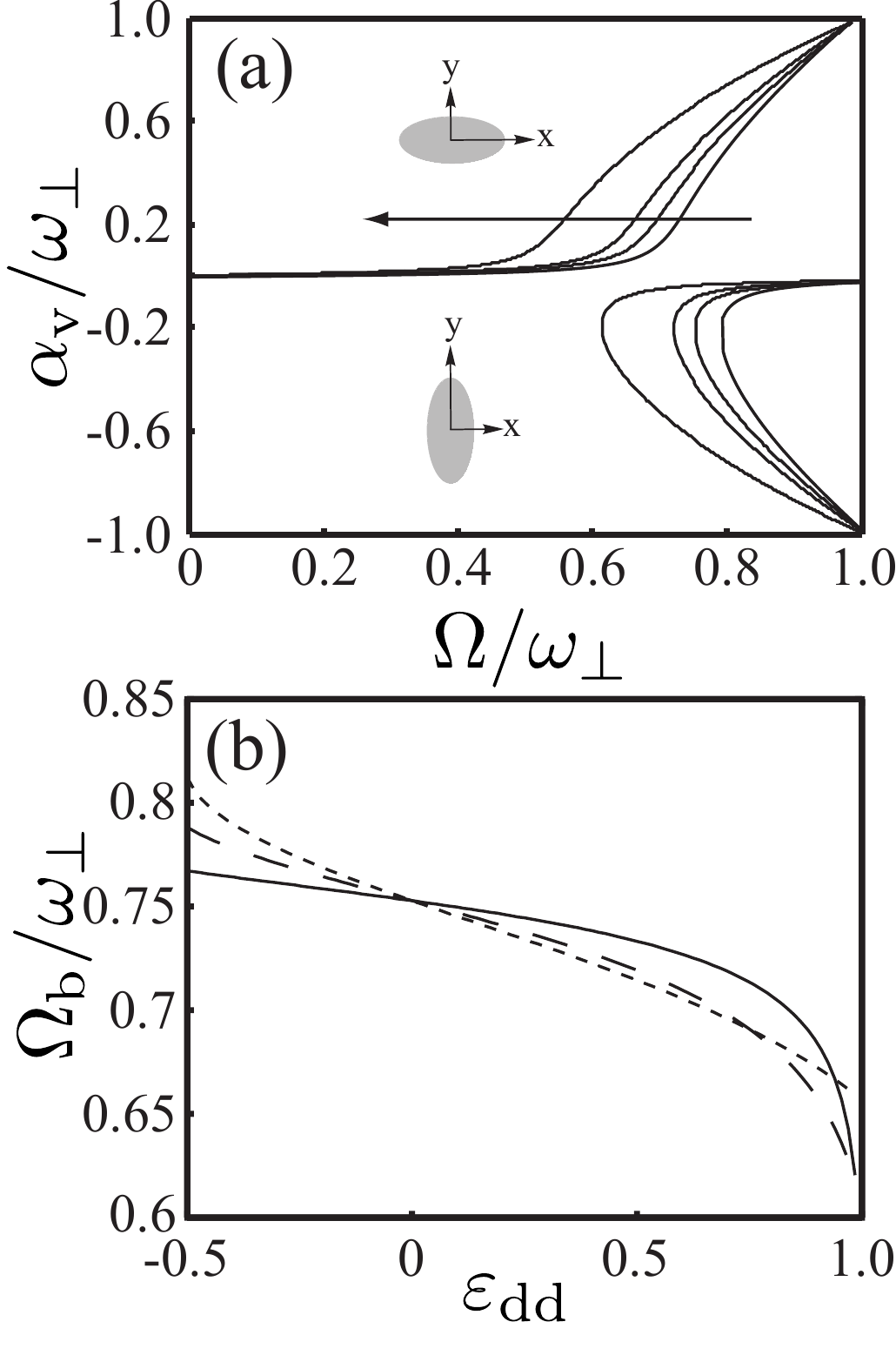}
\caption{ (a) Irrotational velocity field amplitude $\alpha_{\rm v}$ of the rotating frame stationary solutions as a function of the rotation frequency of the trap, $\Omega$, with $\gamma=1$ and $\epsilon=0.025$ and $\edd=-0.49$, $0$, $0.5$ and $0.99$, with increasing $\edd$ denoted by the arrow. Insets illustrate the deformation of the condensate in the $x-y$plane, for $\alpha_{\rm v}>0$ and $\alpha_{\rm v}<0$. (b) Back-bending frequency $\Omega_{\rm b}$ versus $\edd$ for $\epsilon=0.025$ and $\gamma=0.5$ (solid curve), $1.0$ (long dashed curve) and $2.0$ (short dashed curve). Reprinted figure with permission from \cite{Bijnen09}. Copyright 2009 by the American Physical Society.}\label{Fig2}
\end{figure}

Rotating elliptical traps have been created experimentally with lasers and magnetic fields \cite{Madison00,Hodby02}. Following the experiment by Madison {\it
et al.} \cite{Madison00}, below we consider a trap with a weak ellipticity of $\epsilon=0.025$. Figure~\ref{Fig2}(a) \cite{Bijnen09} shows the solutions to Eq.~(\ref{alpha}) for various values of $\edd$ for $\gamma=1$.  As in the non-dipolar case \cite{Recati01,Sinha01}, the solutions become heavily modified for $\epsilon>0$. There is an upper branch solution  ($\alpha_{\rm v}>0$) which extends over the range $0 \le \Omega \le \omega_{\perp}$ and a lower solution  ($\alpha_{\rm v}<0$) which is doubled valued and exists above some critical rotation frequency.  The critical rotation frequency for the lower brach solution is denoted as the back-bending frequency $\Omega_{\rm b}$, which in the limit $\epsilon=0$ can be regarded as the limiting case of the bifurcation frequency and hence we use the same notation for both. Unlike the circular trap, considered in Section~\ref{sec:circular_trap}, there are no $\alpha_{\rm v}=0$ solutions for  $\Omega>0$. In the absence of dipolar interactions increasing the trap ellipticity results in an {\em increase} of the back-bending frequency $\Omega_{\rm b}$. As shown in Figure~\ref{Fig2}(b) dipolar interactions, as in the case of $\epsilon=0$, {\em reduce} ({\em increase}) $\Omega_{\rm b}$ for $\edd > 0$ ($\edd<0$). 

Due to the anisotropy of the dipolar interactions increasing $\edd$ decreases both $\kappa_x$ and $\kappa_y$, i.e. the BEC becomes more
prolate. Under rotation dipolar interactions also increase the deformation of the BEC in the $x-y$ plane, as can be deduced from Figure~\ref{Fig2}(a). Specifically, as $\edd$ is increased, for $\Omega>0$, $ \alpha_{\rm v}$ increases and hence, see Eq.~(\ref{eq:alphadeformation}), the deformation ($\mathcal{D}$) of the dipolar BEC increases.

\subsection{Dynamical stability of stationary solutions \label{SecStability}}
The static solutions derived above are stationary but not necessarily stable. In this section the dynamical stability of the stationary solutions is analyzed. This is done by considering small perturbations in the BEC density and phase of the form $n=n_{\rm eq}+\delta n$ and $S=S_{\rm eq}+\delta S$. By linearizing the dipolar hydrodynamic equations Eqs.~(\ref{cont_eq_d}, \ref{motion_eq_d}), the dynamics of such perturbations can be described as \cite{Bijnen07,Martin08,Bijnen09},
 \begin{eqnarray}
 \label{stability}
\frac{\partial }{\partial t} \left[\begin{array}{c}
 \delta S \\
  \delta n \\
\end{array}
\right] = -\left[\begin{array}{cc}
 {\bf v_c} \cdot \boldsymbol{\nabla} & \frac{g}{m}\left(1+\edd K\right) \\
 \boldsymbol{\nabla} \cdot n_{\rm eq}
\boldsymbol{\nabla} & \boldsymbol{\nabla} \cdot
{\bf v}+{\bf v_c} \cdot \boldsymbol{\nabla}  \\
\end{array}
\right] \left[\begin{array}{c}
 \delta S \\
  \delta n \\
\end{array}
\right] \nonumber \\
 \end{eqnarray}
where ${\bf v_c}={\bf v}-{\bf \Omega}\times{\bf r}$ and the integral operator $K$ is defined as
\begin{eqnarray}
(K \delta n)({\bf r})=-3\frac{\partial^2}{\partial
z^2}\int\frac{\delta n ({\bf r}^{\prime})}{4\pi\left|{\bf r}-{\bf r}^{\prime}\right|}\,{\rm d}{\bf r}^{\prime}-\delta n({\bf r})\label{Koperator}.
\end{eqnarray}
To investigate the stability of the BEC the eigenfunctions and eigenvalues of the operator  in Eq.~(\ref{stability}) can be examined. Dynamical instability arises when one or more eigenvalues $\lambda$ possess a positive real part. While the imaginary parts of the eigenvalues characterise the frequencies of collective modes of the BEC \cite{Castin01}, the magnitudes of any real parts control the rate of growth of unstable modes. If the density and phase fluctuations are expressed as polynomials of degree $N$, the operators in Eq.~(\ref{stability}), including $K$ \cite{Ferrers, Dyson, LevinMuratov}, result in polynomials which are of degree $N$ or less. This enables  Eq.~(\ref{stability}) to be recast as a scalar matrix operator which acts on vectors of polynomial coefficients. Finding the eigenvalues (determining stability) and eigenvectors (characterising modes) is then a simple computational task \cite{Bijnen07,Martin08,Bijnen09,Sinha01}.

\begin{figure}[t]
\centering
\includegraphics[width=7.5cm]{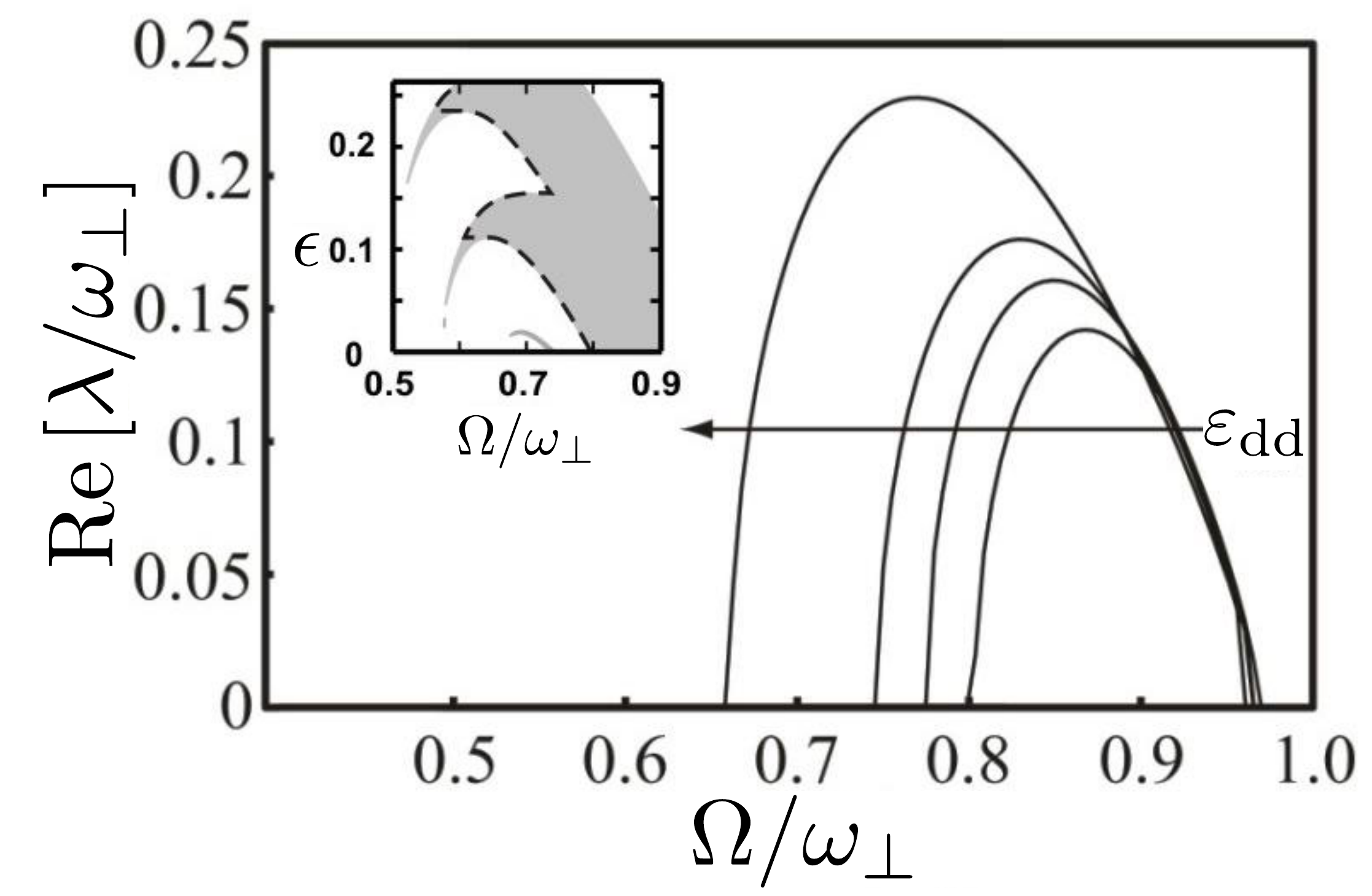}
\caption{The maximum positive real eigenvalues of Eq.~(\ref{stability}) (solid curves) for the upper-branch solutions of $\alpha_{\rm v}$ as a function of $\Omega$, for $\epsilon=0.025$, $\gamma=1$, $N=3$, $\edd=-0.49$, $0$, $0.5$ and $0.99$, with $\edd$ increasing in the direction of the arrow. The inset shows the region of dynamical instability in the $\epsilon-\Omega$ plane for $\edd=0$. The narrow regions, around $\Omega/\omega_{\perp} < 0.6$ with $\epsilon < 0.1$ and $\Omega/\omega_{\perp} < 0.56$ with $\epsilon < 0.25$,  have negligible effect and so only the main instability region is considered (bounded by the dashed line).  Reprinted figure with permission from \cite{Bijnen09}. Copyright 2009 by the American Physical Society.
\label{FigEigs}}
\end{figure}

Below, the stability of the rotating solutions are considered for  $\epsilon>0$ [Figure~\ref{Fig2}(a)]. For $\alpha_{\rm v}<0$ there are two static solutions for $\Omega > \Omega_{\rm b}$. The solution  nearest the $\alpha_{\rm v}=0$ axis is always dynamically stable except when  $\Omega \rightarrow \omega_\perp$, where the BEC is susceptible to a centre-of-mass instability  \cite{Rosenbusch02}. The other solution, in the $\alpha_{\rm v}<0$ half-plane, is always dynamically unstable and hence experimentally irrelevant. The dynamical stability of solutions in the upper half-plane ($\alpha_{\rm v} > 0$) is more interesting. In  Figure~\ref{FigEigs} \cite{Bijnen09} the maximum positive real eigenvalues of the upper-branch solution is plotted as a function of $\Omega$ for a fixed trap geometry and various values of $\edd$. To obtain these results a maximum polynomial perturbation of   $\delta n = x^p y^q z^r$ with $p+q+r \le 3=N$ was considered. 

In Figure~\ref{FigEigs}(inset) an example of the unstable region (for the upper-branch solution) is shown, in the $\epsilon  - \Omega$ plane, for $\edd=0$ and $\gamma=1$. The shaded crescents \cite{Sinha01} denote the regimes of dynamical instability (${\rm Re} (\lambda) >0$). Each crescent corresponds to a single value of $N$,  the total degree of the polynomial perturbation. Each higher value of $N$ adds another crescent from above. The crescents merge for large rotation frequencies, with the eigenvalues being comparatively large. At lower rotation frequencies the crescents become vanishingly thin with comparatively small eignevalues  \cite{Corro07} (at least an order of magnitude smaller than the main instability region). The relative smallness of the eigenvalues in this region indicates that instabilities grow over a much longer time-scale as compared to the main region of instability. For non-dipolar BECs this was numerically investigated by solving the GPE \cite{Corro07}. These numerical results showed that the narrow instability regions have negligible effect when ramping $\Omega$ at rates greater than ${\rm d}\Omega/{\rm d} t = 2 \times 10^{-4}\omega_\perp^2$.  Hence these narrow regions of instability have minimal consequence over the time-scales typically considered in an experiment and can be ignored, with the main region of instability denoted by the dashed line in Figure~\ref{FigEigs}(inset). Experimental trap ellipticities are usually $\epsilon \leqslant 0.1$  and the unstable regime can be quantified with $N=3$ for the perturbations. Denoting the lower bound for the dynamical instability as  $\Omega_{\rm i}$, Figure~\ref{FigEigs} indicates that as  $\edd$ is increased the lower bound for  $\Omega_{\rm i}$ decreases.

As the size of the scalar matrix operator (\ref{stability}) is increased to $N=4,5,\ldots$, the higher lying modes develop real eigenvalues as $\Omega$ is increased. These higher lying modes fall within the region of instability already shown in Figure~\ref{FigEigs} for $N=3$ and so do not alter the range of parameters where instability occurs. As we shall now explain, this result implies that the Thomas-Fermi spectrum does not contain a roton minimum as a function of angular momentum $L$. Ultimately, this is because a roton minimum means, somewhat counter-intuitively, that higher lying modes can have lower energy than lower lying modes. This has important consequences in the presence of fluid flow. For example, Pitaevskii \cite{Pitaevskii84} considered the case of superfluid $^4$He flowing through a pipe with rough walls. If the excitation spectrum of the superfluid at rest is $E(p)$, when it flows at speed $v$ the spectrum is Galilean-shifted $E(p) \rightarrow E(p) -p v$. There is, therefore, a critical velocity $v_{c}=\mathrm{min} \ E(p)/p$, where min means the minimum with respect to $p$, at which excitations have zero energy and can be freely produced, depleting the superfluid. However, because superfluid $^4$He has a roton minimum in its spectrum, the excitations created at $v_{c}$ have a specific momentum $p \approx p_{r}$, where $p_{r}$ is the roton minimum, triggering an instability to the formation of a density wave with wavelength $\propto p_{r}^{-1}$. For a rotating system similar arguments can be made with the Galilean-shifted energy taking the form $E \rightarrow E-L \Omega$ with  angular roton modes  \cite{Ronen07} becoming unstable at some critical rotation frequency. Crucially, in the dynamical instability analysis presented above it was found that the modes become unstable in order as $\Omega$ is increased. This implies that for the regimes considered, an angular-roton minimum at finite angular momentum does not exist. Within the context of the analysis presented ($-0.5 \leq \edd \leq 1$) this is not surprising since roton minima in dipolar BECs do not occur for  $-0.5 \leq \edd \leq 1$. To access regimes where a roton minimum occurs requires a treatment beyond the Thomas-Fermi approximation since the zero-point energy plays a crucial role in determining the properties of a BEC in the presence of strong dipolar interactions.

\subsection{Routes to instability and vortex lattice formation \label{sec:routes}}
For a non-dipolar BEC in the Thomas-Fermi limit,  only the rotation frequency and trap ellipticity determine the stability of the rotating frame solutions. As such adiabatic changes in $\epsilon$ and $\Omega$ can be utilised to trigger dynamical instability.   For non-dipolar BECs this has been realised both experimentally \cite{Hodby02,Madison01} and numerically \cite{Lundh03,Lobo04,Parker06}, with excellent agreement with the hydrodynamic predictions. 
\begin{figure}[b]
\centering
\includegraphics[width=7.5cm]{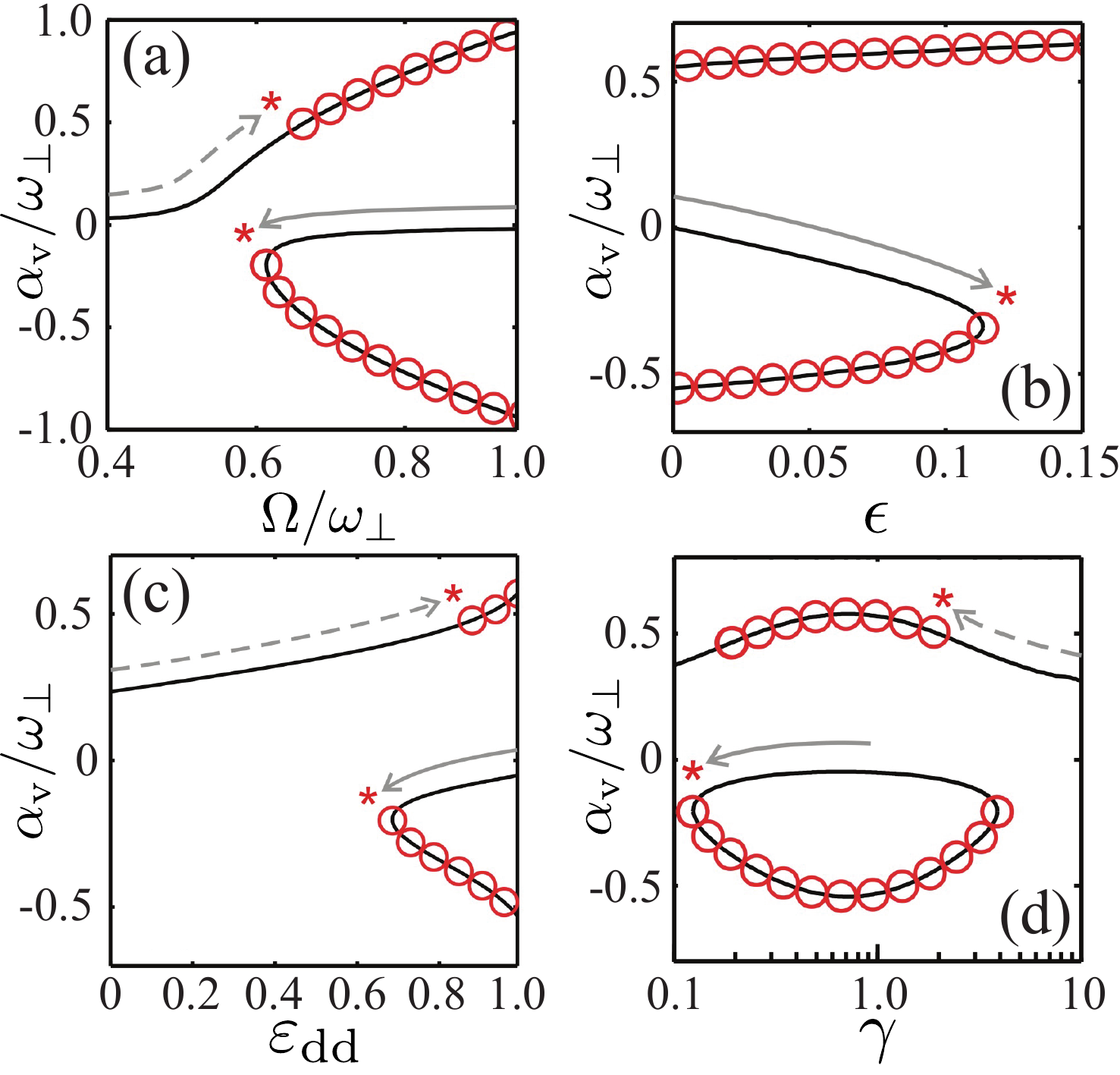}
\caption{Stationary states in the rotating trap characterised by the velocity field amplitude $\alpha_{\rm v}$, determined from Eq.~(\ref{alpha}). Red circles denote dynamically unstable solutions. Adiabatic pathways to instability are denoted by dashed and solid arrows, with the point of instability, for a particular pathway, denoted by an asterisk. Pathways towards a dynamical instability are denoted by a dashed arrow and pathways towards an instability due to the  disappearance of the stationary state are denoted by solid arrows.  In each of the figures the trap rotation frequency $\Omega$ (a), trap ellipticity $\epsilon$ (b), dipolar interaction strength $\edd$ (c) and axial trapping strength $\gamma$ (d) are varied adiabatically, whilst the remaining parameters remain fixed at $\Omega = 0.7 \omega_{\perp}$, $\epsilon = 0.025$, $\edd = 0.99$, and  $\gamma = 1$.  Reprinted with permission from \cite{Bijnen09}. \label{FigRoutes}}
\end{figure}

For dipolar BECs  the static solutions and their instability depend on the rotation frequency $\Omega$, the trap ellipticity $\epsilon$, the trap ratio $\gamma$ and the interaction parameter $\edd$, see Sections \ref{sec:dip_ellip} and \ref{SecStability}. In principle all of these parameters can be adiabatically tuned in time. This provides many routes through parameter space  to induce instability in the system.

Figure~\ref{FigRoutes} \cite{Bijnen09} shows examples of these routes. Specifically, Figure~\ref{FigRoutes} shows the static solutions $\alpha_{\rm v}$ of Eq.\ (\ref{alpha}) as a function of $\Omega$ [Figure~\ref{FigRoutes}(a)] (with $\epsilon=0.025$, $\gamma=1$ and $\edd=0.99$), $\epsilon$ [Figure~\ref{FigRoutes}(b)] (with $\gamma=1$, $\Omega=0.7\omega_{\perp}$ and $\edd=0.99$), $\edd$ [Figure~\ref{FigRoutes}(c)] (with $\epsilon=0.025$, $\gamma=1$ and $\Omega=0.7\omega_{\perp}$) and $\gamma$ [Figure~\ref{FigRoutes}(d)] (with $\epsilon=0.025$, $\Omega=0.7\omega_{\perp}$ and $\edd=0.99$). In each figure the routes towards instability, due to an adiabatic change in the free parameters $\Omega$, $\epsilon$, $\edd$ or $\gamma$, are denoted by grey arrows, with the onset of instability highlighted by asterisks. Two types of instability can arise: (i) a dynamical instability (dashed grey arrows), or (ii) back-bending of the stationary solution such that it ceases to exist (solid grey arrows). For $\alpha_{\rm v} >0$, the instability always arises from dynamical instabilities in the rotating frame stationary solution. For  $\alpha_{\rm v} < 0$ the instability always arises from the rotating frame stationary solution ceasing to exist.

\subsubsection{Does the final state of the system contain vortices?}
In Section~\ref{sec:routes} we considered the routes to instability in a rotating dipolar BEC. This Section addresses the question of whether such instabilities lead to the seeding of a vortex or vortex lattice in the BEC.  We need to consider both the hydrodynamical surface instability of the rotating BEC and the  energetic favourability for a vortex to be supported in the BEC. 

For a non-dipolar BEC  the admittance of a vortex into the BEC becomes energetically favourable when the rotation frequency exceeds the critical frequency $\Omega_{\rm v}$ defined in Eq.\ (\ref{eqn:criticalfrequency}), where in the Thomas-Fermi limit (assuming cylindrical symmetry) $\Omega_{\rm v}$ is given by Eq.\ (\ref{crit_freq}). For typical condensate parameters $\Omega_{\rm v} \sim 0.3 \omega_\perp$. This result is inconsistent with experimental observations of a much higher threshold for vortex lattice formation,  $\Omega \sim 0.7 \omega_\perp$. This discrepancy can be understood by considering the mechanism for vortices to be seeded into the BEC. For example, consider the case where $\Omega$ is increased adiabatically from zero [Figure~\ref{FigRoutes}(a)]. Then the pertinent stationary solution is in the $\alpha_{\rm v} >0$ half-plane, which becomes dynamically unstable when  $\Omega \approx 0.7\omega_\perp$. This instability provides a mechanism for vortices to be seeded into the BEC and ultimately relax into a vortex lattice. Alternatively, for $\Omega >  \Omega_{\rm v}$ the global energy minimum of the BEC is one which contains a vortex or vortex lattice, with the vortex-free solution being a local energy minimum. However, as $\Omega$ is adiabatically increased from zero, for $\Omega < \Omega_{\rm v}$ the global energy minimum is defined by the rotating frame stationary solutions. As such as $\Omega$ passes through $\Omega_{\rm v}$ there is no adiabatic path from the rotating frame stationary solution to the vortex solution, i.e. to move from the local energy minimum of the rotating frame stationary solution to the global energy minimum of the vortex state requires overcoming a significant energy barrier, due to the change in topology of the BEC between the two states. The dynamical instabilities discussed in Section~\ref{sec:routes} provide a mechanism to overcome this {\it topological} barrier and hence seed the formation of vortices and vortex lattices at rotation frequencies $\Omega \approx 0.7 \omega_{\perp}$.

Of course, the details of how a vortex or a vortex lattice forms once a hydrodynamical instability occurs is non-trivial, however the instability is the first step in the process  \cite{Lundh03,Lobo04,Parker06}.  For example, consider a BEC undergoing an adiabatic introduction of  $\Omega$ (from zero), see dashed grey arrow in Figure~\ref{FigRoutes}(a). At a critical rotation frequency, $\Omega_{\rm i}$, the BEC becomes dynamically unstable. This leads to the exponential growth of surface ripples in the BEC  \cite{Madison00,Madison01,Parker06}. Alternatively, consider a BEC undergoing an adiabatic introduction of $\epsilon$ (from zero), see solid grey arrow in Figure~\ref{FigRoutes}(b). Above some critical trap ellipticity the stationary solutions no longer exist and the BEC undergoes large shape oscillations. In each case, and also for the adiabatic introduction of $\edd$  [Figure~\ref{FigRoutes}(c)] or $\gamma$ [Figure~\ref{FigRoutes}(d)], these instabilities provide a mechanism for vortices to nucleate into the condensate. From this a  turbulent vortex state  emerges which then relaxes to a vortex lattice \cite{Parker06,Parker05}.

The rotation frequency at which it becomes energetically favourable for a vortex to be admitted into a  dipolar condensate depends crucially on the trap geometry $\gamma$ and the strength of the dipolar interactions $\edd$ as discussed in Section \ref{subsec:vortexTF} based upon the results in Ref.\ \cite{ODell07}.  There it was assumed that the system was cylindrically-symmetric, however, if we assume a very weak ellipticity $\epsilon=0.025$,  it is expected that the correction to the critical frequency will be correspondingly small.

\begin{figure}[t]
\centering
\includegraphics[width=7cm]{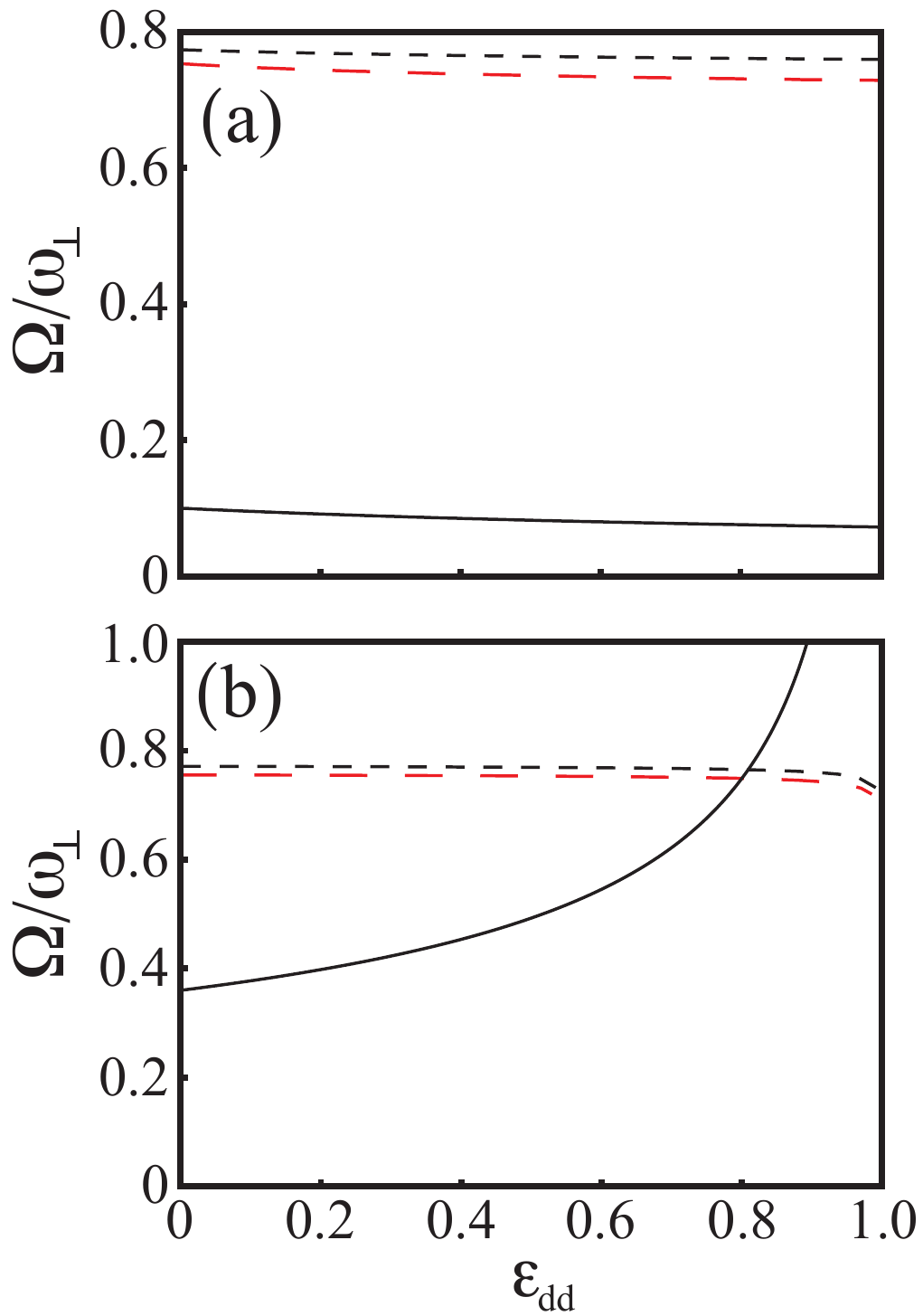}
\caption{The relation between the instability frequencies, $\Omega_{\rm b}$ (long dashed red curve) and $\Omega_{\rm i}$ (short dashed curve), and the critical rotation frequency for vorticity $\Omega_{\rm v}$ (solid curve) for (a) an oblate trap $\gamma=10$ and (b) a prolate trap $\gamma=0.1$. The instability frequencies are based on a trap with ellipticity $\epsilon=0.025$ while $\Omega_{\rm v}$ is obtained from Eq.~(\ref{crit_freq}) under the assumption of a $^{52}$Cr BEC with $150,000$ atoms and scattering length $a_s=5.1$nm in a circularly symmetric trap with $\omega_\perp=2\pi \times 200$Hz$=200$rad/s. Reprinted figure with permission from \cite{Bijnen09}. Copyright 2009 by the American Physical Society. \label{Fig5}}
\end{figure}

Below we consider two regimes for admittance of a vortex or vortex lattice into a dipolar BEC, as a function of rotation frequency and $\edd$. Initially an oblate trap is considered with $\gamma=10$, see Figure~\ref{Fig5}(a) which plots the instability frequencies $\Omega_{\rm i}$ (short dashed curve) and $\Omega_{\rm b}$ (long dashed red curve) as a function of the dipolar interactions $\edd$.  For adiabatic changes in $\Omega$ (vertical path) or $\edd$ (horizontal path), the system becomes unstable when it reaches one of the instability lines. The key feature to note is that the instability frequencies decrease weakly as the dipolar interactions are increased and have the approximate value $\Omega_{\rm i}\approx \Omega_{\rm b} \approx 0.75\omega_\perp$. Also shown in Figure~\ref{Fig5}(a) is the rotation frequency  $\Omega_{\rm v}$ (solid curve) at which it becomes energetically favourable for a single vortex to reside in the dipolar BEC.  Dipolar interactions, in this oblate system, lead to a weak decrease in $\Omega_{\rm v}$, with $\Omega_{\rm v} \approx0.1\omega_{\perp}$ for the parameters considered \cite{ODell07}. These results show us that when the condensate becomes dynamically unstable a vortex state is already energetically favoured, i.e.\ the rotation frequency at which it is energetically preferable to have a vortex in the BEC is much lower than the rotation frequency required to induce an instability. As such, it is expected that in an oblate dipolar BEC a vortex lattice will ultimately form when these instabilities are reached.

For a prolate trap with $\gamma=0.1$, Figure~\ref{Fig5}(b) plots the instability frequencies $\Omega_{\rm i}$ (short dashed curve) and $\Omega_{\rm b}$ (long dashed red curve) and $\Omega_{\rm v}$ (solid curve). The instability frequencies follow a similar trend to the oblate case. However, $\Omega_{\rm v}$ is drastically different, increasing significantly with $\edd$ \cite{ODell07}. Depending on the strength of the dipolar interactions, two regimes are predicted. For  $\edd \leqslant 0.8$ both  $\Omega_{\rm i}$ and $\Omega_{\rm b}$  are larger than $\Omega_{\rm v}$, and hence it is expected that an instability will lead to the admission of vortices into the condensate (and subsequently a vortex lattice). However, for  $\edd \geqslant 0.8$, both $\Omega_{\rm i}$ and $\Omega_{\rm b}$  are lower than $\Omega_{\rm v}$, implying that although an instability in the vortex free solution occurs, a vortex state is not energetically favourable. In this scenario the final state is not clear due to the net attractive dipolar interactions in this prolate configuration, and similar behaviour to non-dipolar BECs with attractive contact interactions ($g<0$) may arise. For non-dipolar BECs with attractive interactions the formation of vortices is also unfavourable with the possible final states including centre-of-mass motion, quadrupole oscillations and higher angular momentum-carrying shape excitations \cite{Wilken98,Mottelson99,Pethick00}. However, the true nature of the final state warrants further investigation.

Numerical results of the dipolar GPE in the quasi-two-dimensional regime \cite{Kumar12, Kumar14} show that, for an adiabatic introduction of the rotation frequency,  the strength of the dipolar interaction influences the rotation frequency at which vortices are admitted into the condensate. In agreement with analysis presented above, it is also found that as the dipolar interaction is increased the rotation frequency required to nucleate vortices is reduced. 

\subsection{Generalisation of Thomas-Fermi analysis}
In principle the analysis presented in Sections~\ref{sec:dip_ellip}, \ref{SecStability} and \ref{sec:routes} can be modified to consider vortex generation through the application of a synthetic magnetic field and extended to investigate how off-axis alignment of the dipoles affects the stationary solutions and their dynamical stability.

\subsubsection{Synthetic magnetic fields}
 In a remarkable experiment in non-dipolar BECs, synthetic magnetic fields have been used to nucleate vortices \cite{Lin09}. The nucleation of vortices in such systems can be analysed via the general methods presented Sections~\ref{sec:dip_ellip} and \ref{SecStability}  \cite{Taylor11}, with favourable agreement being found with experimental results. Although the extension of this calculation to the dipolar case has not yet been pursued in the literature,  the analysis presented  in Sections~\ref{sec:dip_ellip}, \ref{SecStability} and \ref{sec:routes} for dipolar BECs can be modified to include synthetic magnetic fields. Let us outline the calculation: The starting point is,
 \begin{eqnarray}
\frac{\partial n}{\partial t} = -\boldsymbol{\nabla} \cdot \left[
n\boldsymbol{\nu}\right],\\
m \frac{\partial {\bf v}}{\partial t} = -\boldsymbol{\nabla} \left(
\frac{1}{2} m \nu^2 + V + g n+ \Phi \right),
\end{eqnarray}
where the generalized velocity is $\boldsymbol{\nu}={\bf v}-q^*{\bf A^*}/m$, with ${\bf A^*}$ being the synthetic vector potential associated with the synthetic magnetic field. From this it is possible to derive an equivalent set of stationary solutions and determine their stability via,
 \begin{eqnarray}
\frac{\partial }{\partial t} \left[\begin{array}{c}
 \delta S \\
  \delta n \\
\end{array}
\right] = -\left[\begin{array}{cc}
 \boldsymbol{\nu} \cdot \boldsymbol{\nabla} & \frac{g}{m}\left(1+\edd K\right) \\
 \boldsymbol{\nabla} \cdot n_{(0)}
\boldsymbol{\nabla} & \boldsymbol{\nabla} \cdot
\boldsymbol{\nu}+\boldsymbol{\nu} \cdot \boldsymbol{\nabla} \\
\end{array}
\right] \left[\begin{array}{c}
 \delta S \\
  \delta n \\
\end{array}
\right]. \nonumber \\
 \end{eqnarray}
 Such an analysis would provide the basis to quantify under what regimes it is favourable for a vortex to be seeded into a dipolar BEC in the presence of a synthetic magnetic field.

\subsubsection{Off-axis dipole orientation}
In  Sections~\ref{sec:dip_ellip}, \ref{SecStability} and \ref{sec:routes} the analysis was restricted to the case where the dipoles were aligned perpendicular to the plane of rotation. In the absence of rotation it is possible to find Thomas-Fermi solutions when the dipoles are not aligned along one trap axis \cite{Sapina10}. Although non-trivial, it is in principle possible to generalise the work of Sapina {\it et al.} \cite{Sapina10} to include rotation, thereby providing a framework to quantify under what regimes it is favourable for a vortex to be seeded into a dipolar BEC when the dipoles are not perpendicular to the axis of rotation\footnote{To be able to carry out such a calculation, along the lines presented in Sections~\ref{sec:dip_ellip}, \ref{SecStability} and \ref{sec:routes}, one must consider the case where the dipole alignment is stationary in the rotating frame, i.e.\ rotating with the trap.}.

\section{Vortex Lattices \label{sec:lattices}}
To date there have been several examinations of the properties of vortex lattices in quasi-two-dimensional rotating dipolar BECs \cite{Pu06,Cooper05,Zhang05,Cooper06,Cooper07,Malet11,Kumar15b}. Work by Cooper {\it et al.} \cite{Cooper05,Cooper06,Cooper07}  found, by numerical minimization of the interaction energy with the wavefunction constrained to states in the Lowest Landau Level (LLL), that the dipolar interaction could modify the symmetry of the vortex lattice. Specifically, they found new phases could emerge with square, stripe (rectangular) and bubble phases, in contrast to the conventional triangular lattice structure of non-dipolar BECs. These new phases emerge in the regime $a_s \lesssim -0.13 C_{\rm dd} m/\hbar^2$. This result coincided with the work of Zhang and Zhai \cite{Zhang05} who also found that the triangular phase is not favoured when the contact interactions are attractive ($a_s<0$), with stronger dipolar interactions leading to a square then rectangular lattice. Additionally, recent work by Kishor Kumar {\it et al.} \cite{Kumar15b}, using numerical solutions of the three-dimensional purely ($a_s=0$) dipolar GPE, found both triangular and square vortex lattice configurations, with both the strength of the dipolar interactions and the rotation frequency determining the symmetry of the final state.  Each of these investigations assumed that the axis of rotation of the BEC was the same as the alignment direction of the dipoles. Numerical simulations of the dipolar GPE, carried out by Yi and Pu \cite{Pu06} did not find any evidence of this change of lattice structure for such a configuration. However, for dipoles aligned off axis, they did find evidence of a change in the symmetry of the vortex lattice. This work also found evidence for {\it ripply} vortex lattices, where density modulations arise in the vicinity of a vortex. This can be understood in the context of the analysis presented in Section \ref{sec:single_vortices} where we deduced that as the BEC approaches the roton instability density ripples appear around a single vortex. This is consistent with numerical work presented by Jona-Lasinio {\it et al.} \cite{Lasinio13}.

Below we present new results for the structure of a vortex lattice in a dipolar BEC in the quasi-two-dimensional regime. We consider both the case where the dipole alignment is perpendicular to the condensate plane (in reference to Figure~\ref{fig:schematic} $\alpha=0$), and generalise this to include configurations where the alignment has a component into the plane (in reference to Figure~\ref{fig:schematic} $\alpha>0$). The treatment considered draws on the works of  Cooper {\it et al.}  \cite{Cooper05,Cooper07}, Zhang and Zhai \cite{Zhang05} and \cite{Butts99,Mueller02}. For clarity, we shall begin our analysis by considering vortex lattices for the case of a non-dipolar BEC. Although this case has been covered rather extensively in the literature, it will be useful for the reader to provide a comprehensive, self-contained review here. We shall then build on the methodology presented to consider dipolar BECs.

\subsection{Vortex lattice in a non-dipolar BEC \label{sec:vortex_conventional}}
Assuming that a condensate is confined in a cylindrically-symmetric harmonic trap, the energy functional in the rotating frame is given by,
\begin{eqnarray}
E[\psi]&=&\int  \psi({\bf r}) H^{\prime} \psi({\bf r})\,{\rm d}{\bf r} \nonumber \\
&+&\frac{1}{2}\int n({\bf r_1})U({\bf r_1}-{\bf r_2})n({\bf r_2})\,{\rm d}{\bf r_1}{\rm d}{\bf r_2},
\label{eq:E_Psi_general}
\end{eqnarray}
where,
\begin{eqnarray}
H^{\prime}&=&\frac{1}{2m}\left(i\hbar \nabla +\Omega m {\bf {\hat z}} \times \brho \right)^2 \nonumber \\
&+&\frac{m}{2}\left(\omega_{\perp}^2 - \Omega^2\right)\rho^2 +\frac{m}{2}\omega_z^2 z^2. 
\end{eqnarray}
There are two distinct contributions to the total energy: a single-particle energy contribution, $E_0$ (the first term in Eq.~(\ref{eq:E_Psi_general})) and an interaction energy contribution, which in the absence of dipolar interactions is $E_{\rm vdW}$ (the second term in Eq.~(\ref{eq:E_Psi_general})). 

To study the properties of vortex lattices it is appropriate to consider the  quasi-two-dimensional regime. Physically, this corresponds to a situation where the condensate is rapidly rotating so that centrifugal spreading is significant, and where the longitudinal trapping is strong ($\hbar \omega_z \gg gn(0)$). In such circumstances it is appropriate to assume that the longitudinal motion is described by the ground state of the $z$-confinement, so that the condensate wavefunction can be factorized according to Eq.~(\ref{eq:quasi_2d_psi}). The resulting form for the energy functional, in the absence of dipolar interactions is,
\begin{eqnarray}
E&=&\frac{N}{2}\hbar \omega_z +\int \psi^{\star}_{\perp}(\brho)  H_{\perp}^{\prime} \psi^{\star}_{\perp}(\brho)\,{\rm d} \brho \nonumber \\
&+& \frac{g}{2\sqrt{2\pi}l_z} \int  n_{\perp}^2 (\brho)\,{\rm d} \brho,
\end{eqnarray}
where,
\begin{eqnarray}
H^{\prime}_{\perp} &=&-\frac{1}{2m} \left(i\hbar  \nabla_{\perp}+\Omega m {\bf {\hat z}} \times \brho \right)^2 \nonumber \\
&+&\frac{m}{2}\left(\omega_{\perp}^2 -\Omega^2\right)\rho^2. 
\end{eqnarray}

Consider the fast-rotating limit where $\Omega \rightarrow \omega_{\perp}$. This is the point at which the centrifugal spreading due to rotation almost overwhelms the confinement due to the radial trap. In this limit, the single-particle Hamiltonian in quasi-two dimensions tends to $-(i\hbar \nabla_{\perp} +\Omega m{\bf {\hat z}}\times \brho)^2/2m$, up to a constant. The eigenfunctions of this Hamiltonian are well known: they are the Landau level orbitals $u_{m^{\prime},n^{\prime}}(x, y)$ with corresponding eigenenergies $\epsilon_{n^{\prime}} =  \hbar \Omega(n^{\prime} + 1/2)$. The $n$ quantum number labels the Landau level, and may take on any non-negative integer value. Each Landau level is infinitely degenerate since $\epsilon_{n^{\prime}}$ does not depend on $m^{\prime}$ (which also takes on non-negative integer values). 

Although the Landau level orbitals do not necessarily form a complete basis for the full quasi-two-dimensional Hamiltonian, in the limit of weak interactions they are a good approximate basis choice \cite{Wilken98}. In searching for the ground state of the condensate it is assumed that it is adequately described by a superposition of $n^{\prime} = 0$ Landau level orbitals, i.e.  the lowest Landau level (LLL) approximation. Using an unrestricted minimization this assumption implies that \cite{Fetter10},
\begin{eqnarray}
\left(1-\frac{\Omega}{\omega_{\perp}}\right)\lesssim \frac{\sqrt{2\pi} l_z}{2 N a_s}.
\label{eq:LLL_criteria}
\end{eqnarray}
In this limit the two-dimensional ground state wavefunction of the condensate can be written as,
\begin{eqnarray}
\psi_{\perp}(\brho)&=&\sum_{m^{\prime}=0}^{\infty} c_{m^{\prime}} u_{m^{\prime},0}(\brho) \nonumber
\\
&=&\sum_{m^{\prime}=0}^{\infty} \frac{c_{m^{\prime}}}{\sqrt{2 \pi m^{\prime}!}} \left(\frac{x+iy}{l_{\perp}}\right)^{m^{\prime}} \exp\left(-\frac{x^2+y^2}{2l_{\perp}^2}\right), \nonumber \\
\label{eq:Phi_LLL}
\end{eqnarray}
where the right-hand side follows from the explicit form of $u_{m^{\prime},0}(x,y)$. The length scale $l_{\perp} = \sqrt{\hbar /m\Omega}$ characterises the radial extent of the condensate and is effectively equal to the transverse trap length since $\Omega \rightarrow \omega_{\perp}$.

At this point it turns out to be convenient to convert to a complex number representation, rather than working with the components of a two-dimensional vector. To this end $\brho = x{\bf {\hat x}} + y{\bf {\hat y}}$ is mapped onto the complex number $w = x + iy$. Then the ground state wavefunction of Eq.~(\ref{eq:Phi_LLL}) may be written as,
\begin{eqnarray}
\psi_{\perp}(w)=h(w) \exp\left(-\frac{|w|^2}{2l_{\perp}^2}\right),
\label{eq:Phi_w}
\end{eqnarray}
where $h$ is an analytic function of $w$. With this definition, the coefficients of the superposition $c_{m^{\prime}}$ have been absorbed into $h$. It is possible to fully specify $h$ in terms of its roots since it is an analytic function, where each root specifies the location of a vortex core at the corresponding $x$ and $y$ coordinates in the condensate \cite{Mueller02}. 

A vortex lattice ground state corresponds to a situation where the roots of $h$ lie on a lattice. The next step is to construct an analytic function with roots that satisfy this property. Fortunately, there is a well-studied function in complex analysis which will be useful here: the Jacobi theta function $\theta _1(z,\zeta)$. The roots of $\theta_1(z,\zeta)$ are able to describe any regular lattice up to a rotation by making an appropriate choice for the parameter $\zeta $. This means that $h$ may be expressed in terms of $\theta_1$ to obtain a vortex lattice ground state wavefunction. However, in order to make this connection we must introduce a way to describe a regular lattice mathematically.

A two-dimensional lattice may be fully specified by a pair of basis vectors: ${\bf b_1}$ and ${\bf b_2}$. The points of the lattice are obtained from these basis vectors by constructing all possible linear combinations of the form $m_1 {\bf b_1} + m_2{\bf b_2}$ where $m_1$ and $m_2$ are integers, i.e. a two-dimensional Bravais lattice. In general, four real parameters are required to fully specify a lattice: two real components for each of the two basis vectors. However, one of these parameters may be fixed since there is no need  to distinguish between lattices which are equivalent up to a rotation. This is justified because the Hamiltonian which describes the condensate is cylindrically symmetric. In order to fix one of the parameters the first basis vector ${\bf b_1}$ is chosen to be orientated along the $x$-axis.
\begin{figure}[b]
\includegraphics[width=\columnwidth]{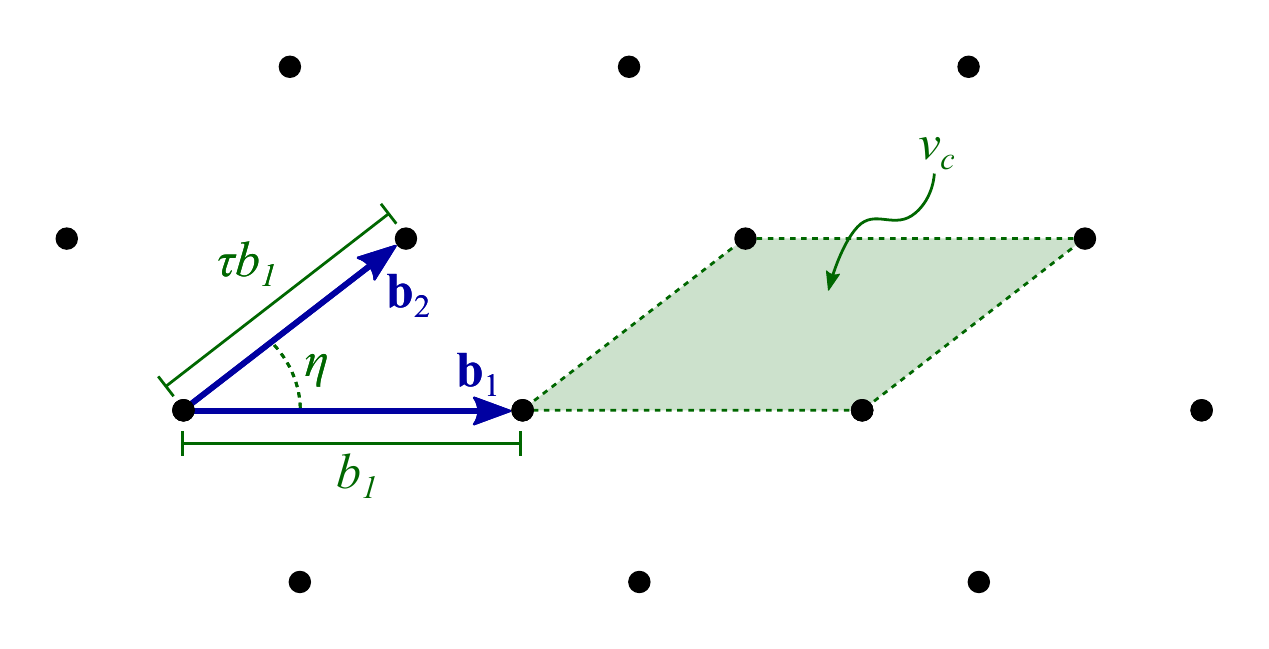}
\caption{Illustration of the Bravais lattice basis vectors and the lattice parameters.}
\label{fig:basis_vectors}
\end{figure}

The three remaining parameters which describe the lattice are depicted in Figure~\ref{fig:basis_vectors}, along with the lattice basis vectors. The first basis vector ${\bf b_1}$ is specified solely in terms of its magnitude, which is denoted by the parameter $b_1$, since its direction is fixed along ${\bf {\hat x}}$. The second basis vector ${\bf b_2}$ is defined with reference to the first basis vector through a rotation and rescaling. This requires two additional parameters: a rotation angle $\eta$ and a scaling factor $\tau$. Writing out the basis vectors explicitly in terms of the parameters $b_1$, $\tau$ and $\eta$, results in ${\bf b_1} =b_1{\bf {\hat x}}$ and ${\bf b_2} =\tau b_1(\cos \eta {\bf {\hat x}} +\sin \eta {\bf {\hat y}})$.

The other parameter that appears in Figure~\ref{fig:basis_vectors} is the area of the unit cell, given by $v_c = b^2_1\tau \sin \eta$. This parameter is redundant in specifying the lattice because it depends on the other parameters: $b_1$, $\tau$ and $\eta$. However, it is useful to mention it because it appears in a number of places in the subsequent analysis.

Another useful concept regarding our mathematical description of the vortex lattice is the reciprocal lattice. It is needed to represent a function which is defined on a lattice as a Fourier series. The reciprocal lattice is obtained from the original lattice by a transformation of the lattice basis vectors. Denoting the basis vectors of the reciprocal lattice as ${\bf q_1}$ and ${\bf q_2}$, we have ${\bf q_1} = 2\pi {\bf b_2} \times {\bf {\hat z}}/v_c = 2\pi b_1 \tau (\sin \eta {\bf {\hat x}} -\cos \eta  {\bf {\hat y}})/ v_c$ and ${\bf q_2} = 2\pi {\bf {\hat z}} \times {\bf b_1}/v_c = 2\pi b_1{\bf {\hat y}}/ v_c$. As for the case of the original lattice, the reciprocal lattice is constructed from its basis vectors by considering all linear combinations of the form $m_1{\bf q_1} + m_2{\bf q_2}$ where $m_1$ and $m_2$ are integers. Each one of these linear combinations is a reciprocal lattice vector, which is denoted in general by ${\bf q}$.

From this, it is possible to describe the two-dimensional density as,
\begin{eqnarray}
n_{\perp}({\brho})=\frac{N}{\pi \chi^2}e^{-\frac{\rho^2}{\chi^2}}g(\brho),
\label{eq:LLL_n_rho}
\end{eqnarray}
where $g(\brho) =  \sum_{\bf q} {\tilde g}_{\bf q} \exp(i{\bf q} \cdot \brho)$ and,
\begin{eqnarray}
g_{\bf q}=\frac{(-1)^{m_1+m_2}e^{-v_c q^2/8\pi}}{\sqrt{\tau \sin \eta}},
\label{eq:fourier_coefs}
\end{eqnarray}
with ${\bf q} = m_1{\bf q_1} + m_2{\bf q_2}$. In other words, the two-dimensional condensate density is the product of a Gaussian envelope and a function $g(\brho) =  \sum_{\bf q} {\tilde g}_{\bf q} \exp(i{\bf q} \cdot \brho)$ which is periodic on the vortex lattice. The Fourier coefficients of $g(\brho)$ are rescaled compared to those in Eq.~(\ref{eq:fourier_coefs}) so that ${\tilde g}_{\bf q} =  g_{\bf q}/ \sum_{\bf v} {\tilde g}_{\bf v} \exp(-iv^2 \chi^2/4)$. The radial extent of the condensate cloud is quantified by the length-scale $\chi = (l_{\perp}^{-2} -\pi v_c^{-1})^{-1/2}$ which is related to the number of vortices in the system.

From this ansatz for the vortex lattice ground state, it is possible to calculate the energy of the condensate as a function of the lattice parameters. Using the fact that $\psi_{\perp}(\brho)$ is in the LLL, the single-particle energy contribution, in the limit of large vortex number, is independent of the vortex lattice parameters. The interaction energy contribution may be rewritten in the following form,
\begin{eqnarray}
E_{\rm vdW}=\frac{N^2 g}{(2 \pi)^{3/2} l_z \chi^2}{\cal I}[n].
\label{eq:E_int_contact}
\end{eqnarray}
Here  ${\cal I}[n]=\pi^2\chi^2/N^2 \int n_{\perp}^2(\brho){\rm d}\brho$ is a dimensionless analogue of the interaction energy contribution. Substituting the ansatz for the vortex lattice ground state from Eq.~(\ref{eq:LLL_n_rho}) gives,
\begin{eqnarray}
{\cal I}(\tau,\eta)=\frac{1}{2} \sum_{{\bf q}, {\bf v}}{\tilde g}_{\bf q}{\tilde g}_{\bf v} e^{-\frac{\chi^2|{\bf q}+{\bf v}|^2}{8}}.
\label{eq:I_full}
\end{eqnarray}
This quantity depends on $\tau$ and $\eta$ implicitly through its dependence on the reciprocal lattice vectors. Considering the limit of large vortex number where $\chi^2{\bf q}^2 \gg 1$, it is possible to simplify this expression significantly. In fact, one may assume that the exponential $e^{-\frac{\chi^2|{\bf q}+{\bf v}|^2}{8}}$ is so sharply peaked at ${\bf q} = -{\bf v}$ that it may be approximated by a Kronecker delta, $\delta_{{\bf q},-{\bf v}}$. This collapses the double sum  to a single sum over ${\bf q}$, which is much simpler to evaluate,
\begin{eqnarray}
{\cal I}(\tau,\eta) &\approx& \frac{1}{2} \sum_{{\bf q}}\left({\tilde g}_{\bf q}\right)^2 \nonumber \\
&=& \frac{1}{2}\sum_{m_1,m_2=-\infty}^{\infty}e^{\pi \left( 2 m_1 m_2 \cot \eta -m_1^2 \tau \csc \eta -\frac{m_2^2 \csc \eta}{\tau}\right)}. \nonumber \\
\label{eq:I_approx}
\end{eqnarray}

In order to minimise ${\cal I}(\tau,\eta)$ as defined in Eq.~(\ref{eq:I_approx}) a numerical treatment is required, which means cutting off the sums over $m_1$ and $m_2$ at some upper and lower bounds. Defining the upper and lower bounds to be at $M$ and $-M$ respectively, with $M$ set to $15$ ensures that ${\cal I}(\tau,\eta)$ is accurate to double precision for values of $\tau$ and $\eta$ greater than about $0.05$. To explore regions in which $\tau$ or $\eta$ is less than $0.05$, $M$ needs to be increased above $15$ to include higher frequency terms in the Fourier series. Fortunately, it is  reasonable to exclude these regions, in the absence of dipolar interactions, because they correspond to an unphysical situation where the vortex lattice begins to collapse onto a line. 
\begin{figure}[t]
\includegraphics[width=\columnwidth]{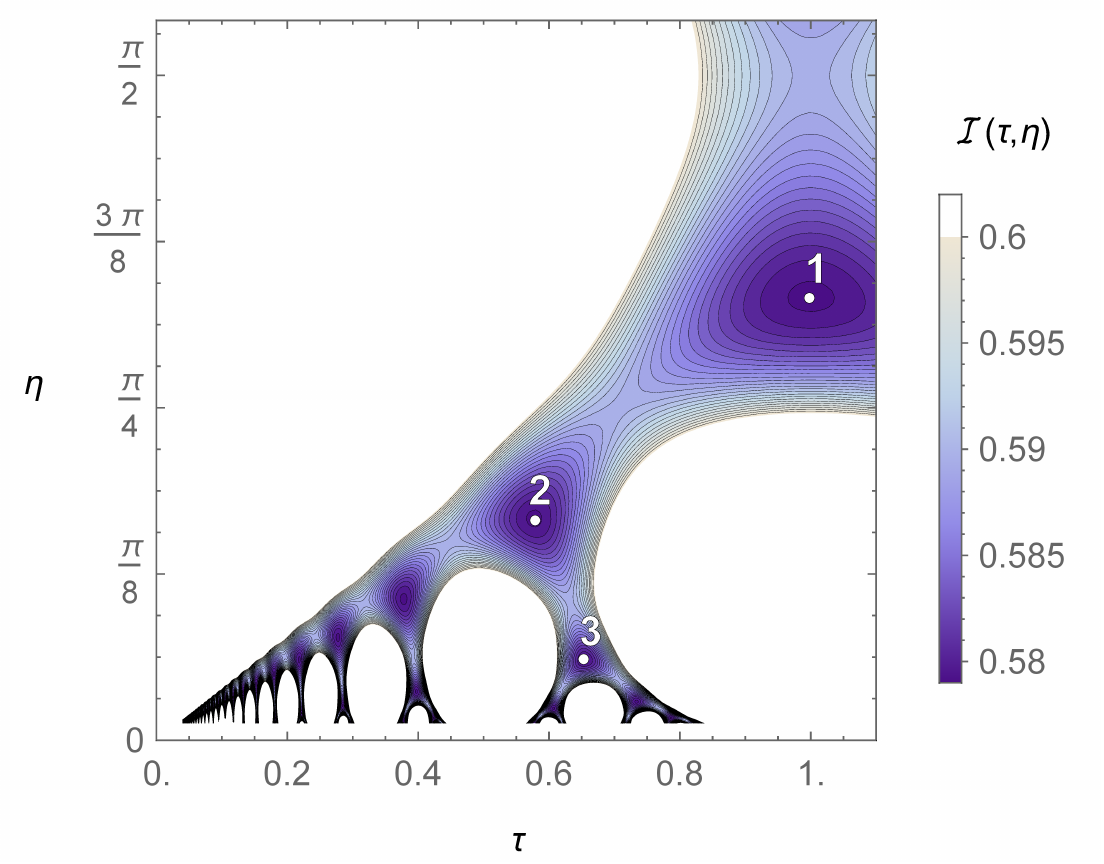}
\caption{Contour plot of ${\cal I}(\tau,\eta)$, a dimensionless analogue of the interaction energy. The white areas are regions where ${\cal I}(\tau,\eta)$ is greater than the cut-off value of $0.6$. The purple areas correspond to regions where ${\cal I}(\tau,\eta)$ approaches its minimum value. Multiple local minima are visible, although they become difficult to see in the lower left-hand region of the plot. The three minima labelled by white numbers are mentioned in the text.}
\label{fig:I_graph}
\end{figure}

\begin{figure}[t]
\includegraphics[width=\columnwidth]{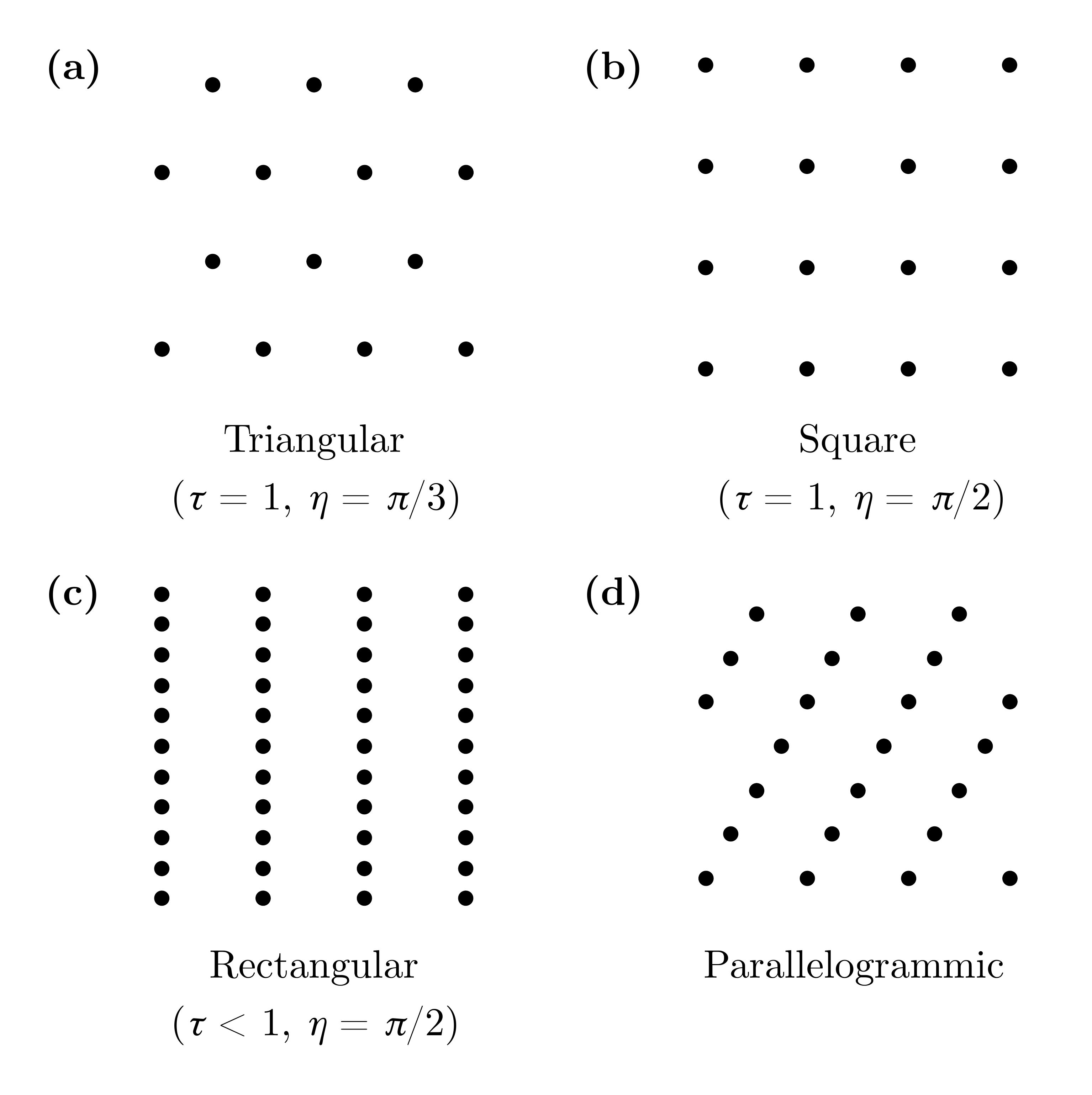}
\caption{The Bravais lattice may be classified as one of five types according to the geometry of the unit cell. The three pertinent Bravais lattice structures considered are triangular, square and rectangular lattices which are all special cases of the parallelogrammic lattice. For the triangular, square and rectangular lattice, we have given the corresponding ($\tau$,$\eta$) values in the so-called standard parametrisation.}
\label{fig:lattices}
\end{figure}

It is illustrative to perform the minimisation of  ${\cal I}(\tau,\eta)$ graphically, by generating a contour plot of ${\cal I}(\tau,\eta)$ as a function of $\tau$ and $\eta$. The result is shown in Figure~\ref{fig:I_graph}. In this plot, the dark purple shading corresponds to regions where ${\cal I}(\tau,\eta)$ approaches its minimum value. The white areas correspond to regions where ${\cal I}(\tau,\eta)$ is greater than the cut-off value, which is in this case set to $0.6$. In Figure~\ref{fig:I_graph} each local minimum has the same value of ${\cal I}$ equal to $0.5797$. However, a  careful consideration shows that each of these solutions in fact corresponds to the same type of lattice; just at a different scale and orientation. The solution labelled `1' in Figure~\ref{fig:I_graph} is the standard parametrisation of the triangular lattice. It is illustrated in Figure~\ref{fig:lattices}, along with the standard parametrisations of the three other types of lattice: square, rectangular and parallelogrammic. In general, the standard parametrisation is defined to be the one for which $\tau$ is closest to $1$. The alternative parametrisations of the same lattice have smaller values of $\tau$ compared to the standard one. For example, the second solution, denoted as `2' in Figure~\ref{fig:I_graph} is an equally valid parametrisation of the triangular lattice. 

The above calculation verifies that a triangular vortex lattice geometry is always favoured in non-dipolar BECs. Of course, this was to be expected based on the results of numerous experiments and previous theoretical studies. It is interesting to note that the triangular lattice geometry is not significantly favoured over other possible lattice geometries. In particular, the energy corresponding to the square lattice geometry at $\tau = 1$ and  $\eta=\pi/2$ (the saddle point in Figure~\ref{fig:I_graph}) is only 1.8\% larger than that of the triangular lattice geometry. It is therefore conceivable that the energy minimum may shift to a non-triangular lattice geometry if the functional form of the interaction energy contribution is altered. With this motivation the above calculation is generalized below to include dipolar interactions.

\subsection{Vortex lattice in a quantum ferrofluid: dipoles perpendicular to the plane of rotation \label{sec:lattice_dipoles_aligned}}
For the case of dipolar BECs the dipolar interaction potential, $U_{\rm dd}({\bf r})$, needs to be included. This is the only modification to the theory that is required to account for the effect of the dipoles.  Assuming that the dipoles are aligned perpendicular to the plane of rotation, i.e. along the $z$-axis,  implies that the cylindrical symmetry of the Hamiltonian is maintained.  In this limit the criterion for being in the LLL becomes
\begin{eqnarray}
\left(1-\frac{\Omega}{\omega_{\perp}}\right) &\lesssim& \frac{\sqrt{2\pi} l_z}{2 N a_s\left(1+\frac{16\pi \edd}{3}\left[1-{\overline \sigma}\sqrt{\frac{9\pi}{8}}\right]\right)} \nonumber \\
&+&\frac{\pi^2 l_z\edd {\overline \sigma}}{a_s N\left(1+\frac{16\pi \edd}{3}\left[1-{\overline \sigma}\sqrt{\frac{9\pi}{8}}\right]\right)^2}, 
\end{eqnarray}
where ${\overline \sigma} =l_z/R_{\perp}$. The above has been obtained by considering the expansion of the ${\tilde U}_{\rm dd}^{\perp}({\bf q})$ in terms of the width of the BEC to first order, i.e. Eq.~(\ref{eq:U_2d_expansion}), and then calculating the dipolar potential to second order in $\rho/R_{\perp}$. 

The additional contribution to the energy functional arises due to the dipolar interactions,
\begin{eqnarray}
E_{\rm dd}=\frac{1}{2}\int n({\bf r_1})U_{\rm dd}({\bf r_1}-{\bf r_2})n({\bf r_2})\,{\rm d} {\bf r_1} {\rm d}{\bf r_2}.
\end{eqnarray}
Adding this to the single-particle and contact interaction energy contributions gives the total energy in the rotating frame for a dipolar BEC,
\begin{eqnarray}
E=E_0 + E_{\rm vdW}+E_{\rm dd}.
\end{eqnarray}
The results obtained in Section \ref{sec:vortex_conventional} for the single-particle ($E_0$) and contact interaction ($E_{\rm vdW}$) energy contributions are unchanged for the dipolar case. All that remains then, is to perform the calculation for $E_{{\rm dd}}$.

Rewriting the dipolar interaction energy contribution in reciprocal space by applying the Fourier convolution theorem leads to,
\begin{eqnarray}
E_{\rm dd}=\frac{1}{2}\frac{1}{(2\pi)^3}\int {\tilde n}({\bf k}){\tilde n}(-{\bf k}) {\tilde U}_{\rm dd}({\bf k})\,{\rm d} {\bf k}.
\end{eqnarray}
To evaluate this in quasi-two dimensions, we substitute the Fourier transform of the quasi-two-dimensional condensate density, ${\tilde n}({\bf k}) = e^{-k_z^2l_z^2/4}{\tilde n}_{\perp}({\bf q})$, such that
\begin{eqnarray}
E_{\rm dd}=\frac{1}{2}\frac{1}{(2\pi)^2}\int {\tilde n}_{\perp}({\bf q}){\tilde n}_{\perp}(-{\bf q}) {\tilde U}^{\perp}_{\rm dd}({\bf q}) \,{\rm d} {\bf q},
\end{eqnarray}
where ${\tilde U}^{\perp}_{\rm dd}({\bf q})$ is given by Eq.~(\ref{eq:dipolar_interaction_2D}), with $\alpha=0$.

Now that  an expression for the dipolar interaction energy in quasi-two dimensions has been derived, it is possible to evaluate the energy assuming that the condensate is in the vortex lattice ground state. Performing the Fourier transform on the vortex lattice condensate density specified in Eq.~(\ref{eq:LLL_n_rho}) results in,
\begin{eqnarray}
{\tilde n}({\bf q})=N\sum_{\bf q'} {\tilde g}_{\bf q'}e^{-\chi^2 |{\bf q'}-{\bf q}|^2/4}.
\end{eqnarray}
This enters into the expression for the dipolar interaction energy, leading to,
\begin{eqnarray}
E_{\rm dd}(\tau,\eta)=\frac{N^2 C_{\rm dd}}{3 \left(2 \pi\right)^{\frac{3}{2}}l_z^3}\sum_{{\bf q},{\bf v}}{\tilde g}_{\bf q} {\tilde g}_{\bf v}e^{-\chi^2 (q^2+v^2)/4}A({\bf v}-{\bf q}), \nonumber \\
\label{eq:Edd_LLL}
\end{eqnarray}
where,
\begin{eqnarray}
A({\bf q})&=&\frac{1}{2\pi}\int e^{-\frac{\chi^2}{l_z^2}\left(u^2+\frac{l_z {\bf q} \cdot {\bf u}}{\sqrt{2}} \right)}\left[2-3\sqrt{\pi} ue^u {\rm erfc}(u)\right]\,{\rm d} {\bf u}\nonumber \\
&=&2\int_0^{\infty} e^{-\frac{\chi^2u^2}{l_z^2}}uI_0\left(\frac{\chi^2 q u}{\sqrt{2}l_z^2}\right)\, {\rm d}u \nonumber \\
&-&3\sqrt{\pi}\int_0^{\infty} e^{-\frac{\chi^2u^2}{l_z^2}}u^2 e^u {\rm erfc}(u)I_0\left(\frac{\chi^2 q u}{\sqrt{2}l_z^2}\right)\, {\rm d} u, \nonumber \\
\end{eqnarray}
and $I_0(\cdot)$ is the modified Bessel function of the first kind. At this point, the integral may be separated into two terms $A_1({\bf q}) + A_2({\bf q})$. The first term has a simple solution,
\begin{eqnarray}
A_1({\bf q})=2\int_0^{\infty}u e^{-\frac{\chi^2 u^2}{l_z^2}}I_0\left(\frac{\chi^2 q u}{\sqrt{2}l_z^2}\right)\,{\rm d}u=\left(\frac{l_z}{\chi}\right)^2 e^{\frac{\chi^2 q^2}{8}}. \nonumber \\
\end{eqnarray}

Returning to the expression for the dipolar interaction energy given in Eq.~(\ref{eq:Edd_LLL}), and substituting the simplified result for $A_1({\bf q})$ gives
\begin{eqnarray}
E_{\rm dd}(\tau,\eta)=\frac{2N^2 C_{\rm dd}}{3 \left(2 \pi\right)^{\frac{3}{2}}l_z \chi^2} {\cal I}(\tau,\eta)+ \frac{N^2 C_{\rm dd}}{3 \left(2 \pi\right)^{\frac{3}{2}}l_z^3} {\cal W}(\tau,\eta), \nonumber \\
\end{eqnarray}
where ${\cal I}(\tau,\eta)$ is as defined in Eq.~(\ref{eq:I_full}) and,
\begin{eqnarray}
{\cal W}(\tau,\eta)=\sum_{{\bf q},{\bf v}} {\tilde g}_{\bf q} {\tilde g}_{\bf v}e^{-\chi^2 ({\bf q}^2+{\bf v})^2/4}A_2({\bf v}-{\bf q}).
\label{eq:W_full}
\end{eqnarray}
This expression shows that the dipolar energy has been separated into two distinct contributions: a local contribution which is proportional to ${\cal I}(\tau,\eta)$ and a non-local contribution which is proportional to ${\cal W}(\tau,\eta)$. In principle, it is now possible to calculate the dipolar interaction energy as a function of the lattice parameters, however the function for the non-local contribution, ${\cal W}(\tau,\eta)$ is not analytically tractable and it must be evaluated numerically. In order to resolve this the difficult integral is expanded in a series of simpler integrals, each of which can be performed analytically. This can be done by expressing the complementary error function, which appears in the integrand, as a power series,
\begin{eqnarray}
{\rm erfc}(u)=1-\frac{1}{\sqrt{\pi}}\sum_{n=0}^{\infty}\frac{(-1)^n u^{2n+1}}{n!(2n+1)}
\end{eqnarray}
and bringing the sum outside of the integral. Additionally, as in the case of the contact interactions it is possible to reduce the double sum, in Eq.~(\ref{eq:W_full}) to a single sum in the limit of large vortex number, resulting in,
\begin{eqnarray}
{\cal W}(\tau,\eta)\approx \sum_{{\bf q}} ({\tilde g}_{\bf q})^2 e^{-\frac{\chi^2 q^2}{2}}\left(A^{\rm a}_2(2{\bf q})+A^{\rm b}_2(2{\bf q})\right), 
\end{eqnarray}  
where,
\begin{eqnarray}
e^{-\frac{\chi^2 q^2}{2}}A^{\rm a}_2(2{\bf q})&=&\frac{3\pi}{8(\beta^2-1)^{\frac{5}{2}}} e^{-\frac{(\beta^2-2)q^2\chi^2}{4(\beta^2-1)}}\nonumber \\
&\times&\left[\beta^4l_z^2 q^2 I_1\left(\frac{\beta^4l_z^2q^2}{4-4\beta^2}\right)\right. \nonumber \\
&-&\left.\left(2\beta^2 +\beta^4 l_z^2q^2-2 \right) I_0\left(\frac{\beta^4l_z^2q^2}{4-4\beta^2}\right)\right] \nonumber \\
\end{eqnarray}
and,
\begin{eqnarray}
& &e^{-\frac{\chi^2 q^2}{2}}A^{\rm b}_2(2{\bf q})=3e^{-\frac{q^2\chi^2}{2(\beta^2-1)}} \times \nonumber \\ &~&\sum_{n=0}^{\infty}\frac{(-1)^n(n+1)}{(\beta^2-1)^{2+n}(2n+1)}L_{n+1}\left(\frac{\beta^4l_z^2q^2}{2-2\beta^2}\right), 
\end{eqnarray}
where $\beta=\chi/l_z$ and $L_n(\cdot)$ is the $n^{\rm th}$ Laguerre polynomial.

\begin{figure}[b]
\includegraphics[width=\columnwidth]{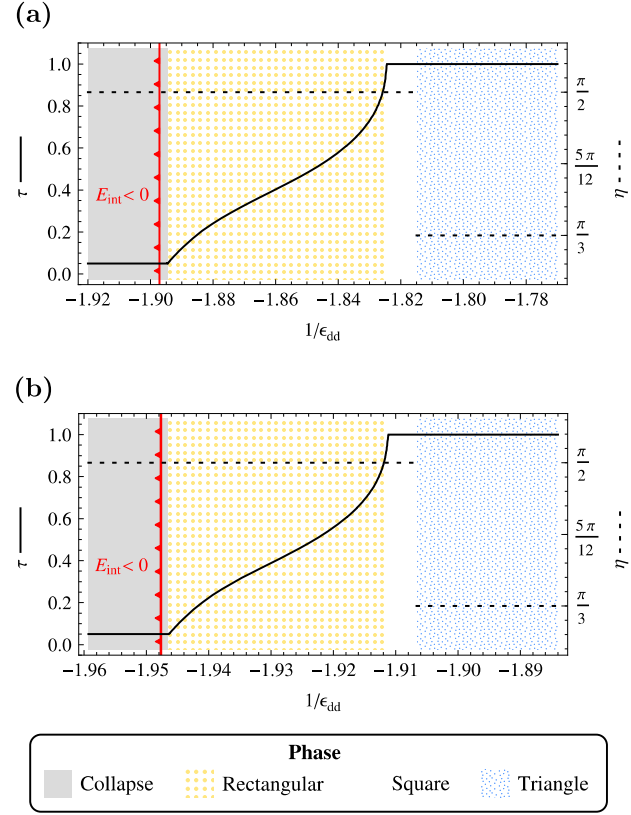}
\caption{Plots showing the optimal values of $\tau$ and $\eta$ in dipolar BECs with on-axis polarisation. Plot (a) assumes the length scales $v_c/(\pi l_{\perp}^2)=1.0191$, $l_{\perp}/\l_z=40$ and $\chi/l_z=292.33$ (b) assumes the length scales $v_c/(\pi l_{\perp}^2)=1.0191$, $l_{\perp}/\l_z=80$ and $\chi/l_z=584.66$. In each plot, the black solid (dashed) line represents the optimal value of $\tau$ ($\eta$) with reference to the scale on the left (right) vertical axis. By classifying the ($\tau$,$\eta$) parameters, four distinct regions in phase space are identified: a collapse phase, rectangular lattice (stripe) phase, square lattice phase, and triangular lattice phase. These regions are indicated by distinct shading. The red line with arrows on the left side specifies the region for which the interaction energy is less than zero.}
\label{fig:dipolar_lattice}
\end{figure}

With the results of Section \ref{sec:vortex_conventional} it is now feasible, from a computational perspective, to numerically minimise the condensate energy, $E(\tau, \eta) = E_0 + E_{\rm int}(\tau, \eta)$, with respect to $\tau$ and $\eta$ to determine the optimal vortex lattice geometry. In minimising the condensate energy, as in Section \ref{sec:vortex_conventional}, only the interaction energy contribution needs to be considered,
\begin{eqnarray}
E_{\rm int}(\tau, \eta)=\frac{N^2 C_{\rm dd}}{3 \left(2 \pi\right)^{\frac{3}{2}}l_z^3} \times \nonumber \\
\left[\left(2+\frac{1}{\edd}\right)\left(\frac{l_z}{\chi}\right)^2 {\cal I}(\tau,\eta)+{\cal W}(\tau,\eta)\right],
\end{eqnarray} 
since the single-particle contribution does not depend on $\tau$ and $\eta$. Assuming $C_{\rm dd} > 0$ so that the factor outside the square brackets is positive the vortex lattice configuration is determined by the the minimization of 
 \begin{eqnarray}
 \left(2+\frac{1}{\edd}\right)\left(\frac{l_z}{\chi}\right)^2 {\cal I}(\tau,\eta)+{\cal W}(\tau,\eta),
 \label{eq:min_vort_dipolar}
 \end{eqnarray}
with respect to $\tau$ and $\eta$.  The results of this minimization are shown in Figure~\ref{fig:dipolar_lattice} for particular choices of length scale parameters which are given in the figure caption.  In both figures, the optimal values of $\tau$ and $\eta$ are plotted against $\edd^{-1}$. The optimal value of $\tau$ is represented by a solid line with reference to the scale on the left vertical axis, while the optimal value of $\eta$ is represented by a dotted line with reference to the scale on the right vertical axis. Taken together, the optimal values of $\tau$ and $\eta$ describe the optimal vortex lattice geometry.
\begin{table*}
\begin{center}
\begin{tabular}{c|c|c}
\hline
Phase & Figure \ref{fig:dipolar_lattice}(a) & Figure \ref{fig:dipolar_lattice}(b) \\
\hline
Collapse & $\edd^{-1} < -1.895$ & $\edd^{-1} < -1.946$ \\
Rectangular & $-1.895 <\edd^{-1} < -1.825$ & $-1.946 <\edd^{-1} < -1.911$ \\
Square & $-1.825 <\edd^{-1} < -1.815$ & $-1.911 <\edd^{-1} < -1.906$ \\
Triangular & $ \edd^{-1} > -1.815$ & $\edd^{-1} > -1.906$ \\
\hline
\end{tabular}
\end{center}
\caption{Definition of the four regions in phase space shown in Figure~\ref{fig:dipolar_lattice}.}
\label{tab:lattice}
\end{table*}%

By comparing the optimal values of $\tau$ and $\eta$ to the standard lattice parameterisations given in Figure~\ref{fig:lattices}, it is possible to classify the geometry of the vortex lattice. For example, at the point $\edd^{-1} = -1.82$ in Figure~\ref{fig:dipolar_lattice}(a), a square lattice is favourable since the optimal values of $\tau$ and $\eta$ at that point are $1$ and $\pi/2$ respectively. Continuing in this way, three types of lattice emerge depending on the value of $\edd^{-1}$: triangular, square or rectangular. Each of these lattice geometries occurs in a distinct region of the phase space, indicated by distinct shading in Figure~\ref{fig:dipolar_lattice}. The fact that the transition to different lattice geometries occurs when $\edd^{-1} \approx -2$  indicates that the transition arises in the regime where the long-range contribution to the interactions [${\cal W}(\tau,\eta)$] dominates, see Eq.~(\ref{eq:min_vort_dipolar}).  Comparing Figures~\ref{fig:dipolar_lattice}(a), with $l_{\perp}/\l_z=40$, and (b), with $l_{\perp}/\l_z=80$, the relative size of each region is comparable for both condensate aspect ratios. There is however an overall translation and scaling difference between the two cases: the regions in  Figure~\ref{fig:dipolar_lattice}(b), where the condensate is more oblate,  are contracted and shifted to the left compared to those in  Figure~\ref{fig:dipolar_lattice}(a), where the condensate is less oblate. A more accurate description of the phase regions is given in Table \ref{tab:lattice} in terms of inequalities.

In addition to the three pattern-shaded regions, there is also a solid grey-shaded region for which the condensate is in the so-called collapse phase. In the collapse phase, the optimal value of $\tau$ tends to zero and the vortex lattice analysis begins to break down. Physically, this phase corresponds to a situation where the vortices are arranged in densely-packed lines, with the spacing between the lines being larger than the extent of the condensate in the $x-y$ plane. Since the unit cell of the lattice extends beyond the boundaries of the condensate in this situation, the vortex lattice can be considered to have {\it collapsed}. When the system enters the collapse phase, there are also signs that the analysis becomes invalid. Since $\tau$ approaches zero in this phase, the expression for interaction energy becomes inaccurate because only enough terms in the Fourier decomposition to consider values of $\tau$ greater than about $0.05$ are included (as in Section \ref{sec:vortex_conventional} $M = 15$). 

Apart from the stability of the vortex lattice, we may also assess the stability of the condensate itself by looking at the sign of the interaction energy. If the interaction energy is negative, then it can approximately be regarded that the condensate as being prone to collapse. The region of phase space for which interaction energy is negative is the area to the left of the red line in Figures~\ref{fig:dipolar_lattice} (a) and (b). Interestingly, the interaction energy becomes negative at roughly the value of $\edd^{-1}$ where the optimal value of $\tau$ approaches zero. This suggests that there may be a link between the collapse of the condensate and the collapse of the vortex lattice.

In this section  the vortex lattice geometry in dipolar BECs for the special case of on-axis polarisation has been analysed. The results show that three lattice geometries are possible, depending on the value of $\edd^{-1}$ and the values of the length scales $\l_{\perp}$, $l_z$ and $\chi$. In general,  a triangular lattice geometry is favoured in regions where the local interaction contribution dominates, as was seen in the non-dipolar case. However, in regions where the non-local interaction contribution becomes significant, the favoured lattice geometry changes from triangular to square or rectangular. Below a certain value of $\edd^{-1}$ (corresponding to reasonably strong, attractive contact interactions) the vortex lattice and the condensate appear to collapse concurrently.

The results obtained above qualitatively agree with previous results of Zhang and Zhai \cite{Zhang05} and Cooper {\it et al.} \cite{Cooper05,Cooper06}. Zhang and Zhai also find that the lattice geometry undergoes a transition from triangular $\rightarrow$ square $\rightarrow$ rectangular $\rightarrow$ collapse as the value of $\edd^{-1}$ decreases. Cooper {\it et al.} also find the same transitions between lattice geometries. However, they do not find that the lattice collapses after passing through the rectangular lattice phase. Instead, they observe a bubble phase - a different kind of periodic vortex structure in which the vortices are arranged around {\it bubbles} of high particle density. Such states do not occur in the above analysis since they fall outside the scope of the analytic treatment used.

Yi and Pu \cite{Pu06} also conducted a similar study of vortex lattice geometry based on numerical simulations of the GPE, however their results are not in agreement with those obtained above, nor with those of Zhang and Zhai \cite{Zhang05} and Cooper {\it et al.} \cite{Cooper05,Cooper06}. They only observe triangular lattice geometries in their simulations, and conclude that the square and rectangular lattice geometries do not exist. Possible explanations for this discrepancy include that the particular parameter values  chosen for the simulations do not fall in the square and rectangular lattice regions and/or the simulations were not in the LLL regime.

\subsection{Vortex lattice in a quantum ferrofluid: dipoles not perpendicular to the plane of rotation}
\label{sec:lattice_dipoles_nonaligned}
Although it is no longer appropriate to assume that the wavefunction envelope has cylindrical symmetry for the case of off-axis polarisation, there is still another useful symmetry that can be exploited: reflection symmetry. This reflection symmetry occurs about the $x-z$ plane -- the plane which contains both the polarisation vector and the axis of rotation.  In order to derive a new ansatz for the vortex lattice ground state which assumes reflection symmetry, only  minor modifications need to be made to the  derivation given in Section \ref{sec:lattice_dipoles_aligned}. Specifically, it is necessary to introduce two new variational parameters: $\lambda$ and $\zeta$.

The parameter $\lambda$ is required to describe the deviation of the condensate cloud from cylindrical symmetry. It is  the ratio of the width of the condensate cloud along the $y$-direction divided by the width along the $x$-direction. If the density profile of the cloud is expressed in the form $\exp[-x^2/l_x^2-y^2/l_y^2]$, then the aspect ratio is written as $\lambda=l_y/l_x$. For a cylindrically symmetric BEC, the width of the cloud along the $x$- and $y$-directions must be the same, which implies that $\lambda=1$. For $\alpha>0$ [see Figure~\ref{fig:schematic}], since dipolar BECs elongate along the direction of polarisation, it is expected that  $l_x > l_y$ and hence $0<\lambda <1$.

The other new parameter, $\zeta$, is required to allow the vortex lattice to adopt any orientation with respect to the polarisation direction. It is defined to be the angle between the first lattice basis vector, ${\bf b_1}$, and the projection of the dipole polarisation onto the plane of rotation. For a cylindrically symmetric BEC,  the energy will be independent of $\zeta$. However, for  non-cylindrically symmetric dipolar BECs ($\alpha >0$), it is conceivable that the energy may depend on $\zeta$.

By modifying the derivation of the ansatz for the vortex lattice ground state to incorporate the new parameters, it is possible to show that the condensate density must be of the following form,
\begin{eqnarray}
n_{\perp}({\brho})&=&\frac{2N\lambda}{\pi \chi^2\left(1+\lambda^2\right)}e^{-\frac{2}{1+\lambda^2}\left(\frac{\lambda^2x^2+y^2}{\chi^2}\right)} \nonumber \\
&\times&\sum_{\bf q} \overline{g}_{\bf q} e^{i{\bf q}\cdot {\hat R}_{\zeta}\brho},
\end{eqnarray}
where ${\hat R}$ represents the standard two-dimensional rotation operator, $\chi = [(l_x l_y)^{-1} -\pi v_c^{-1}]^{-1/2}$  and,
\begin{eqnarray}
\overline{g}_{\bf q}=\frac{g_{\bf q}}{\sum_{\bf v} g_{\bf v} \exp\left(-\frac{\chi^2(1+\lambda^2)}{8}\left[{\bf v} \cdot {\hat R}_{\zeta} {\bf {\hat x}}/\lambda^2 +{\bf v} \cdot {\hat R}_{\zeta} {\bf {\hat y}}\right]\right)}. \nonumber \\
\end{eqnarray}

From this starting point is possible to generalize the approach presented in Section \ref{sec:lattice_dipoles_aligned} to the case where $\alpha > 0$. The total energy can be broken into two components, 
\begin{eqnarray}
E(\tau,\eta,\lambda,\zeta)&=&E_0(\lambda)+E_{\rm int}(\tau,\eta,\lambda,\zeta).
\end{eqnarray}
The non-interacting component of the energy is given by,
\begin{eqnarray}
E_0(\lambda)&=& N\left(\hbar \Omega+ \frac{1}{2}\hbar \omega_z\right) \nonumber \\
&+&\frac{N\hbar \chi^2 \left(\omega_{\perp}^2-\Omega^2\right)}{2\Omega \left(l_x^2+l_y^2\right)}\left(\frac{1+\lambda^2}{2\lambda}\right)^2,
\end{eqnarray} 
which in the limit $\Omega \rightarrow \omega_{\perp}$ tends to a constant. Following the same procedure as in Section \ref{sec:lattice_dipoles_aligned}, using Eq.~(\ref{eq:LLL_n_rho}), the interaction energy is given by,
\begin{eqnarray}
E_{\rm int}(\tau,\eta,\lambda,\zeta)&=&\frac{N^2 C_{\rm dd}}{6 (2\pi)^{3/2} l_z\chi^2}\left[2 \left(\frac{\chi}{l_z}\right)^2 {\cal W}(\tau,\eta,\lambda,\zeta)\right. \nonumber \\
&+&\left.\left(1+3\cos(2\alpha)+\frac{2}{\edd}\right){\cal I}(\tau,\eta,\lambda)\right], \nonumber \\
\end{eqnarray}
where,
\begin{eqnarray}
{\cal I}(\tau,\eta,\lambda) \approx  \frac{\lambda^2}{1+\lambda^2} \sum_{{\bf q}}\left({\tilde g}_{\bf q}\right)^2
\end{eqnarray}
and,
\begin{eqnarray}
{\cal W}(\tau,\eta,\lambda,\zeta)&\approx& \sum_{{\bf q}} ({\tilde g}_{\bf q})^2 e^{-\chi^2\frac{\left(1+\lambda^2\right)}{2}\left[\left(\frac{q_x^{(-\zeta)}}{\lambda}\right)^2+q_y^{(-2\zeta)}\right]} \nonumber \\
&\times& A_2\left(\frac{2q_x^{(-\zeta)}}{\lambda},2q_y^{(-\zeta)}\right), 
\end{eqnarray}
where $q_x^{(-\zeta)}$ ($q_y^{(-\zeta)}$) represents the $x$-component ($y$-component) of ${\hat R}_{-\zeta} {\bf q}$ and,
\begin{eqnarray}
A_2(q_x,q_y)&=&\frac{3}{2\sqrt{\pi}}\int_0^{\infty}du\,\int_0^{2\pi}du_{\phi}\, u^2e^{u^2} {\rm erfc}(u) \nonumber \\
&\times& \left(\sin^2\alpha \cos^2 u_{\phi}-\cos^2\alpha\right) \nonumber \\
&\times& \exp\left[-\frac{\chi^2\left(1+\lambda^2\right)}{2l_z^2}\left(\frac{u^2\cos^2 u_{\phi}}{2}\right)\right] \nonumber \\
&\times& \exp\left[-\frac{\chi^2\left(1+\lambda^2\right)}{2l_z^2}u^2\sin^2 u_{\phi}\right] \nonumber \\
&\times& \exp\left[-\frac{\chi^2\left(1+\lambda^2\right)}{2l_z^2}\left(\frac{l_zq_xu\cos u_{\phi}}{\sqrt{2}\lambda}\right)\right]  \nonumber \\
&\times& \exp\left[-\frac{\chi^2\left(1+\lambda^2\right)}{2l_z^2}\left(\frac{l_zq_yu\sin u_{\phi}}{\sqrt{2}}\right)\right].  \nonumber \\
\end{eqnarray}

Although there are severe computational limitations surrounding the minimisation of the above  expression for the condensate energy, it is still  possible to make some meaningful calculations if only triangular and square lattices are considered. For example, it is possible to address the question of whether there is a transition between triangular and square lattices. By considering the minimization at two points in parameter space: $(\edd^{-1}=0.9,\alpha=\pi/2)$ and $(\edd^{-1}=0.95,\alpha=\pi/2)$ we find evidence for a transition. In the limit  $\Omega \rightarrow \omega_{\perp}$ ($E_0(\lambda) \rightarrow N[\hbar \Omega+ \hbar \omega_z/2]$) Table \ref{tab:lattice_off}  shows  $E_{\rm int}(\tau,\eta,\lambda,\zeta)$ for the two vortex lattices at the two points in parameter space considered.  Looking at the results, there is a phase transition in the vortex lattice geometry from triangular to square as a function of  $\edd$. It is also found that the variational parameter $\zeta$ is essentially irrelevant, since the minimum value of $E_{\rm int}(\tau,\eta,\lambda,\zeta)$ is found to be the same for any choice of $\zeta$. This suggests that the orientation of the vortex lattice is unaffected by the broken cylindrical symmetry due to the off-axis polarisation. However, the optimal value of $\lambda$ does deviate from the cylindrically symmetric value of $1$. At both points  considered, the optimal value is  less than $1$, which indicates, as expected, that the dipolar BEC is elongated along the direction of polarisation. Interestingly, the optimal value of $\lambda$ does not depend on the lattice geometry.
\begin{table}[t]
\begin{center}
\begin{tabular}{c|c|c|c}
\hline
$\edd^{-1}$ & Phase & Optimal $\lambda$ & ${\tilde E}_{\rm int}$ \\
\hline
$0.9$ & Square & $0.93$ & $-4979.49$ \\
$0.9$ & Triangular & $0.93$ & $-4788.31$\\
\hline
$0.95$ & Square & $0.78$ & $-1377.22$ \\
$0.95$ & Triangular & $0.78$ & $-1426.81$\\
\hline
\end{tabular}
\end{center}
\caption{Results of the minimisation of ${\tilde E}_{\rm int}=(3(2\pi)^{\frac{3}{2}} l_z^3)/(N^2C_{\rm dd}) E_{\rm int}$ with respect to $\lambda$ and $\zeta$, for
triangular and square lattices, at two values of $\edd$ with $\alpha=\pi/2$. The specific parameters used are $v_c/(\pi l_xl_y)=1.0191$, $\sqrt{l_xl_y}/l_z=40$ and $\chi/l_z=292.33$. The optimal value of $\zeta$ is not included because ${\tilde E}_{\rm int}$  was found to be independent of $\zeta$. The results show that the optimal value of $\lambda$ is the same for both triangular and square lattice geometries.}
\label{tab:lattice_off}
\end{table}

Numerical studies \cite{Pu06}, based on the dipolar GPE,  suggest that for $\alpha>0$ the vortex lattice can undergo a phase transition from triangular structure to a non-triangular structure as $\edd$ is increased. This is  consistent with analysis presented above, however, to our knowledge there has not, to date, been a thorough study of vortex lattice structures for the regime of $\alpha>0$.

\subsection{Vortex lattices in two-component dipolar Bose-Einstein condensates}
The theoretical study of non-dipolar two-component BECs \cite{Mueller02,Kasamatsu03} has shown how interspecies interactions $g_{12}$ can influence the vortex lattice structure of the two components. The analysis presented in Section \ref{sec:lattices} can be adapted to analyse the vortex lattice structure of two component condensates. Doing this Mueller and Ho \cite{Mueller02} were able to quantify various regimes through the parameter $\beta=g_{12}/\sqrt{g_1g_2}$, where $g_{12}$ is the interspecies interaction parameter and $g_{1(2)}$ is the intraspecies interaction parameter for component $1$ $(2)$ of the two-component BEC. This treatment and subsequent numerical analysis \cite{Kasamatsu03} shows that for $\beta<0$ the two components overlap and a single triangular vortex lattice arises. For $\beta>0$ the two components separate to form interlaced triangular vortex lattices. As $\beta \sim 1$ the triangular lattice distorts to form square or rectangular arrays. This theoretical analysis is consistent with experiment \cite{Schweikhard04}.

It is also possible to consider two-component dipolar BEC systems, with both  interspecies and intraspecies contact and dipolar interactions. Work by Shirley {\it et al.} \cite{Shirley14} showed that such systems (where one of the components has zero dipolar interactions), under rotation, exhibit a rich phase diagram, which includes triangular vortex lattices, square vortex lattices, vortex sheets (where half quantum vortices of one component align in a winding sheet, which is interwoven with a sheet in the other component \cite{Kasamatsu09a}), half quantum vortex chains (where vortices, alternating between each component, line up along a chain) and half quantum vortex molecules (where a vortex in a given component pairs up with a vortex in the other component). This analysis is consistent with further studies \cite{Zhao13,Ghazanfari14,Zhang16} and has been extended to consider how dipole alignment in the plane of rotation \cite{Zhang16} and component-dependent optical lattices \cite{Wang16} influences the phase diagram. 

\section{Summary and Outlook \label{sec:summary}}
\subsection{Summary}
The aim of this review was to take the reader on a journey, starting with the fundamental concepts and methodologies used to understand the properties of dipolar BECs and then show how these have been applied to understand the properties of vortices and vortex lattices in these systems.  Throughout, dipolar interactions are seen to enrich the physical properties of the system and vortices therein.  The journey started in earnest in Section \ref{sec:Quant_Ferro} where we met the dipolar interactions; these introduce a long-range and anisotropic component to the interactions, making a significant departure from conventional $s$-wave interactions which appear in the theory as isotropic contact interactions.  In Section \ref{sec:vortex_free} we saw that the dipolar interaction significantly modifies the fundamental stationary solutions of a dipolar condensate, including the introduction of collapse instabilities dependent on the shape of the boundary relative to the polarization direction.  In Section \ref{sec:single_vortices} we found that dipolar interactions can alter the energy and structure of a single vortex. Specifically, in quasi-two-dimensional systems when the dipole alignment is in the plane of the condensate, the vortex core is no longer circularly symmetric. Additionally, density ripples appear in the vicinity of the vortex core, as the roton instability is approached, due to the roton mixing with the ground state of the system.  In Section \ref{sec:dynamics} we found that the interaction between vortices can be altered by the absence of dipoles in the vortex core, introducing an additional long-range and anisotropic contribution to the vortex-vortex interaction. This can, for instance, lead to the suppression of the annihilation of vortex-antivortex pairs and induce the co-rotational dynamics of vortex-vortex pairs to become anisotropic.  In Section \ref{sec:gen_vortices} we summarised the methods for generating vortices in condensates, and discussed the role of dipolar interactions in these processes.  Concentrating on the properties and instabilities of rotating condensates, dipolar interactions were shown to significantly alter the regimes of stability and the critical rotation frequencies for vortices to be nucleated.  This also allowed us to identifiy routes to vortex formation under rotation.  Finally, in Section \ref{sec:lattices} we found that dipolar interactions lead to new and exotic vortex lattice phases; whereas lattices in non-dipolar condensates are well-known to follow a triangular pattern, dipolar interactions can support rectangular, square and bubble phases.

Of course any journey is a compromise between taking an efficient route and a scenic path, i.e. in this case a compromise between completing the review, in a {\it timely} manner, and detailing every contribution to the field.  Unfortunately our path has been fairly efficient and as such we have omitted several other aspects relating to vortices and vortex lattices in quantum ferrofluids. Below we, all too briefly, provide a snapshot of some of the scenery we have missed along the way and avenues for further exploration.

\subsection{Outlook}
\subsubsection{Dipolar Bose-Einstein condensates in toroidal traps}
The experimental study of persistent superfluid flow in BECs confined to toroidal traps \cite{PRL99_260401,PRL106_130401,PRA86_013629,PRL110_025301,PRL110_025302,PRA88_063633,PRL111_205301,PRL111_235301,PRL113_045305,PRL113_135302,NJP16_013046,Nature506_200} has matured significantly over the last decade.  As such, ring shaped BECs in toroidal traps have been the subject of many experimental and theoretical investigations \cite{PRA66_053606,EuroLett46_275,JPhysB34_L113,SciRep2_352,PRA64_063602,PRA74_061601} focusing on persistent currents \cite{PRL99_260401,PRL110_025301,PRA88_051602}, weak links \cite{PRL106_130401,PRL110_025302}, formation of matter-wave patterns by rotating potentials \cite{PRA86_023832}, solitary waves \cite{JPhysB34_L113,PRA79_043602}, and the decay of the persistent current via phase slips \cite{PRA86_013629,PRA80_021601,JPhysB46_095302}. In these studies the transference of angular momentum from optical fields  \cite{PRL99_260401,PRL110_025302} or stirring with an optical potential \cite{PRL110_025302,PRA88_063633} is used to generate persistent flow. 

Within the context of this review we consider a persistent flow in a toroidal BEC as a giant vortex state.  One might  suppose that beyond studying the density profile and stability of a dipolar condensate within a toroidal trap \cite{Abad10,Adhikari12} that dipolar interactions have a limited influence on the superflow properties in such a geometry. This assumption arises since to a close approximation the wavefunction can be considered to have the form $\psi({\bf r})=\sqrt{n({\bf r})} \exp[i q_{\rm v} \theta]$, where $\theta$ is the azimuthal angle and  $q_{\rm v}$ is the vortex charge (charaterising the persistent flow) around the toroid, and the density is independent of $q_{\rm v}$. As such, when considering the energy difference between $q_{\rm v}$ and $q_{\rm v}+1$ the dipolar interactions play no role. 

Despite this observation work has been carried out on the properties  of dipolar BECs in toroidal traps focusing on the generation of persistent flows via the He-McKellar-Wilkens or Aharonov-Casher effect \cite{Wood16} and the properties of two-component dipolar BECs in concentrically coupled toroidal traps \cite{Zhang15}. In the former case it was shown that for atomic dipolar BECs, where the dipolar interaction is mediated via a magnetic dipole moment, that although it is theoretically possible to induce persistent flow, via the Aharonov-Casher effect \cite{Aharonov84,Anandan82,Cimmino89}, the strength of electric  field required is prohibitive. In the case of polar molecules, with significant electric dipole moments, the He-McKellar-Wilkens effect \cite{He93,Wilkens94,Lepoutre12} could ultimately be used to generate a persistent flow in a dipolar BEC in a toroidal geometry. For the case of a two-component dipolar BECs \cite{Zhang15} in concentrically coupled toroidal traps, various vortex structures can arise depending on the strength of the dipolar interactions and the rotation frequency. The interesting vortex structures predicted include polygonal vortex clusters and vortex necklaces.

\subsubsection{Fractional quantum Hall physics in dipolar Bose-Einstein condensates}
In Section \ref{sec:lattices} we considered the LLL regime to investigate vortex lattice structures in dipolar BECs. Ultimately, in the limit $\Omega \rightarrow \omega_{\perp}$, a rotating BEC is predicted to make a quantum phase transition to a highly correlated, non-superfluid, fractional quantum Hall groundstate. This  state emerges when the LLL meanfield vortex lattice melts, i.e.\ when the number of vortices ($N_{\rm v}$) in the BEC is the same as  or greater than the number of atoms (N). This regime occurs approximately when $\Omega/\omega_{\perp} \sim 0.999$. In the absence of dipolar interactions theoretical evidence for such a transition has come from exact two-dimensional groundstate calculations for a small number of bosons with a large angular momentum \cite{Wilken98,Wilken00,Cooper01}. For a review of quantum Hall physics in rotating BECs see Ref. \cite{Viefers08}. 

The natural question which arises is: do dipolar interactions influence this phase transition and the properties of the highly correlated state? Work by Rezayi {\it et al.} \cite{Rezayi05} showed that at a filling factor of $N/N_{\rm v}=3/2$, with $N=18$, dipolar interactions support an incompressible fluid ground state which possesses non-Abelian statistics for the quasiparticle excitations. Additionally, dipolar interactions in lattice systems have been shown to increase the gap between the ground state and the first excited state \cite{Hafezi07}, for $N/N_{\rm v}=1/2$. It is expected that this increase in the gap will be maintained in the thermodynamic limit and hence relevant for experiments. Numerical simulations by Chung and Jolicoeur \cite{Chung08} showed that at $N/N_{\rm v}=1$ the ground state is a Moore-Read paired state, as is the case for bosons with purely contact  interactions. This state is destabilized when the contact interactions are small enough, i.e.\ dipolar interactions alone can not support this state. For $N/N_{\rm v}=1/3$ a composite fermion sea emerges, where each boson is bound with three vortices. The robustness of fractional quantum Hall states in artificial gauge fields, in the presence of dipolar interactions, has been investigated by Gra{\ss} {\it et al.} \cite{Grab11}.

\subsubsection{Dipolar fermions \label{sec:current_fermions}}
There is considerable interest in the properties of dipolar Fermi gas systems. The partially attractive nature of the dipolar interaction in single-component dipolar Fermi gases opens up the possibility of BCS pairing resulting in superfluid states in three dimensions \cite{Baranov02,Baranov04,Baranov04a,Zhao10,Baranov08} and two dimensions \cite{Brunn08,Cooper09,Levinson11} at sufficiently low temperatures. In these systems, the anisotropy of the superfluid order parameter causes a major change in properties  as compared to the case of a two-component BCS superfluid, dominated by van der Waals interactions, where the  superfluid order parameter is isotropic ($s$-wave). It is expected that the anisotropic gap will lead to significant differences in the properties of single and multiple vortex states in these systems. For example, Levinson {\it et al.} \cite{Levinson11} have proposed a scheme to construct a topological $p_x+ip_y$ superfluid phase in a quasi-two-dimensional single component dipolar Fermi gas  which can support vortices which carry zero energy Majorana modes on their cores \cite{Gurarie07,Read00,Stern08}. However, to date, there has been very little research into the properties of vortices and vortex lattices of such states.  Assuming that such a state can be experimentally achieved there is a significant opportunity to revisit much of what has been discussed in this review within the context of vortices and vortex lattices in the BCS state of a dipolar Fermi gas.  

There have been extensive studies of the properties of rotating single-component fermionic quantum ferrofluids \cite{Baranov05c,Osterloh07,Baranov08a,Eriksson12,Jheng13,Ancilotto15} away from the BCS superfluid transition. These studies have primarily focused on the emergence of fractional quantum Hall states in the $\Omega \rightarrow \omega_{\perp}$ regime. Specifically, it has been shown \cite{Baranov05c,Osterloh07} that for a filling fraction of $\nu=2\pi l_{\perp}^2 n_0=1/3$ the many-body state is well described by the Laughlin wave function with a significant gap between the ground and the excited states.  Further studies \cite{Osterloh07,Baranov08a,Jheng13} have shown that as the filling fraction is reduced further ($\nu < 1/7$) Wigner crystal \cite{Wigner34} states may emerge. To our knowledge, these studies have all been carried out in the regime where the plane of rotation is perpendicular to the orientation of the dipoles. As such an interesting question to ask may be how do such states change when the dipole orientation has a component in the plane of rotation.

\subsubsection{Berezinskii-Kosterlitz-Thouless transition}
In  a strictly  homogeneous two-dimensional system at finite temperature, long range phase coherence cannot be established and condensation will not  occur.   Superfluidity can still occur at very low temperatures where quasi-long-range order exists \cite{Hadzibabic11,Bagnato91}, but even this is destroyed when the temperature is raised through the Berezinskii-Kosterlitz-Thouless  transition  \cite{Berezinski72,Kosterlitz73} at which point spatial correlations change from power law  to exponential  decay. Physically, this transition can be thought of as occurring when virtual vortex-antivortex pairs unbind and there is a proliferation of free vortices. The  Berezinskii-Kosterlitz-Thouless  and  BEC  transition have been studied in trapped ultracold gases  via observations of phase defects \cite{Habzibabic06}, vortices \cite{Tung10}  and changes in the density profile due to the onset of superfluidity \cite{Claude09}. The Berezinskii-Kosterlitz-Thouless transition  in the presence of dipolar interactions has been studied using, for a homogeneous system, Monte Carlo methods \cite{Filinov10}, the mean-field Hartree-Fock-Bogoliubov-Popov model \cite{Ticknor12}, the dipolar XY-model \cite{Vasiliev14} and most recently renormalized Hartree-Fock theory \cite{Wu16}.  A simple picture underpinning our understanding of the Berezinskii-Kosterlitz-Thouless transition comes from asking the question: what is the energy required to separate a vortex-antivortex pair? For vortex-antivortex pairs in a uniform two-dimensional non-dipolar BEC (of two-dimensional background density $n_0$), the energy of a pair, separated by a distance $d$ and calculated from hydrodynamical arguments, is approximately given by $V(d)=2\pi\hbar^2n_0 \ln(d/\xi)/m$.  The critical temperature associated with the transition is given by the relation $2\pi \hbar^2 n_0/(mk_BT)=4$.  This is calculated by determining the average distance between the pairs,
\begin{eqnarray}
\langle d^2 \rangle =\frac{\int_{\xi}^{\infty} d^3 e^{-\beta V(d)}\,{\rm d} d}{\int_{\xi}^{\infty}  d e^{-\beta V(d)}\,{\rm d} d}=\xi^2\frac{\Gamma-2}{\Gamma-4},
\label{eq:BKT}
\end{eqnarray}  
where we have used the short-hand notation $\Gamma=2\pi \hbar^2 n_0/(mk_BT)$.  This result diverges at $\Gamma=4$, signifying the Berenzinksii-Kosterlitz-Thouless transition. 

In this simple model the inclusion of dipolar interactions adds three complications. The first complication is that as dipolar interactions are introduced $\xi$ will change. The second complication is, as seen in Section \ref{sec:dynamics}, the interaction between a vortex and an antivortex is modified in the presence of dipolar interactions due to the absence of dipoles in the vortex cores. As such the energy scaling of the pair with separation, i.e. $V(d)$, is changed. The third complication arises if the dipole alignment has some component in the two-dimensional plane of the gas. In this case the interaction between the vortex-antivortex pair is no longer just a function of the distance between them, it also depends on the in-plane angle of the pair relative to the polarization direction, i.e. $V(d) \rightarrow V(d, \eta)$ [see Figure~\ref{fig:angle}(b)]. The above generalisations can be incorporated into Eq.~(\ref{eq:BKT})  by considering the following analysis  \cite{Camp}. To quantify if dipolar interactions have a significant effect on the Berezinskii-Kosterlitz-Thouless transition temperature we consider a simple model for the interaction between a vortex-antivortex pair. Specifically, the vortices carry a dipole moment $\propto(0,\sin \alpha, \cos \alpha)$. If the separation between the vortices is ${\bf d}=d(\cos \eta, \sin \eta,0)$ then the interaction between the vortex and the anti-vortex can be written as
\begin{eqnarray}
\frac{V(d,\eta)}{k_BT}=\Gamma \ln \left(\frac{d}{\xi}\right)-\lambda \left(\frac{\xi}{d}\right)^3 f(\alpha,\eta).
\end{eqnarray}
For simplicity, we have assumed that the interaction between vortices arising from the absence of dipoles in their cores is $\propto 1/d^3$. Given the results in Section~\ref{subsubsection:vortexpotential}, which finds that due to the absence of dipoles in a vortex core $\Phi_{\rm dd}$ has a term $\propto \ln(d)/d^3$, this represents something of a simplification. From this 
\begin{eqnarray}
\langle d^2 \rangle &=&\frac{\int_{\xi}^{\infty} d^{3-\Gamma}\,{\rm d} d \int_0^{2\pi} \exp\left[\lambda f(\alpha,\eta)\left(\frac{\xi}{d}\right)^3\right] {\rm d} \eta}{\int_{\xi}^{\infty} d^{1-\Gamma}\,{\rm d} d \int_0^{2\pi} \exp\left[\lambda f(\alpha,\eta)\left(\frac{\xi}{d}\right)^3\right] {\rm d}\eta} \\
&=&\xi^2 \left[\frac{\sum_{n=0}^{\infty} a_n \lambda^n \left(\Gamma +3n-4\right)^{-1}}{\sum_{n=0}^{\infty} a_n \lambda^n \left(\Gamma +3n-2\right)^{-1}}\right], \label{eq:BKT_dip}
\end{eqnarray} 
where
\begin{eqnarray}
a_n=\frac{1}{n!}\int_0^{2\pi} f^n(\alpha,\eta) {\rm d}\eta.
\end{eqnarray}
When $\lambda=0$ the result obtained from Eq.~(\ref{eq:BKT}) is regained, i.e. $\langle d^2 \rangle=\xi^2(\Gamma-2)/(\Gamma-4)$, with the  Berezinskii-Kosterlitz-Thouless  transition arising from the divergence at $\Gamma=4$. However, the inclusion of  the additional power law interaction between the vortex and antivortex, arising from the absence of dipoles in the region of the cores, does not fundamentally change this result. Specifically, the mean-square separation, given by Eq.~(\ref{eq:BKT_dip}), diverges first at $\Gamma=4$, i.e. this simple analysis implies that although the dipolar interaction can change the interaction between vortex-antivortex pairs the long-range hydrodynamic interaction always wins, suggesting that the Berezinskii-Kosterlitz-Thouless  transition is unaffected by dipolar interactions. This is consistent with more detailed analysis presented in \cite{Filinov10,Vasiliev14} which found, at best, a weak dependence of the Berezinskii-Kosterlitz-Thouless  transition on the strength of the dipolar interactions. For example, working within the XY-model Vasiliev \textit{et al} \cite{Vasiliev14} concluded that polarization of the system (i.e.\ the generation of additional vortex-antivortex pairs between any two points) screens out the long-range interactions. However, in our opinion there is still room for more work in this area such as considering finite-size effects and the true form of the {\it dipolar} interaction between a vortex-antivortex pair in order to firmly establish the role of dipolar interactions on this phase transition in physical systems.

\subsubsection{Vortex lattices in the supersolid phase}
A supersolid phase is characterised by the spontaneous formation of a periodic structure in a system which also supports superfluidity, i.e.\ both on- and off-diagonal long-range order in the single particle density matrix. For more than half a century considerable effort has been expended theorizing about the possibility of supersolidity in condensed matter systems  \cite{Gross57,Andreev69,Thouless69,Legget70,Chester70}. In particular, the investigation of a supersolid phase in  $^4$He \cite{Balibar12,Nozieres06,Nozieres09} has been the primary focus of the more recent work. However, the most credible claim for the experimental observation  of a supersolid phase in $^4$He \cite{Kim04} has now been withdrawn \cite{Kim12}. The relatively recent realization of dipolar cold atoms and molecules now provides an alternative venue for investigating supersolid phases. Apart from the strongly correlated dipolar systems  \cite{Buchler07,Astrakharchik07,Cinti2010,Macia12,Macia14,Lu15} cited in Section \ref{sec:Clas_Ferro},  extended Bose-Hubbard lattice models are also thought to support supersolidity \cite{Goral02,Kovizhin05,Sengupta05,Scarola05,Scarola06,Torre06,Schmidt08,Sen08,Trefzger09,Pollet10,Dutta15}. The common ingredient in both sets of investigations is the presence of long-range interactions. There have also been various works focusing on how an artificial gauge field in a dipolar BEC influences the boundaries between Mott-insulator and supersolid regimes \cite{Sachdeva10,Sachdeva12}, and how staggered fluxes lead to supersolid phases with staggered vortex phases \cite{Tieleman11}. However, there does not appear to have been much investigation into the structural properties of vortices \cite{Sachdeva14} and vortex lattices in the supersolid state.

\subsubsection{The Onsager vortex phase transition}
By studying a  two-dimensional point vortex model Onsager predicted  that  negative  temperature states may be relevant for two-dimensional fluids \cite{Onsager49}.    While intended as a model of two-dimensional fluids in general, Onsager noted that the model was potentially particularly relevant for two-dimensional superfluids, whose vortices have quantized circulation and uniform size. Simulations of the two-dimensional GPE have shown how it is possible to dynamically evolve to the negative temperature Onsager vortex state \cite{Simula14,Billam14}.  Starting with a random configuration of vortices and antivortices one might expect the vortices and antivortices to evaporate from the BEC via pairwise annihilation and re-thermalization of the emitted sound, simply resulting in a BEC with an increased temperature. However, Simula {\it et al.} \cite{Simula14} showed that only some vortices annihilate, and the remaining vortices self-organise into two ordered clusters of like-sign circulation, which represent the Onsager vortices. This outcome was found to be the result of  the evaporative heating of quantized vortices via vortex-antivortex pair annihilation, leaving the remaining vortices to re-thermalize to a state with higher energy per vortex. This process drives the vortex component of the superfluid to ever higher energies, leading to the Onsager vortex phase transition. The final Onsager vortex state emerges as a clustering of vortices and antivortices. The dynamical process which leads to this final state is non-trivial, but is underpinned by the interaction between the constituent vortices and antivortices in the system. Introducing dipolar interactions, as shown in Section \ref{sec:dynamics}, can significantly modify this interaction and as such may result in significant changes to the Onsager vortex phase transition. As such we suspect that dipolar interactions may offer the opportunity to significantly modify the Onsager vortex phase transition.

\section{Acknowledgements}
AMM acknowledges support by the Australian Research Council (Grant No. DP150101704), DOD acknowledges support from the Natural Sciences and Engineering Research Council (Canada), and NGP acknowledges support by the Engineering and Physical Sciences Research Council (Grant No. EP/M005127/1).   We thank A.L. Fetter for providing extensive feedback regarding this review.

\end{document}